\documentclass{article}
\usepackage[english]{babel}
\usepackage[utf8]{inputenc}
\usepackage{johd}
\usepackage{amsmath}
\usepackage{amsfonts}
\usepackage{todonotes}
\usepackage{bbm}
\usepackage[ruled,linesnumbered]{algorithm2e}
\RestyleAlgo{ruled}
\usepackage{graphicx}
\usepackage{appendix}
\usepackage{enumitem}

\newtheorem{prop}{Proposition}

\usepackage{tikz}
\usetikzlibrary{shapes.geometric, arrows, shadows.blur, shapes.multipart}
\usetikzlibrary{arrows.meta}
\tikzstyle{blank} = [rectangle, minimum width=5cm, minimum height=1cm, text centered, draw=white]
\tikzstyle{process} = [rectangle, minimum width=1.7cm, minimum height=1cm, text centered, draw=black, fill=orange!20, blur shadow={shadow blur steps=5}]
\tikzstyle{arrow} = [thick,-{Latex[length=2.5mm]}, >=stealth]
\tikzstyle{arrow2} = [thick, -{Latex[length=2.5mm]}, >=stealth, dashed]

\pgfdeclarelayer{bg}    
\pgfsetlayers{bg,main}  

\newcommand{\Pa}{\mathrm{pa}}
\newcommand{\Fa}{\mathrm{fa}}
\newcommand{\Ch}{\mathrm{ch}}
\newcommand{\calD}{\mathcal{D}}
\newcommand{\calN}{\mathcal{N}}
\newcommand{\Do}{\mathrm{do}}

\title{Bayesian Causal Inference in Doubly Gaussian DAG--probit Models}

\author{Rasool Tahmasbi$^{1}$$^{*}$, Keyvan Tahmasbi$^{b}$ \\
        \small $^{a}$University of Colorado, Boulder, USA \\
        \small $^{b}$Department of Statistics, Shahid Beheshti University, Tehran, Iran \\
        \small $^{*}$Corresponding author: Rasool Tahmasbi; \tt{Rasool.Tahmasbi@Colorado.edu}
}

\date{} 

\begin{document}

\maketitle

\begin{abstract} 
\noindent 
We consider modeling a binary response variable together with a set of covariates for two groups under observational data. The grouping variable can be the confounding variable (the common cause of treatment and outcome), gender, case/control, ethnicity, etc. Given the covariates and a binary latent variable, the goal is to construct two directed acyclic graphs (DAGs), while sharing some common parameters. The set of nodes, which represent the variables, are the same for both groups but the directed edges between nodes, which represent the causal relationships between the variables, can be potentially different. For each group, we also estimate the effect size for each node. We assume that each group follows a Gaussian distribution under its DAG. Given the parent nodes, the joint distribution of DAG is conditioanlly independent due to the Markov property of DAGs. We introduce the concept of Gaussian DAG–probit model under two groups and hence doubly Gaussian DAG–probit model. To estimate the skeleton of the DAGs and the model parameters, we took samples from the posterior distribution of doubly Gaussian DAG–probit model via MCMC method. We validated the proposed method using a comprehensive simulation experiment and applied it on two real datasets. Furthermore, we validated the results of the real data analysis using well-known experimental studies to show the value of the proposed grouping variable in the causality domain.

\end{abstract}

\noindent\keywords{Bayesian Causal Inference; Causal Effects; $\Do$ Calculus; Graphical Models; MCMC; Modified Cholesky Decomposition; Observational study; }\\


\section{Introduction}
The goal of etiological research is to uncover causal effects, whilst prediction research aims to predict an outcome with the best accuracy. Causal and prediction research usually require different methods, and yet their findings may be meaningless if the right method is not used. Machine-learning (ML) systems have made astounding progress in analyzing data patterns, but ML algorithms cannot tell whether a crowing rooster makes the sun rise, or the other way around (\cite{Pearl2020AI}). So by just training a model on historical data we cannot say anything about the causes or the direction of causation, which is crucial as well. More granular distiction between the causality asnd prediction methods can be found in \cite{ramspek2021prediction} and \cite{gische2021forecasting}.

Type of data availibility is also important for picking the right causality model. The gold standard data for answering the "what if" questions, which are related to the causal effect discovery, is \emph{randomized} data. However, it is often not feasible to run randomized experiments due to the ethical issues, costs, or running time. In this case, causal inferences can be obtained from \emph{observational} data (\cite{hernan2010causal}).

Estimating causality from observational data is essential in many data science questions but can be a challenging task.Some mathematical models such as Directed Acyclic Graph (DAG) and Structural Equation Model (SEM) are important tools to infer the causal effects based on observational data. DAGs have been extensively used to construct statistical models embodying conditional independence relations in graphical models (\cite{lauritzen1996graphical}) while causal DAGs will be instrumental to define the notion of causal effect (\cite{pearl2000models}). Some models such as DoWhy (\cite{dowhypaper}) assume that the causal graph is known and the goal is to estimate the effect sizes.

Gaussian graphical models (GGM) are also used in many causal discovery problems. GGM is a statistical framework defined with respect to a graph, in which the nodes index a collection of jointly Gaussian random variables and the edges represent the conditional independence relations among the variables. A number of papers have studied covariance estimation in the context of GMM selection. For example, \cite{chen2016asymptotically} estimated the covariate-adjusted GMM using asymptotic theory. They showed that for each finite subgraph, their estimator is asymptotically normal and efficient. \cite{cai2013covariate} introduced a sparse high dimensional multivariate regression model for studying conditional independence relationships among a set of genes adjusting for possible genetic effects. They presented a covariate-adjusted precision matrix estimation method using a constrained $\ell_1$ minimization.

The problem of latent variables, where all the variables (both observed and
latent) are jointly Gaussian, is studied by \cite{Venkat2012}. \cite{Venkat2012} studied the case when the graph is sparse and there are a few additional latent components in the DAG structure. They proposed a convex program based on $\ell_1$ and nuclear norm regularized maximum likelihood for latent-variable graphical model selection. \cite{wu2017graphical} proposed a method for learning latent variable graphical models via $\ell_1$ and trace penalized D-trace loss. Gaussian latent variables also have many applications in the generative models such as variational autoencoder (VAE) models, introduced by \cite{kingma2013auto}.
Learning individual-level causal effects (ILCE) from observational data, which is important for the policy makers, is studied by \cite{louizos2017causal}. Examples of ILCE include understanding the effect of medications on a
patient's health, or of teaching methods on a student’s chance of graduation.
Their approach is based on the VAE to estimate the causal effects of the latent variables on large datasets. For a comprehensive review on theoretical properties and optimalities of the estimation of structured covariance and precision matrices, see \cite{cai2016estimating} and the references therein.

Motivated by the probit regression model, \cite{guo2015graphical} introduced the probit graphical model. \cite{castelletti2021bayesiana} extended the probit graphical models by introducing DAG--probit models. For DAG--probit model, they considered a binary response which is potentially affected by a set of continuous variables. Their model assumes that there is just one DAG.

In reality, there are many cases when we should model our data for different groups, separately. The grouping variable can be gender, different ethnicities, or case/control studies. For example, is well known that the physiological differences between men and women affect drug activity, including pharmacokinetics and pharmacodynamics. If we pick a model that does not properly take genders into account, then the results will be bias, a serious problem which is called \emph{gender bias in research} (see \cite{holdcroft2007gender} and \cite{aragon2023gender}). As another example, our metabolism changes with age (\cite{pontzer2021daily}) and therefore a method that can model the age differences is preferred. The results of comparing the outcome between two groups can sometimes be confounded and even reversed by an unrecognised third variable. This concept is known as \emph{Simpson's Paradox} and confounding variable can be gender, ethnicity, etc.

In this paper, we introduce the \emph{doubly Gaussian DAG–probit model} in Section \ref{sec:methodology} by allowing groups to have different DAGs while some of the model parameters are shared between the groups. Allowing a model with the flexibility to have different structures for different groups can potentially take care of Simpson paradox and can reduce gender bias in research. The proposed model is drived for binary grouping variables but can be extended for any non-binary grouping variable as well. The shared parameteres can be the common edges between the two groups, a common cut-off parameter or nodes variance. As an example, \cite{qiao2020doubly} studied the network of EEG (electroencephalogram) signals on alcoholic and control groups. They expressed that since the graphical structures for alcoholic and non-alcoholic groups share some common edges, it is advantageous to jointly estimate two networks. For estimating the heritability using geneome data, it is common to assume that the effect of SNPs (single-nucleotide polymorphism) are the same for all individuals (check our paper \cite{evans2018comparison}). Therefore, we assume that the effect of a gene for all individuals (even in different groups) is the same and changes in the phenotyps are due the differences in the gene network, which enables us to jointly estimate this parameter from all groups.

Choice of priors can speed up the MCMC algorithm and simplify the posterior formula. Our priors for the model parameters are presented in Section \ref{sec:bayesian}. After computing the posteriors for the parameters, we present our MCMC algorithm in Section \ref{sec:MCMC}. \cite{dawid2022effects} expressed and contrasted two distinct problem areas for statistical causality: studying the likely effects of an intervention (effects of causes) and studying whether there is a causal link between the observed exposure and outcome in an individual case (causes of effects). The effect of interventions in terms of the observed probabilities using \emph{do calulus} is computed in section \ref{sec:causal_effect} and causes of effects can be estimated from the last step of our MCMC Algotithm.

To assess the performance of the proposed method, a comprehensive simulation study is performed in Section \ref{sec:simulation}. We provide several evaluation metrics to check the MCMC algorithm and model performance. We apply or method on two well-known real datasets and illustrate the results in Section \ref{sec:real_data}. We use the genome network data of breast cancer for the first example and compare our results with the well-known published genetics papers to validated our outputs in Section \ref{sec:breast_cancer}. For the second real data analysis, we study the impact of airborne particles on the cardiovascular mortality rate (CMR) in Section \ref{sec:cmr}. The finals points for discussion are presented in Section \ref{sec:conclusion}. Some proof of the posteriors together with more simulation results are provided in Appendix.


\section{Methodology}\label{sec:methodology}

Let $\calD = (V,E)$ be a DAG, where $V = \{1, \dots, q\}$ is a set of vertices (or nodes) and $E \subseteq V \times V$ a set of edges whose elements are $(u, v) \equiv u \to v$, such that if $(u, v) \in E$
then $(v, u) \notin E$. In addition, $\calD$ contains no cycles. For a given node $v$, if $u \to v \in E$, we say that $u$ is a \emph{parent} of
$v$ (conversely $v$ is a \emph{child} of $u$). The parent set of $v$ in $\calD$ is denoted by $\Pa(v)$, the set of children by $\Ch(v)$. Moreover, we denote by $\Fa(v) = v \cup \Pa(v)$ the \emph{family} of $v$ in $\calD$. Finally, we say that a DAG is complete if all vertices are joined by an edge. For further theory and notation on DAGs we refer to \cite{lauritzen1996graphical}.

A DAG can be expressed by a probabilistic model via conditional dependence structure between its random variables. We consider a collection of random variables $(X_1, \dots , X_q)$ and assume that their joint probability density function $f(\boldsymbol{x})$ is Markov w.r.t. $\calD$, so that it admits the following factorization
\begin{equation}
    f(x_1,\dots,x_q) = \prod_{j=1}^q f\left(x_j| \boldsymbol{x}_{\Pa(j)}\right).
    \label{eq1}
\end{equation}

\subsection{Gaussian Graphical Models}
Graphical models provide a suitable language to decompose many complex real-world problems through conditional independence constraints. Gaussian Graphical Models (GGMs) are extensively used in many research areas, such as genomics, proteomics, neuroimaging, and psychology, to study the partial correlation structure of a set of variables. This structure is visualized by drawing an undirected network, in which the variables constitute the nodes and the partial correlations the edges. 

GGMs are tightly linked to precision matrices. Suppose $\boldsymbol{X}=(X_1, \dots ,X_q)'$ follows a multivariate Gaussian distribution $\mathcal{N}_q(\boldsymbol{\mu}, \boldsymbol\Sigma)$ with dimention $q$. Without loss of generality, we can assume the mean of $X_i$ is zero. The precision matrix $\boldsymbol\Omega =\boldsymbol\Sigma^{-1}$ with $\boldsymbol\Omega=[\omega_{ij}]_{i,j=1,\dots,q}$ describes the graphical structure of its corresponding Gaussian graph. If the $(i, j)$th entry of the precision matrix $\omega_{ij}$ is equal to zero, then $X_i$ and $X_j$ are independent conditioning on all other variables $X_k, k\neq i, j$. Correspondingly, no edge exists between $X_i$ and variable $X_j$ in the graphical structure of Gaussian graphical model. If $\omega_{ij}\neq 0$, then $X_i$ and $X_j$ are conditionally dependent and they are therefore connected in the graphical structure. Partial correlation can be written in terms of the precision matrix. For the precision matrix $\boldsymbol{\Omega}$, the partial correlation between two variables $X_i$ and $X_j$ given all other nodes, i.e., 
 $V\setminus\{i,j\}$ is given by
\begin{equation}
    \rho_{ij} :=\rho_{ij \,|\, V \setminus \{i,j\}} = -\frac{\omega_{ij}}{\sqrt{\omega_{ii} \omega_{jj}}}.
\label{eq:partial}
\end{equation}
\cite{lafit2019partial} used a partial correlation screening approach for controlling the false positive rate in sparse GMMs.
Under Gaussian assumption with mean zero, i.e., $X_1,\dots,X_q | \boldsymbol\Omega \sim 
 \calN_q(\boldsymbol{0},\boldsymbol\Omega^{-1})$, we can rewrite (\ref{eq1}) as
\begin{equation}
    f(x_1,\dots,x_q|\boldsymbol\Omega) = \prod_{j=1}^q f_{\calN}\left({x}_j|\mu_{j}(\boldsymbol{x}_{\Pa(j)}),\sigma^2_{j} \right),
\label{eq:likex}
\end{equation}
where $f_{\calN}(.|\mu,\sigma^2)$ denotes the normal density having mean $\mu$ and variance $\sigma^2$. In this section, we will show how to compute $\mu_{j}(\boldsymbol{x}_{\Pa(j)})$ and $\sigma^2_{j}$.

For a given DAG, we can always reorder the nodes using topological sorting to have a \emph{parent ordering} of the nodes which numerically relabels the nodes so that if $i\to j$ ($i$ is a child of $j$) then $i<j$, for any $i,j\in V$ (\cite{knuth1997art}). Clearly node $1$ will not have any children.

Equation (\ref{eq:likex}) can be written as a \emph{structural equation model}
\begin{equation}
    \boldsymbol{L}' \boldsymbol{X} = \boldsymbol{\epsilon},
\label{eq:sem}
\end{equation}
where $\boldsymbol{X}=(X_1, \dots ,X_q)'$ and $\boldsymbol{L}=[L_{ij}]_{i,j\in 1, \dots, q}$ is a lower-triangular matrix of coefficients with $L_{ij}\in \mathbb{R}$, $L_{ii}=1$ and $L_{ij}\neq 0$ if and only if $i\rightarrow j$. $\boldsymbol{L}$ is a lower-triangular matrix because of the parent ordering. So $L_{ij}=0$ if and only if $i>j$ or $i\not\rightarrow j$. Moreover, $\boldsymbol{\epsilon}$ is a $q\times 1$ vector of error terms, $\boldsymbol{\epsilon}\sim \calN_q(\boldsymbol{0},\boldsymbol{D})$, where $\boldsymbol{D} = \mathrm{diag}(\boldsymbol\sigma^2)$ and $\boldsymbol\sigma^2$ is the $q\times 1$ vector of conditional variances whose $j$-th element is 
$\sigma^2_j = \mathrm{Var}(X_j | \boldsymbol{x}_{\Pa(j)},\boldsymbol\Omega)$.

Taking variance on both sides of (\ref{eq:sem}), we can show that
\begin{equation}
    \boldsymbol{\Omega} = \boldsymbol{L}\boldsymbol{D}^{-1}\boldsymbol{L}',
    \label{eq:decom}
\end{equation}
which is the \emph{modified Cholesky decomposition} of $\boldsymbol{\Omega}$ (see \cite{pourahmadi2007cholesky}). In fact any positive definite matrix $\boldsymbol\Omega$ can be uniquely decomposed as $\boldsymbol\Omega=\boldsymbol{L}\boldsymbol{D}^{-1}\boldsymbol{L}'$, where $\boldsymbol{L}$ is a lower triangular matrix with unit diagonal entries, and $\boldsymbol{D}$ is a diagonal matrix with positive diagonal entries (see, e.g., \cite{golub2013matrix}). \cite{rothman2010new} presented a new approach for the Cholesky-based covariance regularization in high dimensions. \cite{guo2011joint} considered an automated approach using the lasso to estimate sparse graphical models by selecting sets of edges common to all groups, as well as group-specific edges.

Since the joint probability density function $f(\boldsymbol{x})$ is Markov w.r.t. $\calD$, we can use multivariate Gaussian conditional distribution to compute $\mu_{j}(\boldsymbol{x}_{\Pa(j)})$ and $\sigma^2_{j}$ in (\ref{eq:likex}). For any given $q\times q$ matrix $\boldsymbol{A}$, denote by $\boldsymbol{A}_{\preceq j \succeq}$ the $|\Pa(j)|\times|\Pa(j)|$ submatrix of $\boldsymbol{A}$ with indexes coresponding to $\Pa(j)$, $\boldsymbol{A}_{\preceq j \succ}$ the $|\Pa(j)|\times 1$ submatrix of $\boldsymbol{A}$, and $\boldsymbol{A}_{\prec j \succ}:=\boldsymbol{A}_{jj}$ where the parent nodes, $\Pa(j)$, can be extracted from $\calD$. Therefore,
\begin{equation}
    \mu_{j}(\boldsymbol{x}_{\Pa(j)}) = -\boldsymbol{L}_{\preceq j \succ} \boldsymbol{x}_{\Pa(j)},
    \label{eq:mu-j}
\end{equation}
where $\boldsymbol{L}_{\preceq j \succ} = - \boldsymbol\Sigma_{\preceq j \succeq}^{-1}  \boldsymbol\Sigma_{\preceq j \succ}$ and
\begin{equation}
    \sigma^2_{j} = \boldsymbol\Sigma_{\prec j \succ} -  \boldsymbol\Sigma_{\preceq j \succ}'  \boldsymbol\Sigma_{\preceq j \succeq}^{-1}  \boldsymbol\Sigma_{\preceq j \succ}.
    \label{eq:sigma-j}
\end{equation}
The proof is based on the Gaussian conditional distributions of node $j$ given its parent nodes, $\Pa(j)$. For the cases when $\Pa(j)$ is empty set, we set $\mu_{j}(\boldsymbol{x}_{\Pa(j)})=\boldsymbol{0}$ and $\sigma^2_{j}=\boldsymbol\Sigma_{\prec j \succ}$.

\subsection{DAG--Probit Model}
\label{sec:DAG-Probit-Model}

DAG--Probit model is defined in \cite{castelletti2021bayesiana}. They assumed that $X_1$ is a latent variable and the binary variable $Y\in\{0,1\}$ is observed. For a given threshold $\theta\in \mathbb{R}$, define
\begin{equation}
    Y = \left\{
      \begin{array}{ll}
        0 & \mathrm{ if ~} X_1 < \theta, \\
        1 & \mathrm{ if ~} X_1 \geq \theta. \\
      \end{array}
    \right.
\label{eq:theta}
\end{equation}
Without loss of generality, they assumed that the variance of $X_1$ is 1, i.e., $\sigma^2_1=1$. In reality, $X_1$ is not observed and we are interested in distribution of $(Y, X_2, \dots, X_q)$.
By including the latent variable $X_1$ in the model and using (\ref{eq:likex}) and (\ref{eq:theta}), the joint density of $(Y,X_1, \dots, X_q)$ becomes
\begin{equation}
    f(y, x_1,\dots,x_q| \calD, \boldsymbol{D}, \boldsymbol{L}, \theta) = \prod_{j=1}^q f_{\calN}\left({x}_j|\mu_{j}(\boldsymbol{x}_{\Pa(j)}),\sigma^2_{j} \right) . \mathbbm{1}(\theta_{y_i-1} < x_1 \leq \theta_{y_i}),
\label{eq:likeyx}
\end{equation}
where the notation $\theta_{-1}:=-\infty, \theta_0:=\theta$ and $\theta_1:=\infty$ is used.

For a sample size of $n$ with observations $(y_i, x_{i,1}, \dots , x_{i,q})$, the \emph{augmented} likelihood can be written as
\begin{align}
    f(\boldsymbol{y}, \boldsymbol{X}| \calD, \boldsymbol{D}, \boldsymbol{L}, \theta) &=
    \prod_{i=1}^n f(y_i, x_{i,1}, \dots , x_{i,q} | \calD, \boldsymbol{D}, \boldsymbol{L}) . \mathbbm{1}(\theta_{y_i-1} < x_{i,1} \leq \theta_{y_i}) \nonumber \\
    & = \prod_{j=1}^q f_{\calN_n}\left(\boldsymbol{X}_j| - \boldsymbol{X}_{\Pa(j)} \boldsymbol{L}_{\preceq j \succ}, \sigma^2_{j} \boldsymbol{I}_n \right) . \prod_{i=1}^n \mathbbm{1}(\theta_{y_i-1} < x_{i,1} \leq \theta_{y_i})
    \label{eq:likeyx_bold}
\end{align}
where $\boldsymbol{y}=(y_1,\dots,y_n)'$, $\boldsymbol{X}:=(\boldsymbol{X}_1, \dots, \boldsymbol{X}_q)$ is the $n\times q$ augmented data matrix, $\boldsymbol{X}_s$ is the submatrix of $\boldsymbol{X}$ with columns $s$ and  $\boldsymbol{X}_{-1}:= (\boldsymbol{X}_2, \dots, \boldsymbol{X}_q)$.

\subsection{Doubly Gaussian DAG-Probit Model}
\label{sec:Doubly-Gaussian-DAG-Models}

We assume there are two groups of data, $\boldsymbol{X}^{(1)}$ and $\boldsymbol{X}^{(2)}$ with the coresponding DAG's $\calD^{(1)} = (V,E^{(1)})$ and $\calD^{(2)} = (V,E^{(2)})$. The set of vertices are the same for both groups but potentially they can have different sets of edges.
For each group $k \in \{1,2\}$,  $\boldsymbol{X}^{(k)}$ follows a Gaussian distribution with sample size $n_k$. Furthermore, we also have observed the binary random vector $\boldsymbol{y}^{(k)}$ of size $n_k$, related to the latent variables $X_1^{(k)}$ defined in (\ref{eq:theta}). We assume that $\sigma_j^2$ is equal for both groups, i.e., $\mathrm{Var}(X_j | \boldsymbol{x}^{(1)}_{\Pa(j)})=\mathrm{Var}(X_j | \boldsymbol{x}^{(2)}_{\Pa(j)})$, therefore $\boldsymbol{D}:=\boldsymbol{D}^{(1)}=\boldsymbol{D}^{(2)}$. The skeleton of $\calD^{(k)}$ can have effect on the $\boldsymbol{L}^{(k)}$ as well, because $L_{ij}=0$ if and only if $i\not\rightarrow j$, so they can be different as well as the precision matrices $\boldsymbol\Omega^{(1)}$ and $\boldsymbol\Omega^{(2)}$.


The augmented likelihood for the doubly Gaussian DAG-probit model can be written as
\begin{align}
     & f\left(\boldsymbol{y}^{(1)}, \boldsymbol{y}^{(2)}, \boldsymbol{X}^{(1)}, \boldsymbol{X}^{(2)} \big| \calD^{(1)}, \calD^{(2)}, \boldsymbol{D}, \boldsymbol{L}^{(1)}, \boldsymbol{L}^{(2)}, \theta\right)
     \nonumber \\
     = & \prod_{k=1}^2 f\left(\boldsymbol{y}^{(k)},  \boldsymbol{X}^{(k)} \big| \calD^{(k)},  \boldsymbol{D}, \boldsymbol{L}^{(k)}, \theta\right)
     \nonumber \\
     = & \prod_{j=1}^q f_{\calN_{n_1}} \left(\boldsymbol{X}^{(1)}_j \Big| \hat{\boldsymbol{\mu}}^{(1)}, \sigma^2_{j} \boldsymbol{I}_{n_1} \right) f_{\calN_{n_2}}\left(\boldsymbol{X}^{(2)}_j \Big| \hat{\boldsymbol{\mu}}^{(2)}, \sigma^2_{j} \boldsymbol{I}_{n_2} \right)
    \times \prod_{i=1}^{n_1} \mathbbm{1}(\theta_{y_i-1} < x_{i,1}^{(1)} \leq \theta_{y_i}) \prod_{i=1}^{n_2} \mathbbm{1}(\theta_{y_i-1} < x_{i,1}^{(2)} \leq \theta_{y_i}),
    \label{eq:like_doubly}
\end{align}
where $\hat{\boldsymbol{\mu}}^{(k)} := - \boldsymbol{X}^{(k)}_{\Pa(j)} \boldsymbol{L}^{(k)}_{\preceq j \succ}$ and we assumed the cut-off parameter $\theta$ is the same for both groups.


\section{Bayesian Inference}\label{sec:bayesian}
This section concerns priors for ($\calD^{(1)}, \calD^{(2)}, \boldsymbol{D}, \boldsymbol{L}^{(1)}, \boldsymbol{L}^{(2)}, \theta$) and we review different models in the litrature. We try to pick the conjugate priors to make the equations simpler and the computations faster. In section \ref{sec:MCMC}, we will compute the posteriors for each of the parameters.

\subsection{Prior on DAG $\calD^{(k)}$}

Random graph famously studied by 
 Erd{\H{o}}s and R{\'e}nyi \cite{erdHos1960evolution}. For generating random graphs on DAGs, we refer to \cite{karrer2009random}.
Let $\boldsymbol{A}^\calD$ be the 0--1 adjacency matrix of the skeleton of the parent ordering DAG $\calD$ whose $(i,j)$th element is denoted by $\boldsymbol{A}_{ij}^\calD$. Clearly, $\boldsymbol{A}_{ij}^\calD=0$ for $i\leq j$ because of the parent ordering and no self--loops.

Assuming the probability of edge inclusion is $\xi$, we assign a Bernoulli prior independently to each element $\boldsymbol{A}_{ij}^\calD$, that is
\begin{equation}
    f(\boldsymbol{A}^\calD) = \xi^{|\boldsymbol{A}^\calD|} (1-\xi)^{\frac{q(q-1)}{2}-|\boldsymbol{A}^\calD|},
\label{eq:prior_DAG}
\end{equation}
where $|\boldsymbol{A}^\calD|$ denotes the number of edges in the skeleton. \cite{cao2019posterior} also used 
independent identically distributed Bernoulli random variables as the prior on the probability of edges which  corresponds to an Erd{\H{o}}s--R{\'e}nyi type of distribution.

For the prior on DAG's $\calD^{(k)}$, we independentely set 
\begin{equation}
f(\calD^{(k)})\propto f(\boldsymbol{A}^{\calD^{(k)}}),
\label{eq:prior_DAG_k}
\end{equation}
where $k \in \{1,2\}$.

\subsection{Prior on the Modified Cholesky Decomposition}

Since $\boldsymbol{L}_{ij}=0$ eighter if there is no edge from $i$ to $j$ or $i > j$, so a Gaussian DAG model restricts $\boldsymbol{\Sigma}$ (and  $\boldsymbol{\Omega}$) to a lower-dimensional space by imposing sparsity constraints encoded in $\calD$ on $\boldsymbol{L}$. This constraint is important for both frequentist and Bayesian methods.

On the frequentist side, a variety of penalized likelihood methods for sparse estimation of $\boldsymbol{L}$ exist in the literature; see the refrences in \cite{cao2019posterior}. Some of those methods, such as those in \cite{shojaie2010penalized, yu2017learning}, constrain the sparsity pattern in $\boldsymbol{L}$ to be banded. We assume no constraints on the sparsity pattern in this paper. \cite{gaskins2013nonparametric} used a nonparametric prior for covariance estimation. 

On the Bayesian side, the first class of priors on the restricted space of covariance matrices corresponding to a Gaussian DAG model was initially developed in \cite{geiger2002parameter,smith2002parsimonious}. 
\cite{li2019bayesian}  reparameterized the likelihood of a matrix of Gaussian graphical model and obtained the full conditional distribution of the parameters in Cholesky factor. Using the asymptotic distribution of all parameters in the Cholesky factor, they opbtained a shrinkage Bayesian estimator for large precision matrix. \cite{cao2019posterior} considered a hierarchical Gaussian DAG model with DAG-Wishart priors on the covariance matrix and independent Bernoulli priors for each edge in the DAG. A standard choice of conjugate prior is the Wishart distribution, i.e., $\boldsymbol{\Omega} \sim W_q(a,\boldsymbol{U})$ having expectation $a\,\boldsymbol{U}^{-1}$, where $a>q-1$. The priors in \cite{geiger2002parameter} can be considered as analogs of the G-Wishart distribution for concentration graph models. \cite{ben2011high} introduced a class of DAG-Wishart distributions with multiple shape parameters. Their class of distributions is defined for arbitrary DAG models and offers a flexible framework for Bayesian inference in Gaussian DAG models, and generalizes previous Wishart--based priors for DAG models. 

For the cases when $\calD$ is complete, \cite[Supplemental Section B]{ben2011high} derived the Hyper Morkov propertirties of the DAG--Wishart as
\begin{equation}
    \sigma_j^2 \sim \mathrm{I\text{-}Ga}\left(\frac{a_j}{2}-\frac{|\mathrm{pa}(j)|}{2}-1,\frac{1}{2} \boldsymbol{U}_{\prec j \succ}\right),
    \label{eq:prior_D_complete}
\end{equation}
and
\begin{equation}
    \boldsymbol{L}_{\preceq j \succ} | \sigma_j^2 \sim \calN_{|\Pa(j)|} \Big( \boldsymbol{0}, \sigma_j^2 \, \boldsymbol{U}^{-1}_{\preceq j \succeq} \Big),
    \label{eq:prior_L_complete}
\end{equation}
for the Cholesky parameters, where $a_j = a + q -2j + 3$ and $\mathrm{I\text{-}Ga}(\alpha,\beta)$ is an Inverse--Gamma distribution with shape $\alpha>0$ and rate $\beta>0$ having expectation $\beta/(\alpha-1)$. When $\calD$ is not complete, by setting $\boldsymbol{U}=g \boldsymbol{I}_q$ in equations (\ref{eq:prior_D_complete}) and (\ref{eq:prior_L_complete}), \cite[Supplementary]{castelletti2021bayesiana} showed that
\begin{equation}
    \sigma_j^2 \sim \mathrm{I\text{-}Ga}\left(\frac{a_j}{2},\frac{g}{2}\right),
    \label{eq:prior_D}
\end{equation}
\begin{equation}
    \boldsymbol{L}_{\preceq j \succ} | \sigma_j^2 \sim \calN_{|\Pa(j)|} \left( \boldsymbol{0}, \frac{1}{g}\sigma_j^2 \boldsymbol{I}_{|\Pa(j)|} \right),
    \label{eq:prior_L}
\end{equation}
where $a_j = a + |\Pa(j)| - q +1$ and $\boldsymbol{I}_q$ is the identity matrix and $g>0$ is a hyperparameter.

For the case where there are 2 sparse DAGs, we need to properly define the priors for $\boldsymbol{D}$, $\boldsymbol{L}^{(1)}$ and $\boldsymbol{L}^{(2)}$. Lets define the prior for $\sigma_j^2$ as
\begin{equation}
    \sigma_j^2 \sim \mathrm{I\text{-}Ga}\left(\frac{a_j^{(1)}+a_j^{(2)}}{2}, \frac{g_1 + g_2}{2}\right),
\label{eq:prior_Dk}
\end{equation}
where $a_j^{(k)} = a + |\Pa^{(k)}(j)| - q +1$.

The prior distribution for $\boldsymbol{L}^{(k)}$, as discussed in Section \ref{sec:Doubly-Gaussian-DAG-Models}, depends on the skeleton of $\calD^{(k)}$ too. Therefore, the proir for $\boldsymbol{L}^{(k)}$ becomes 
\begin{equation}
    \boldsymbol{L}^{(k)}_{\preceq j \succ} | \sigma_j^2 \sim \calN_{|\Pa^{(k)}(j)|} \left( \boldsymbol{0}, \frac{1}{g_k}\sigma_j^2 \boldsymbol{I}_{|\Pa(j)|} \right),
    \label{eq:prior_Lk}
\end{equation}
where $\Pa^{(k)}(j)$ is the parent nodes for node $j$ in DAG $\calD^{(k)}$.

In this paper, we use equations (\ref{eq:prior_Dk}) and  (\ref{eq:prior_Lk}) as the priors. We also set $a=q$ and $g_k=1/n_k$ for our simulations in section \ref{sec:simulation}.


\subsection{Prior Distribution on $\theta$}
We assign a flat improper prior to the threshold $\theta\in \mathbb{R}$, i.e., $f(\theta)\propto 1$. This choice of prior will make the posterior of $\theta$ proper, as well. For the proof check \cite[Proposition 4.1]{castelletti2021bayesiana}.

\begin{figure}[t]
\centering
\begin{tikzpicture}[align=center, node distance=2cm]
    \node (blank) [blank] {};
    \node (calD1) [process, left of=blank, xshift=-2cm] {Propose $\calD'^{(1)}$};
    \node (calD2) [process, right of=blank, xshift=2cm] {Propose $\calD'^{(2)}$};
    \node (sigma2_j) [process, below of=blank] {$\sigma^2_j | \boldsymbol{X}_{-1}^{(1)},\boldsymbol{X}_{-1}^{(2)}, \calD^{(1)}, \calD^{(2)} \xrightarrow[\sigma^2_1=1]{\text{for all $j$}} \boldsymbol{D}$};
    \node (Sigma1) [process, fill=blue!20, left of=sigma2_j, xshift=-4cm] {$\boldsymbol{\Sigma}^{(1)}$};
    \node (Sigma2) [process, fill=blue!20, right of=sigma2_j, xshift=4cm] {$\boldsymbol{\Sigma}^{(2)}$};
    \node (L1) [process, below of=blank, yshift=-2cm, xshift=-4cm] { $\boldsymbol{L}_{\preceq j \succ}^{(1)} | \sigma^2_j, \boldsymbol{X}^{(1)}, \calD^{(1)} \xrightarrow{\text{for all $j$}} \boldsymbol{L}^{(1)}$};
    \node (L2) [process, below of=blank, yshift=-2cm, xshift=4cm] {$\boldsymbol{L}_{\preceq j \succ}^{(2)} | \sigma^2_j, \boldsymbol{X}^{(2)}, \calD^{(2)} \xrightarrow{\text{for all $j$}} \boldsymbol{L}^{(2)}$};
    \node (X1) [process, below of=L1] {$\boldsymbol{X}_1^{(1)} | \boldsymbol{y}^{(1)}, \boldsymbol{X}_{-1}^{(1)}, \boldsymbol{L}^{(1)}_{\preceq 1 \succ}, \calD^{(1)}, \theta$};
    \node (X2) [process, below of=L2] {$\boldsymbol{X}_1^{(2)} | \boldsymbol{y}^{(2)}, \boldsymbol{X}_{-1}^{(2)}, \boldsymbol{L}^{(2)}_{\preceq 1 \succ}, \calD^{(2)}, \theta$};
    \node (theta) [process, below of=blank, yshift=-6cm] {  Propose $\theta'$  };
    \node (causal1) [process, fill=blue!20, left of=theta, xshift=-3cm, yshift=0cm] {Post intervention $Y^{(1)}$};
    \node (causal2) [process, fill=blue!20, right of=theta, xshift=3cm, yshift=0cm] {Post intervention $Y^{(2)}$};
    \draw (calD1) [arrow]  -- (sigma2_j);
    \draw (calD2) [arrow]  -- (sigma2_j);
    \draw (calD1) [arrow]  -- (L1);
    \draw (calD2) [arrow]  -- (L2);
    \draw (calD1.west) [arrow] 
    .. controls ++(180:1) and ++(70:1) .. (-7.2,-1.5)
    .. controls ++(-110:1) and ++(110:1) .. (-7,-6)
    .. controls ++(-60:1) .. (theta.north west);
    \draw (calD2.east) [arrow]
    .. controls ++(0:1) and ++(110:1) .. (7.2,-1.5)
    .. controls ++(-70:1) and ++(70:1) .. (7,-6)
    .. controls ++(-120:1) .. (theta.north east);    
    \draw (sigma2_j) [arrow]  -- (L1.north east);
    \draw (sigma2_j) [arrow]  -- (L2.north west);
    \draw (calD1) [arrow] to[out=180,in=120] (X1.north west);
    \draw (calD2) [arrow] to[out=0,in=60] (X2.north east);
    \draw (L1) [arrow]  -- (X1);
    \draw (L2) [arrow]  -- (X2);
    \draw[arrow] (L1.south east) to[out=-20,in=90] (theta) node[yshift=2.5cm, xshift=-.7cm] {$j=1$};
    \draw[arrow] (L2.south west) to[out=-160,in=90] (theta) node[yshift=2.5cm, xshift=.7cm] {$j=1$};
    \begin{pgfonlayer}{bg}
    \draw (calD1.south) [arrow2, blue!50]  -- (causal1);
    \draw (sigma2_j) [arrow2, blue!50]  -- (Sigma1);
    \draw (L1.north west) [arrow2, blue!50]  -- (Sigma1.south);
    \draw (X1.south) [arrow2, blue!50]  -- (causal1);
    \draw (theta.west) [arrow2, blue!50]  -- (causal1);    
    \draw (Sigma1.south west) [arrow2, blue!50]  -- (causal1.north west);    
    \draw (calD2.south) [arrow2, blue!50]  -- (causal2);
    \draw (sigma2_j) [arrow2, blue!50]  -- (Sigma2);
    \draw (L2.north east) [arrow2, blue!50]  -- (Sigma2.south);
    \draw (X2.south) [arrow2, blue!50]  -- (causal2);
    \draw (theta.east) [arrow2, blue!50]  -- (causal2);
    \draw (Sigma2.south east) [arrow2, blue!50]  -- (causal2.north east);    
    \end{pgfonlayer}
 
\end{tikzpicture}
\caption{Proposed MCMC scheme for doubly Gaussian DAG--probit models. In the proposed method, $\boldsymbol{D}$ and $\theta$ will be estimated jointely using all the observed data, $\{\boldsymbol{y}^{(1)}, \boldsymbol{y}^{(2)}, \boldsymbol{X}_{-1}^{(1)}, \boldsymbol{X}_{-1}^{(2)}\}$, while $\calD^{(k)}$, $\boldsymbol{L}^{(k)}$ and $\boldsymbol{X}^{(k)}_1$ will be estimated separetely for each group $k\in\{1,2\}$. The blue nodes and blue dashed edges are related to the causal effect estimation.}
\label{fig:MCMC}
\end{figure}
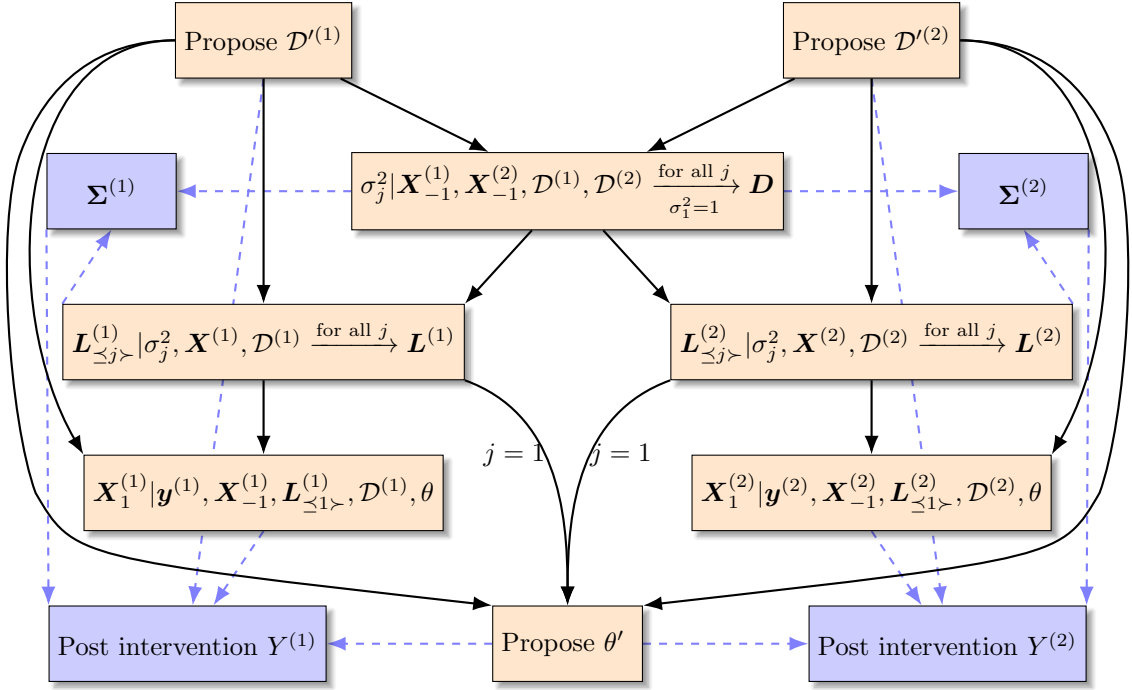

\section{MCMC for Doubly Gaussian DAG--Probit Models}
\label{sec:MCMC}
We assume there are two groups of data, $\boldsymbol{X}^{(1)}$ and $\boldsymbol{X}^{(2)}$ with the coresponding DAGs $\calD^{(1)}$ and $\calD^{(2)}$. We first present the full likelihood distribution of the model gevin the input data $\left\{\boldsymbol{y}^{(1)}, \boldsymbol{y}^{(2)}, \boldsymbol{X}_{-1}^{(1)}, \boldsymbol{X}_{-1}^{(2)}\right\}$ and then find the posterior distributions for the model parameters. The goal is to estimate some of the model parameters using the joint distribution. The schematic view of the algorithm is presented in Figure \ref{fig:MCMC}. We will provide more info on the MCMC algorithm in Section \ref{sec:mcmc_alg}.



\subsection{Full Posterior Distribution}
The full posterior distribution is used to find the posteriors or acceptance rates for $\calD^{(k)}$, $\boldsymbol{L}^{(k)}$, $\boldsymbol{D}$, $\boldsymbol{X}_1^{(k)}$  and $\theta$ in the sebsequent sections.

Given $\left\{\boldsymbol{y}^{(1)}, \boldsymbol{y}^{(2)}, \boldsymbol{X}_{-1}^{(1)}, \boldsymbol{X}_{-1}^{(2)}\right\}$ as the input, the full posterior distribution is
\begin{align}
    & f\left( \calD^{(1)}, \calD^{(2)}, \boldsymbol{L}^{(1)}, \boldsymbol{L}^{(2)}, \boldsymbol{D}, \boldsymbol{X}_1^{(1)}, \boldsymbol{X}_1^{(2)}, \theta \big| \boldsymbol{y}^{(1)}, \boldsymbol{y}^{(2)}, \boldsymbol{X}_{-1}^{(1)}, \boldsymbol{X}_{-1}^{(2)} \right) \nonumber \\
   & \propto \prod_{k=1}^2 f\left( \boldsymbol{y}^{(k)},  \boldsymbol{X}^{(k)} \Big| \calD^{(k)}, \boldsymbol{L}^{(k)}, \boldsymbol{D}, \theta \right)
    f\left(\boldsymbol{L}^{(k)} \big| \boldsymbol{D},  \calD^{(k)} \right) f\left( \calD^{(k)} \right) \times f\left(\boldsymbol{D} \big|  \calD^{(1)}, \calD^{(2)} \right),
\label{eq:posterior}
\end{align}
where the first term is defined in (\ref{eq:like_doubly}), $f\left(\boldsymbol{D} \big|  \calD^{(1)}, \calD^{(2)} \right)$ and $f(\boldsymbol{L}^{(k)} \big| \boldsymbol{D}, \calD^{(k)} )$ are defined in equations (\ref{eq:prior_Dk}) and (\ref{eq:prior_Lk}), and $f(\calD^{(k)})$ is defined in (\ref{eq:prior_DAG_k}). The proof of the full posterior distribution is presented in Appendix \ref{sec:appendix_full_posterior}.


\subsection{Update of $\calD^{(k)}$}
The first step at each MCMC iteration is to propose two new DAG's $\mathcal{D'}^{(1)}$ and $\mathcal{D'}^{(2)}$ from suitable proposal distributions $q(\calD'^{(1)}|\calD^{(1)})$ and $q(\calD'^{(2)}|\calD^{(2)})$. Given the current DAG $\calD^{(k)}$, we construct $\calD'^{(k)}$ by randomly selecting one of the following operators: $\texttt{Insert}(i\to j)$, $\texttt{Delete}(i\to j)$ or $\texttt{Reverse}(i\to j)$. The operator $\texttt{Insert}$ adds the edge $i\to j$ with probability $\xi$ if this edge does not exist in $\calD^{(k)}$,  $\texttt{Delete}$ removes the exsisting edge $i\to j$ with probability $1-\xi$, and the operator $\texttt{Reverse}$ changes the direction of an existing edge, i.e., converting $i\to j$ to $i\leftarrow j$. $\texttt{Reverse}(i\to j)$ operator is equivalent to removing $i\to j$ followed by adding the reverse edge, i.e., $\texttt{Reverse}(i\to j) := \texttt{Delete}(i\to j) + \texttt{Insert}(j\to i)$.

All of these operators must not violate the DAG asumptions, especially they should not create any loop, so we check if the proposed DAG is valid or not at each MCMC iteration. If the proposed DAG has loop, we simply reject it and continue proposing new DAGs until we end up with a valid one. This process ensures that the proposed DAG is valid.

The next step is to accept or reject the proposed valid DAG $\calD'^{(k)}$. We use the algorithm proposed by \cite{wang2012efficient} for Bayesian model determination in Gaussian graphical models under G-Wishart prior distributions. This algorithm is based on PAS (Partial Analytic Structure) algorithm proposed by \cite{godsill2001relationship}, which is based on the reversible jump proposal schemes and takes into account the partial analytic structure of the DAG model. 

It is important to mention that $\texttt{Insert}(i\to j)$ and $\texttt{Delete}(i\to j)$ operators just make changes in the $\Pa(j)$, so the Cholesky parameters under the new and old DAGs differ only with respect to their $j$-th component, but for the $\texttt{Reverse}(i\to j)$ operator, both $\Pa(i)$ and $\Pa(j)$ will change. As a result, the acceptance probabilities for these two cases are different and we present them separetely.
 

\subsubsection{Acceptance Probability for $\texttt{Insert}$ and $\texttt{Delete}$ Operators}

Under $\texttt{Insert}(i\to j)$ and $\texttt{Delete}(i\to j)$ operators, the acceptance probability for $\calD'^{(k)}$ for $j\in \{2,\dots,q\}$ is given by
\begin{equation}
    \alpha_{\calD'^{(k)}} = \min \left\{ 1, \frac{m(\boldsymbol{X}_j^{(k)}| \boldsymbol{X}_{\Pa_{\calD'}(j)}^{(k)}, \calD'^{(k)})} {m(\boldsymbol{X}_j^{(k)}| \boldsymbol{X}_{\Pa_{\calD}(j)}^{(k)}, \calD^{(k)})} 
    . \frac{f(\calD'^{(k)})}{f(\calD^{(k)})}
    . \frac{q(\calD^{(k)}|\calD'^{(k)})}{q(\calD'^{(k)}|\calD^{(k)})}
    \right\},
\label{eq:update_calD}
\end{equation}
where for $\calD \in \{ \calD^{(1)}, \calD^{(2)}, \calD^{'(1)}, \calD^{'(2)} \}$,
\begin{align}
    & m(\boldsymbol{X}_j^{(k)}| \boldsymbol{X}_{\Pa_{\calD}(j)}^{(k)}, \calD) \nonumber \\
    & = (2\pi)^{-\frac{n_k}{2}}
    \frac{|\boldsymbol{T}_j^{(k)}|^{1/2}}{|\bar{\boldsymbol{T}}_j^{(k)}|^{1/2}} \,.\, 
    \frac{\Gamma(a_j^\calD/2+n_k/2)} {\Gamma(a_j^\calD/2)}
    \left[ \frac{1}{2} g_k \right]^{a_j^\calD/2}
    \left[ \frac{1}{2} \left(g_k + \boldsymbol{X}'^{(k)}_j \boldsymbol{X}^{(k)}_j - \hat{\boldsymbol{L}}'^{(k)}_j \bar{\boldsymbol{T}}^{(k)}_j \hat{\boldsymbol{L}}^{(k)}_j \right) \right]^{-(a_j^\calD+n_k)/2},
\label{eq:m1_Xj}
\end{align}
and
\begin{align}
    \boldsymbol{T}_j^{(k)} & = g_k \boldsymbol{I}_{|\Pa_\calD(j)|}, \nonumber \\
    \bar{\boldsymbol{T}}^{(k)}_j & = \boldsymbol{T}_j^{(k)} + \boldsymbol{X}'^{(k)}_{\Pa_\calD(j)} \boldsymbol{X}^{(k)}_{\Pa_\calD(j)}, \nonumber \\
    \hat{\boldsymbol{L}}^{(k)}_j & = \Big(\bar{\boldsymbol{T}}^{(k)}_j\Big)^{-1} \boldsymbol{X}'^{(k)}_{\Pa_\calD(j)} \boldsymbol{X}^{(k)}_j,
\label{eq:Tj}
\end{align}
with $a_j^\calD=a + |\Pa_\calD(j)|-q+1$.

For $j=1$, because we fixed $\sigma^2_1=1$, we have
\begin{align}
    m(\boldsymbol{X}_1^{(k)}| \boldsymbol{X}_{\Pa_{\calD}(1)}^{(k)}, \calD)
    = (2\pi)^{-\frac{n_k}{2}}
    \frac{|\boldsymbol{T}_1^{(k)}|^{1/2}}{|\bar{\boldsymbol{T}}_1^{(k)}|^{1/2}} \,.\, 
    \exp\left\{ -\frac{1}{2} \left( \boldsymbol{X}'^{(k)}_1 \boldsymbol{X}^{(k)}_1 - \hat{\boldsymbol{L}}'^{(k)}_1 \bar{\boldsymbol{T}}^{(k)}_1 \hat{\boldsymbol{L}}^{(k)}_1 \right) \right\}.
\label{eq:m1_X1}
\end{align}
The proof for both $j=1$ and $j>1$ cases is given in Appendix \ref{sec:proof_acceptance_probability_calD_1parent}.


\subsubsection{Acceptance Probability for $\texttt{Reverse}$ Operator}

For $\texttt{Delete}(i\to j)$ operator, both the parents for the nodes $i$ and $j$ will change. So, the acceptance probability for $\calD'^{(k)}$ becomes
\begin{equation}
    \alpha_{\calD'^{(k)}} = \min \left\{ 1, \frac{m(\boldsymbol{X}_i^{(k)}| \boldsymbol{X}_{\Pa_{\calD'}(i)}^{(k)}, \calD'^{(k)}) \, m(\boldsymbol{X}_j^{(k)}| \boldsymbol{X}_{\Pa_{\calD'}(j)}^{(k)}, \calD'^{(k)})}
    {m(\boldsymbol{X}_i^{(k)}| \boldsymbol{X}_{\Pa_{\calD}(i)}^{(k)}, \calD^{(k)}) \, m(\boldsymbol{X}_j^{(k)}| \boldsymbol{X}_{\Pa_{\calD}(j)}^{(k)}, \calD^{(k)})}
    . \frac{f(\calD'^{(k)})}{f(\calD^{(k)})}
    . \frac{q(\calD^{(k)}|\calD'^{(k)})}{q(\calD'^{(k)}|\calD^{(k)})}
    \right\},
    \label{eq:update_calD2}
\end{equation}
where $m(.)$ is defined in (\ref{eq:m1_Xj}) and (\ref{eq:m1_X1}) for $j>1$ and $j=1$, respectively. For the proof, check Appendix \ref{sec:proof_acceptance_probability_calD_2parent}.

Finally, using equation (\ref{eq:prior_DAG_k}), the term 
$f(\calD'^{(k)})/f(\calD^{(k)})$ defined in equations (\ref{eq:update_calD}) and (\ref{eq:update_calD2}) is equal to $\xi/(1-\xi)$, $(1-\xi)/\xi$ and 1, respectively for $\texttt{Insert}$, $\texttt{Delete}$ and $\texttt{Reverse}$ operators with $\xi$ as the probability of edge inclusion. Let's denote by $\mathbb{O}^\calD$ the set of \emph{valid} operators in $\calD$. The probability of transition from $\calD$ to $\calD'$ is then equal to $q(\calD'|\calD) = 1/|\mathbb{O}^\calD|$. For transitioning from DAG $\calD$ with $q$ nodes and $|E|$ edges to $\calD'$, the number of  $\texttt{Delete}$, $\texttt{Reverse}$ and $\texttt{Insert}$ operators are $|E|$, $|E|$ and $q(q-1)/2-|E|$ respectively, but some of the $\texttt{Reverse}$ and $\texttt{Insert}$ operators might not be valid, because they can create loops. Since $\calD$ and $\calD'$ only differ at most in one edge, so $q(\calD^{(k)}|\calD'^{(k)})/q(\calD'^{(k)}|\calD^{(k)})\approx 1$ for sparse $\calD$.

\subsection{Update of $\boldsymbol{L}^{(k)}$}
For $j\in \{2, \dots, q\}$, we have
\begin{equation}
    \boldsymbol{L}_{\preceq j \succ}^{(k)} | \sigma^2_j, \boldsymbol{X}^{(k)}, \calD^{(k)} \sim \calN_{|\Pa^{(k)}(j)|} \left(-\hat{\boldsymbol{L}}'^{(k)}_j, \sigma^2_j (\bar{\boldsymbol{T}}^{(k)}_j)^{-1} \right).
\label{eq:update_Lk}
\end{equation}
Moreover, for node $1$ we have
\begin{align}
    \boldsymbol{L}_{\preceq 1 \succ}^{(k)} |  \boldsymbol{X}^{(k)}, \calD^{(k)} \sim \calN_{|\Pa^{(k)}(1)|} \left(-\hat{\boldsymbol{L}}'^{(k)}_1,  (\bar{\boldsymbol{T}}^{(k)}_1)^{-1} \right).
\end{align}

The proofs for $j=1$ and $j>1$ are given in Appendix \ref{sec:proof_posterior_L}.

\subsection{Update of $\boldsymbol{D}$}
We update $\boldsymbol{D}$ using information from both groups. For $j\in \{2, \dots, q\}$,
\begin{equation}
    \sigma^2_j | \boldsymbol{X}^{(1)}, \boldsymbol{X}^{(2)}, \calD^{(1)}, \calD^{(2)} \sim \mathrm{I\text{-}Ga}\left(\frac{\alpha^{(1)} }{2}+\frac{\alpha^{(2)}}{2},\frac{\beta^{(1)}}{2}+\frac{\beta^{(2)}}{2}\right),
\label{eq:update_D}
\end{equation}
where
$\alpha^{(k)} = a^{(k)}_j + n_k$ and
$\beta^{(k)} = g_k + \boldsymbol{X}'^{(k)}_j\boldsymbol{X}^{(k)}_j - 
\hat{\boldsymbol{L}}'^{(k)}_j \bar{\boldsymbol{T}}^{(k)}_j \hat{\boldsymbol{L}}^{(k)}_j$ for $k\in\{1,2\}$.
For node $1$, we set $\sigma^2_1=1$. The proof is given in Appendix \ref{sec:proof_posterior_D}.

\subsection{Update of $\boldsymbol{X}_1^{(k)}$}
Since $\boldsymbol{X}_1^{(k)}$ is the latent variable, we need to get a sample from its posterior distribution, which follows
\begin{equation}
    \boldsymbol{X}_1^{(k)} | \boldsymbol{y}^{(k)}, \boldsymbol{X}_{-1}^{(k)}, \boldsymbol{L}^{(k)}_{\preceq 1 \succ}, \calD^{(k)}, \theta \sim \calN_{n_k} \left(-\boldsymbol{X}^{(k)}_{\Pa(1)} \boldsymbol{L}^{(k)}_{\preceq 1 \succ}, \boldsymbol{I}_{n_k} \right),
\label{eq:update_Xk}
\end{equation}
truncated at $\theta$ where $n_k$ is the sample size of the $k$-th group. Clearly this equation is not a function of $\boldsymbol{D}$ because $\sigma^2_1=1$. For the proof, check \cite{castelletti2021bayesiana}.

\subsection{Update of $\theta$}
To update the cut-off $\theta$, we propose a new $\theta'$ using Metropolis Hastings method with the proposal distribution $\theta'|\theta \sim \calN(\theta, \sigma^2_0)$, so the transition kernel becomes $q(\theta'|\theta)=f_\calN(\theta'|\theta, \sigma^2_0)$ where $\sigma^2_0$ is a hyperparameter. We accept $\theta'$ with probability
\begin{equation}
    \alpha_\theta = \min\{1,r_\theta\},
\label{eq:update_theta}
\end{equation}
where
\begin{equation}
    r_\theta = \frac{\prod_{i=1}^{n_1}\Psi\big(y^{(1)}_i,\theta' | -\boldsymbol{x}^{(1)}_{\Pa(1)} \boldsymbol{L}^{(1)}_{\preceq 1 \succ}, 1 \big)}
    {\prod_{i=1}^{n_1}\Psi\big(y^{(1)}_i,\theta | -\boldsymbol{x}^{(k)}_{\Pa(1)} \boldsymbol{L}^{(k)}_{\preceq 1 \succ}, 1 \big)} \,
    \frac{\prod_{i=1}^{n_2}\Psi\big(y^{(2)}_i,\theta' | -\boldsymbol{x}^{(2)}_{\Pa(1)} \boldsymbol{L}^{(2)}_{\preceq 1 \succ}, 1 \big)}
    {\prod_{i=1}^{n_2}\Psi\big(y^{(2)}_i,\theta | -\boldsymbol{x}^{(2)}_{\Pa(1)} \boldsymbol{L}^{(2)}_{\preceq 1 \succ} , 1 \big)}\,
    \frac{f_\calN(\theta|\theta', \sigma^2_0)}{f_\calN(\theta'|\theta, \sigma^2_0)},
\label{eq:r_theta}
\end{equation}
and $\Psi\big(y,\eta|\mu,\sigma^2\big)= |y - \Phi(\eta|\mu,\sigma^2\big)|$. In fact, $\Psi\big(y,\eta|\mu,\sigma^2\big)$ is either the CDF or the survival function of a $\calN(\mu,\sigma^2)$ ditsribution for $y=0$ and $y=1$, respectively. The proof is presented in Appendix \ref{sec:proof_posterior_theta}.

\subsection{Initial DAGs for $\calD^{(k)}$}
\label{sec:initial_dag}
The number of different combinations to initialize $\calD^{(k)}$ is equal to $2^{q(q-1)/2}$. Choosing appropriate initial values for $\calD^{(1)}[0]$ and $\calD^{(2)}[0]$ can exponentially speed up the convergence of MCMC algorithm. To speed up the computations, we propose the following method.

We showed in Equation (\ref{eq:decom}) that $\boldsymbol{\Omega}$ can be uniquely decomposed into
\begin{equation}
    \boldsymbol{\Omega} = \boldsymbol{L}\boldsymbol{D}^{-1}\boldsymbol{L}',
\end{equation}
where the non-zero elements of $L$ correspond to the edges in $\calD$. Unfortunately, we cannot estimate $\boldsymbol{L}$ given $\boldsymbol{X}_{-1}$ because $\boldsymbol{X}_1$ is a latent variable. But, $\boldsymbol{\Omega}_{-1}^{-1}$ is the covariance matrix of $\boldsymbol{X}_{-1}$, and can be decomposed into
\begin{equation}
    \boldsymbol{\Omega}_{-1} = \boldsymbol{L}_{-1}\boldsymbol{D}^{-1}_{-1} \boldsymbol{L}'_{-1}
    \label{eq:decom_1}
\end{equation}
using modified Cholesky decomposition method. So, we can initialize $\calD$ by binarizing the non-zero elements of the estimated $\boldsymbol{L}_{-1}$ matrix. The following proposition shows the relation between $\boldsymbol{L}_{-1}$ and $\boldsymbol{L}$, and states that it is the coresponding submatrix of  $\boldsymbol{L}$.

\begin{prop}
Let's reorder the matrix $\boldsymbol{X}$ as $\boldsymbol{X}^* =(\boldsymbol{X}_{-1}, \boldsymbol{X}_1)$ and let $\text{cov}(\boldsymbol{X}^*) = \boldsymbol{\Omega}^{-1}$ with $\boldsymbol{\Omega} = \boldsymbol{L}\boldsymbol{D}^{-1}\boldsymbol{L}'$. If we partition $\boldsymbol{L}$ into
\begin{align}
\boldsymbol{L} =
    \left( \begin{array}{cc}
        \boldsymbol{L}_{11} & \boldsymbol{0} \\
        \boldsymbol{L}_{2,1} & \boldsymbol{L}_{22}
    \end{array} \right)
\end{align}
then $\boldsymbol{L}_{-1} = \boldsymbol{L}_{11}$, where $\boldsymbol{L}_{-1}$ is the lower-triangular matrix from the decomposition of $\boldsymbol{\Omega}_{-1}$. 
\end{prop}

For the proof check Appendix \ref{sec:proof_proposition1}. So we use the non-zero elements of 
\begin{align}
\hat{\boldsymbol{L}} :=
    \left( \begin{array}{cc}
        \boldsymbol{0} & \boldsymbol{0} \\
        \boldsymbol{0} & \boldsymbol{L}_{-1}
    \end{array} \right)
\label{eq:init_dag_final}
\end{align}
for initializing $\calD\in \{\calD^{(1)}, \calD^{(2)} \}$.

\subsection{MCMC Algotithm}
\label{sec:mcmc_alg}

The MCMC Scheme for this method is depicted in Fig \ref{fig:MCMC}. Using the samples from the posteriors, we estimate the model parameters at the end of the algorithm. The proposed MCMC is presented in Algorithm \ref{alg:mcmc}.

\SetKwInput{KwOutput}{Output}
\SetKwInput{KwInput}{Input}
\begin{algorithm}
\label{alg:mcmc}
\caption{MCMC algorithm for the proposed method on doubly Gaussian--probit DAG models.}
\KwInput{$\left\{\boldsymbol{y}^{(1)}, \boldsymbol{y}^{(2)}, \boldsymbol{X}_{-1}^{(1)}, \boldsymbol{X}_{-1}^{(2)}\right\}$}
\KwOutput{$T$ samples from posterior distribution (\ref{eq:posterior})}
Initialize $\calD^{(1)}[0]$ and $\calD^{(2)}[0]$ with (\ref{eq:init_dag_final}), set the cut-off $\theta[0]=0$, and the latent variables $\boldsymbol{x}_1^{(1)}[0], \boldsymbol{x}_1^{(2)}[0] \sim \calN(0,1)$ truncated at $\theta[0]$\;
\For{$t=1, \dots, T$}{
    \For{$k \in \{1,2\}$}{
        Given $\mathcal{D}^{(k)}[t-1]$, randomly select a valid operator\; 
        Sample $\mathcal{D'}^{(k)}[t]$ from $q(\calD'^{(k)}|\calD^{(k)})$ with acceptance probability $\alpha_{\calD'^{(k)}}$ defined in (\ref{eq:update_calD}) for the $\texttt{Insert}$ and $\texttt{Delete}$ operators or with acceptance probability defined in (\ref{eq:update_calD2}) if the selected operator is $\texttt{Reverse}$\;
    }
    Sample $\boldsymbol{D}[t]$ from $I\text{-}Ga(.,.)$ defined in (\ref{eq:update_D}) using information from both groups\;
    \For{$k \in \{1,2\}$}{
        Sample $\boldsymbol{L}^{(k)}[t]$ from (\ref{eq:update_Lk})\;
        Sample $\boldsymbol{X}_1^{(k)}[t]$ from (\ref{eq:update_Xk})\;
    }
    Update $\theta[t]$ with acceptance probability $\alpha_\theta = \min\{1,r_\theta\}$ defined in (\ref{eq:update_theta})\;
    \For{$k \in \{1,2\}$}{
        Compute the post intervention $\mathbb{E} \left(Y^{(k)} \,| \, \Do(X^{(k)}_s=\tilde{x}^{(k)}),\boldsymbol{\Sigma}^{(k)}, \theta \right)$ from (\ref{eq:post_intervention_Y})\;
    }
}
The elements of $\hat{\calD}^{(k)}$ can be estimated by comparing $\hat{\xi}^{(k)}(i\to j)$ defined in (\ref{eq:mcmc_end_edge}) with a threshold\;
\end{algorithm}

For each group $k\in\{1,2\}$, the posterior probabilities of edge inclusion can be computed via
\begin{align}
    \hat{\xi}^{(k)}(i\to j) = \frac{1}{T-B} \sum_{t=B}^T \mathbbm{1}_{i\to j}\left(\calD^{(k)}[t]\right),
\label{eq:mcmc_end_edge}
\end{align}
where $\mathbbm{1}_{i\to j}\left(\calD^{(k)}[t]\right)$ is 1 if there is an edge ${i\to j}$ in the $\calD^{(k)}[t]$, and 0 otherwise. We can compare $\hat{\xi}^{(k)}(i\to j)$ with a threshold, say 0.5, to estimate the final $\hat{\calD}^{(k)}$.

\section{Causal Effects}
\label{sec:causal_effect}
Given two disjoint sets of variables, $X$ and $Y$, the \emph{causal effect} of $X$ on $Y$, denoted by $f(y|\, \Do(x))$, is a function from $X$ to the space of probability distributions on $Y$. The goal of $\Do(x)$ calculus is to generate probabilistic formulas for the effect of interventions in terms of the observed probabilities. For each realization $x$ of $X$, $f(y |\,\Do(x))$ gives the probability of $Y=y$ induced by deleting all equations corresponding to variables in $X$ and substituting $X=x$ in the remaining equations (\cite{pearl2009causality}). 
In fact, $\Do(.)$ operator marks an action or an intervention in the model. In an algebraic model, we replace certain functions with a constant $X=x$, and in a graph we remove edges going into the target of intervention, but preserve edges going out of the target.
The graph corresponding to the reduced set of equations is a subgraph of $\calD$ from which all arrows entering $x$ have been pruned.

The effect of interventions of $\Do(X_i=x_i)$ for $s\in \{2,\dots,q\}$ can be expressed in a simple \emph{truncated factorization} formula as following
\begin{equation}
    f(x_1,\dots, x_q\,|\, \Do(X_s=\tilde{x}) = \left\{
       \begin{array}{ll}
        \prod_{j=1, j\neq s}^q f(x_j\,|\,\Pa(x_j)) \, \big|_{x_s=\tilde{x}}  & \mathrm{ if ~} x_s =\tilde{x}, \\
        0 & \mathrm{ otherwise}. 
      \end{array}
    \right.
    \label{eq:post_intervention}
\end{equation}
This quation reflects the removal of the term $f(x_i|\Pa(x_i))$ from the product of (\ref{eq1}), since $\Pa(x_i)$ no longer influences $X_i$ and can be seen as a  transformation between the pre- (\ref{eq1}) and post-intervention (\ref{eq:post_intervention}) distributions.

For the latent variable $X_1$, the post-intervention distribution can be calculated as 
\begin{equation}
    f(x_1\,|\, \Do(X_s=\tilde{x})) = \int f(x_1\,|\,\tilde{x}, \boldsymbol{x}_{\Pa(s)}) f(\boldsymbol{x}_{\Pa(s)}) d\boldsymbol{x}_{\Pa(s)}.
    \label{eq:post_intervention_X1}
\end{equation}
For more details on the post-intervention distribution, take a look at the Theorem 3.2.2 in \cite{pearl2009causality}.

Using the conditional Gaussian distribution, \cite{castelletti2021bayesiana} showed that
\begin{equation}
    f(x_1\,|\, \Do(X_s=\tilde{x}), \boldsymbol{\Sigma}) = f_\calN\big( x_1\,|\,\gamma_s \tilde{x}, \sigma^2_\Do \big),
    \label{eq:post_intervention_X1_normal}
\end{equation}
where
\begin{align*}
\sigma^2_\Do & = \frac{\delta_1^2}{1-(\boldsymbol{\gamma}'\boldsymbol{T}^{-1}\boldsymbol{\gamma})/\delta_1^2}, \\
\delta_1^2 & = \boldsymbol{\Sigma}_{1|\Fa(s)},\\
(\gamma_s,\boldsymbol{\gamma}')' & =  \boldsymbol{\Sigma}_{1,\Fa(s)} (\boldsymbol{\Sigma}_{\Fa(s),\Fa(s)})^{-1},\\
\boldsymbol{T} & = (\boldsymbol{\Sigma}_{\Fa(s),\Fa(s)})^{-1} + \frac{1}{\delta_1^2} \boldsymbol{\gamma}\boldsymbol{\gamma}'.
\end{align*}

Inspired by the Bartlett’s decomposition, \cite{silva2009hidden} called the set $(\delta_1^2, \gamma_s,\boldsymbol{\gamma}')$ as the Bartlett parameters of $\boldsymbol{\Sigma}$. Bartlett’s decomposition, defined in \cite{brown1994inference}, allows the definition of its density function by the joint density of $(\delta_1^2, \gamma_s,\boldsymbol{\gamma}')$. The closed form of the distribution of the corresponding Bartlett parameters is presenetd in \cite[Lemma 1]{silva2009hidden} which are related to the equations (\ref{eq:prior_D_complete}) and (\ref{eq:prior_L_complete}).

Since $\boldsymbol{X}_1$ is a latent variable,  the post-interventionaverage for the observed binary variable $Y$, $\mathbb{E}(Y \,| \, \Do(X_s=\tilde{x}),\boldsymbol{\Sigma}, \theta)$ can be computed by
\begin{align}
    \mathbb{E}(Y \,|\, \Do(X_s=\tilde{x}),\boldsymbol{\Sigma}, \theta)
    = & \mathbb{P}(Y=1 \,|\, \Do(X_s=\tilde{x}),\boldsymbol{\Sigma}, \theta) \nonumber\\
    = & \mathbb{P}(X_1>\theta \,|\, \Do(X_s=\tilde{x}),\boldsymbol{\Sigma}) \nonumber \\
    = & 1-\Phi\Big(\frac{\theta-\gamma_s \tilde{x} }{\sqrt{\sigma^2_\Do} }\Big),
\label{eq:post_intervention_Y}
\end{align}
where $\Phi(.)$ is the  c.d.f. of a standard normal distribution. Figure (\ref{fig:MCMC}) shows the input parameters and steps to compute (\ref{eq:post_intervention_Y}) for $Y^{(k)}$, where we estimate it by substituting $\boldsymbol{\Sigma}^{(k)}$ and $\theta$ with their simulated values at each step of the MCMC discussed in Section \ref{sec:MCMC}.

\section{Simulation}
\label{sec:simulation}
We ran a comprehensive simulation to assess the performance of the proposed method. We ran the simulations for sample sizes $n_1,n_2\in \{50, 100, 200, 500, 1000\}$, number of nodes $q\in\{10, 20, 30, 40, 50\}$, and probability of edge inclusion $\xi\in\{0.1, 0.2, 0.3, 0.4\}$. The number of runs/replications for each scenario is 25. We did not run our algorithm for $(q=40, \xi=0.4)$, $(q=50, \xi=0.3)$ and $(q=50, \xi=0.4)$ because sample sizes larger than 1,000 are needed for bigger DAGs.

For each run, we applied Algorithm \ref{alg:mcmc} with $T=5,000$ MCMC iterations and discarded the first $B=1,000$ burn-in iterations. We also
set $g_k = 1/n_k$ and $a = q$ in the prior on the Cholesky parameters of (\ref{eq:prior_Dk}), and $\sigma^2_0 = 0.5$ for the proposal density of the cut-off in (\ref{eq:r_theta}).

The receiver operating characteristic (ROC) curve is a good method for showing the performance of a binary classification problem. In our case, the binary classification is to assess if an edge $i\to j$ exists. To plot ROC, we need to compute the \emph{sensitivity} and \emph{specificity} indexes, which can be computed from the actual and predicted conditions. To simplify the plots and have an overall evaluation metric, we concatinated the lower elements of $\calD^{(1)}$ and $\calD^{(2)}$ as the actual condition, and concatinated the lower elements of $\hat{\calD}^{(1)}$ and $\hat{\calD}^{(2)}$ estimated form (\ref{eq:mcmc_end_edge}) as the predicted condition. The results are plotted in Figure \ref{fig:sim_roc1} for different scenarios with $\xi=0.1$. We added ROC plots for $\xi\in \{0.2, 0.3, 0.4\}$ in Appendix \ref{sec:Appendix_Simulation_results}. The area under curve (AUC) is also computed and presented in Table \ref{tbl:auc} for $\xi=0.1$. AUC gets better as the sample size increases. A bigger sample size is also needed to improve the AUC if the number of nodes gets bigger. We added AUC tables for $\xi\in \{0.2, 0.3, 0.4\}$ in Appendix.


\begin{table}[!t]
\centering
\caption{AUC computed from the average ROC curves in Figure \ref{fig:sim_roc1} for different  sample sizes and DAG sizes, $q$. The probability of edge inclusion is $\xi=0.1$ for this table.}
\begin{tabular}{|rr|ccccc|}
  \hline
  &  & \multicolumn{5} {c|} {$q$} \\ 
  $n_1$ & $n_2$ & 10 & 20 & 30 & 40 & 50 \\ 
  \hline
 50 & 50 & 0.9843 & 0.9756 & 0.9628 & 0.9468 & 0.8653 \\ 
  50 & 100 & 0.9823 & 0.9751 & 0.9701 & 0.9598 & 0.9184 \\ 
  100 & 100 & 0.9906 & 0.9853 & 0.9777 & 0.9707 & 0.9554 \\ 
  100 & 200 & 0.9912 & 0.9843 & 0.9774 & 0.9704 & 0.9617 \\ 
  200 & 200 & 0.9991 & 0.9886 & 0.9826 & 0.9712 & 0.9725 \\ 
  500 & 500 & 0.9944 & 0.9951 & 0.9802 & 0.9804 & 0.9785 \\ 
  1000 & 500 & 0.9975 & 0.9942 & 0.9828 & 0.9817 & 0.9787 \\ 
  1000 & 1000 & 0.9956 & 0.9948 & 0.9836 & 0.9815 & 0.9824 \\ 
  \hline
\end{tabular}
\label{tbl:auc}
\end{table}

\begin{figure}
\centering
\includegraphics[width=\textwidth]{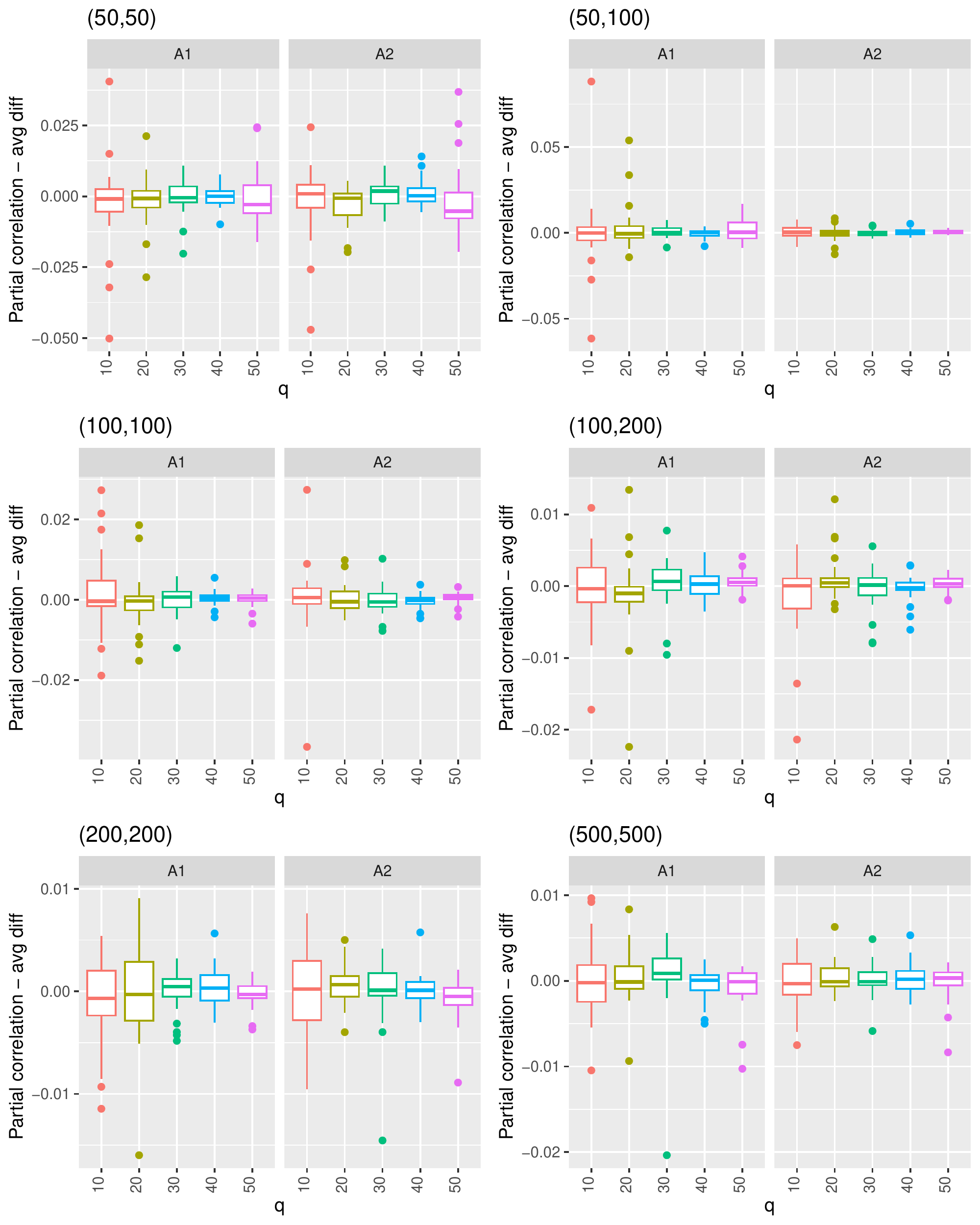}
\caption{Boxplot for the difference between the estimated and true partial correlations defined in (\ref{eq:mcmc_end_partial}) for $T=5000$ iterations and $\xi=0.1$. As the sample size increases, the average difference decreases.}
\label{fig:sim_Partial_mean_edge0.1}
\end{figure}

At each step $t$ of the MCMC iteration, we compute the partial correlations $\hat{\rho}_{ij}^{(k)}[t]; k=1,2$ defined in (\ref{eq:partial}) using the estimated precision matrix $\hat{\boldsymbol\Omega}^{(k)}[t]$ and compare them with the true partial correlations. We take avarage over all the iterations after discarding the burn-in iterations ($B$), i.e., for each run we compute
\begin{align}
    \frac{1}{T-B} \sum_{t=B}^T \frac{1}{q(q-1)/2} \sum_{i<j} \rho_{ij}^{(k)}[t] - \hat{\rho}_{ij}^{(k)}[t].
\label{eq:mcmc_end_partial}
\end{align}
The boxplot for equation (\ref{eq:mcmc_end_partial}) and different runs are plotted in Figure \ref{fig:sim_Partial_mean_edge0.1} for $\xi=0.1$. In almost all the cases, the difference is less than $5\%$ and the average is around zero, depicting the stability of the proposed method. We removed some of the boxplots for big $(n_1,n_2)$ to save space. Similar to (\ref{eq:mcmc_end_partial}), we also depicted the same boxplot for the absolute differences, i.e.,
\begin{align}
    \frac{1}{T-B} \sum_{t=B}^T \frac{1}{q(q-1)/2} \sum_{i<j} \left|\rho_{ij}^{(k)}[t] - \hat{\rho}_{ij}^{(k)}[t] \right|,
\label{eq:mcmc_end_partial_abs}
\end{align}
in Figure \ref{fig:sim_Partial_meanabs_edge0.1} for $\xi=0.1$. We added more plots for $\xi>0.1$ in Appendix \ref{sec:Appendix_Simulation_results}. We can see that the average absolute difference converges to zero as the sample size increases.

\begin{figure}
\centering
\includegraphics[width=\textwidth]{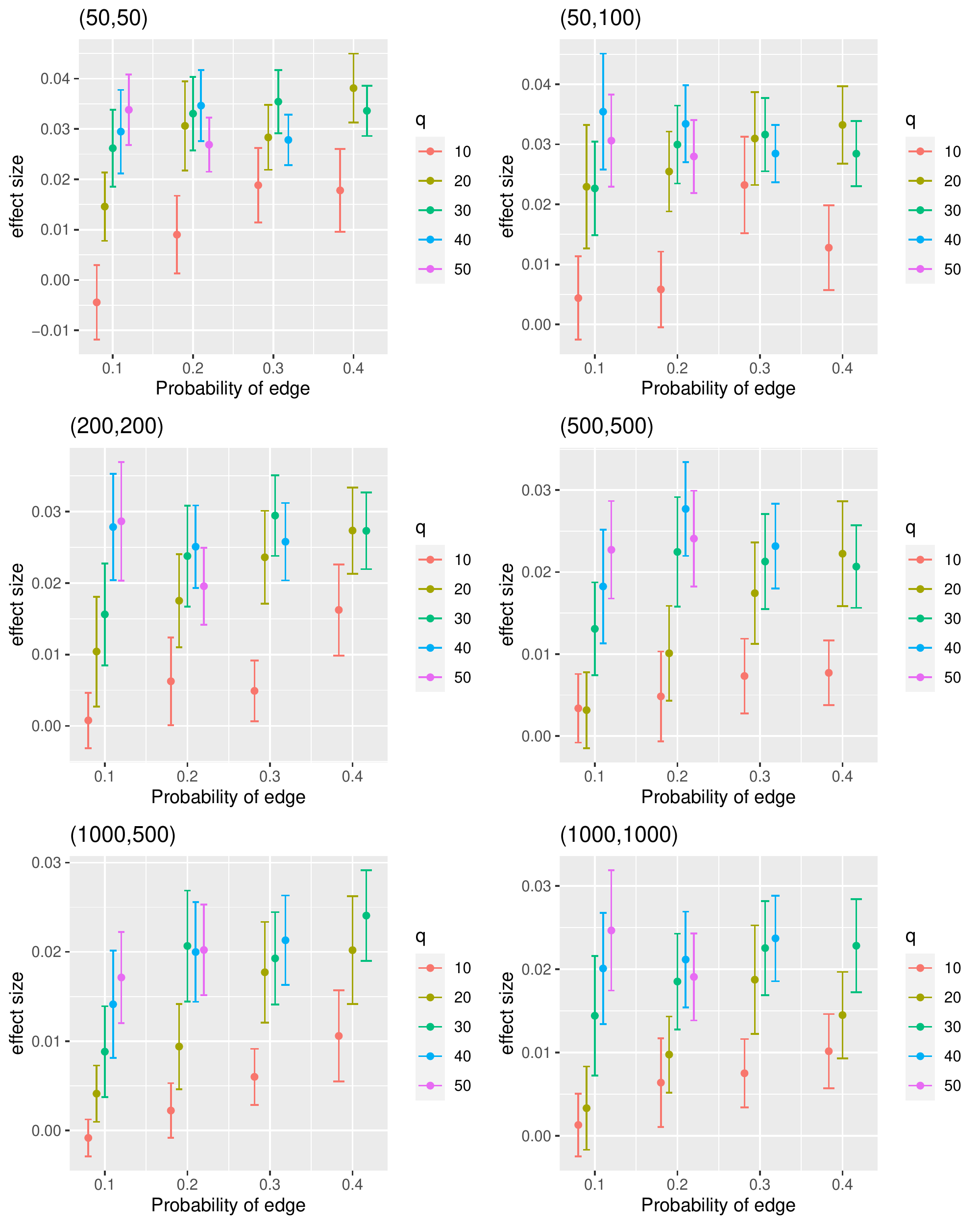}
\caption{Errorbar for the difference between the estimated and true effect sizes defined in (\ref{eq:post_intervention_Y}) for $T=5000$ iterations. The error range are almost the same for different sample sizes but it decreases for smaller DAGs.}
\label{fig:sim_effect_size}
\end{figure}

In order to show the stability of the estimated post-intervention values defined in (\ref{eq:post_intervention_Y}), the difference between the actual effect size and the estimated effect size is quantified for all the incoming edges to node 1. The actual effect size is computed using $(\boldsymbol{\Sigma}^{(k)},\theta)$ and the predicted effect size is computed form the $(\hat{\boldsymbol{\Sigma}}^{(k)},\hat{\theta})$, i.e.,
\begin{equation}
    \mathbb{E}(Y \,|\, \Do(X_s=\tilde{x}),\boldsymbol{\Sigma}, \theta) - \mathbb{E}(Y \,|\, \Do(X_s=\tilde{x}),\hat{\boldsymbol{\Sigma}}, \hat{\theta}),
\end{equation}
where the results are ploted in Figure \ref{fig:sim_effect_size}.

\begin{figure}
\centering
\includegraphics[width=\textwidth]{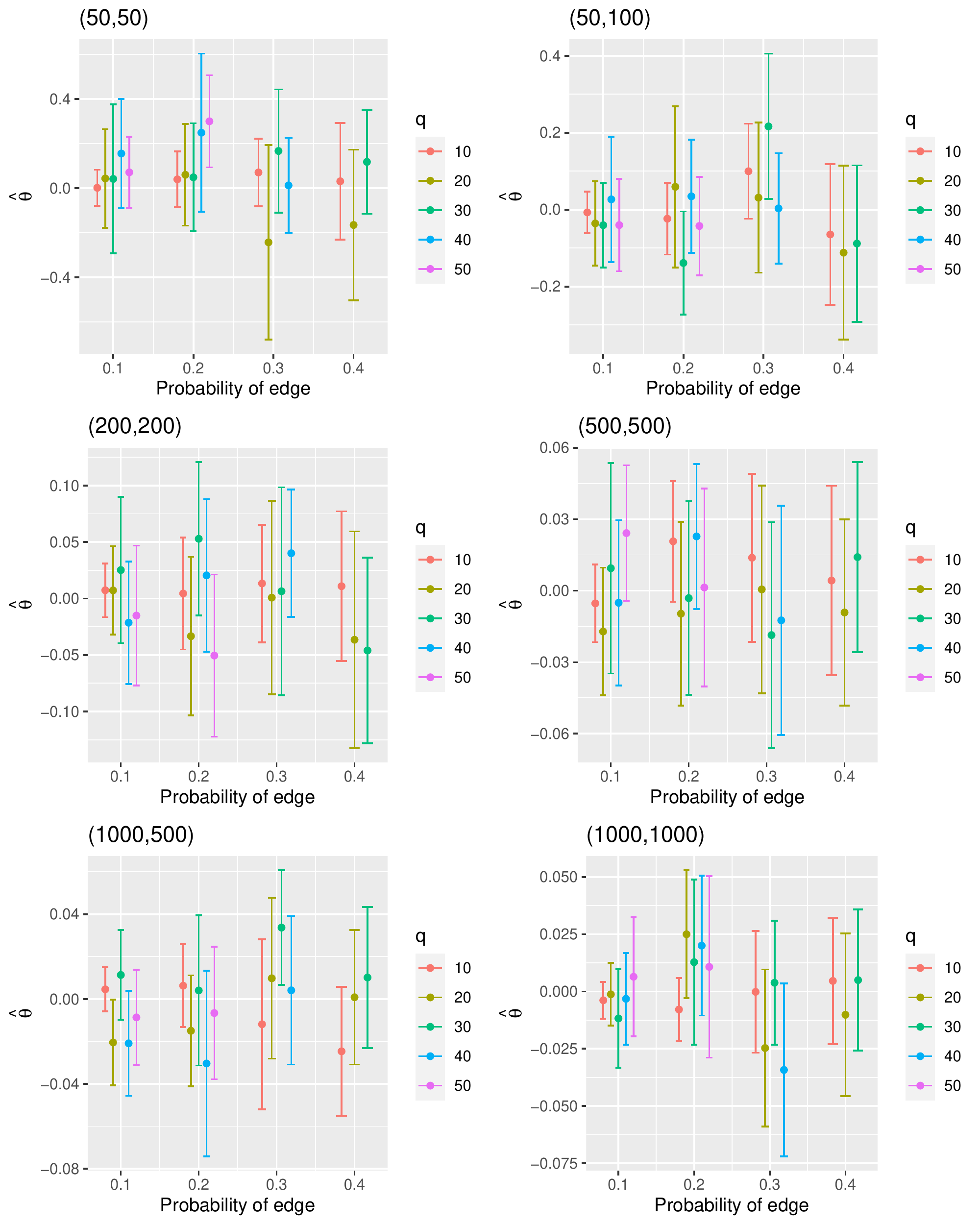}
\caption{Estimated cut-off parameter together with $95\%$ confidence interval for different scienarios with for $T=5,000$ iterations.}
\label{fig:sim_theta}
\end{figure}

The estimated cut-off parameter $\theta$ is also unbiased and its variance shrinks quickly as sample size increases. The estimated $\theta$ together with the $95\%$ confidence intervals for different scenarios are plotted in Figure \ref{fig:sim_theta}.

\begin{figure}
\centering
\includegraphics[width=\textwidth]{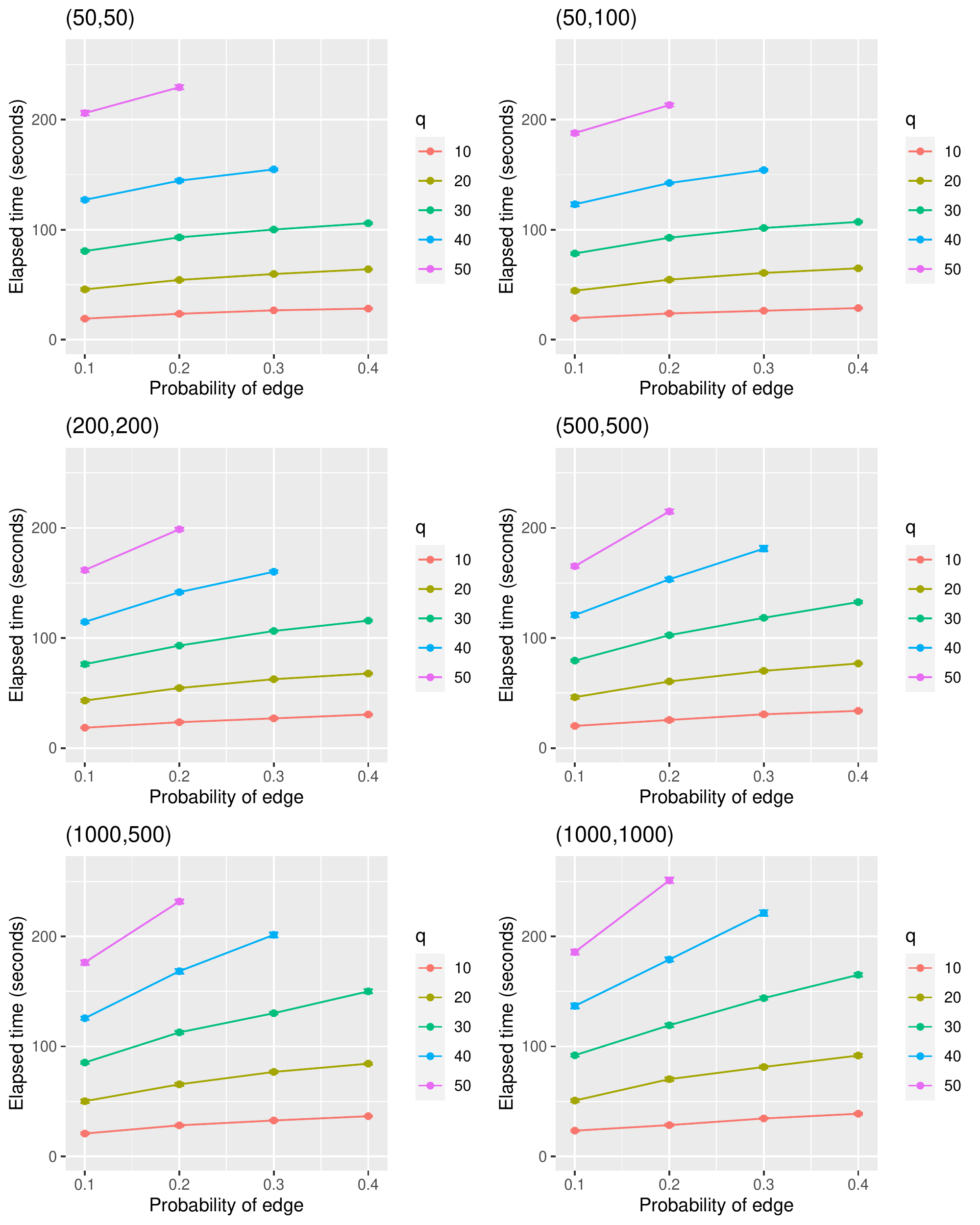}
\caption{Average computation time  for 5,000 replications with 1,000 burn-in. We removed $(n_1,n_2)\in \{ (100,100), (100,200)\}$ to save space.}
\label{fig:sim_time}
\end{figure}

We ran all of the senarios on a AMD RYZEN 7 with 8-Core 3.6 GHz CPU. The average running time for each scenario is plotted in Figure \ref{fig:sim_time}. In each panel, the computational time increases as the number of nodes, $q$, incresaes. Probability of edge, $\xi$, also has a direct effect on the average time. It seems both $q$ and $\xi$ parameters increase the time in a non-linear fashion, because both of them increases the size of $\Pa(.)$ and make it slower to compute equations in (\ref{eq:Tj}). For all the panels, the sample sizes $(n_1,n_2)$ has a little effect on the computational time, which makes it feasible to run the proposed algorithm on large sample sizes.

\begin{figure}
\centering
\includegraphics[width=\textwidth]{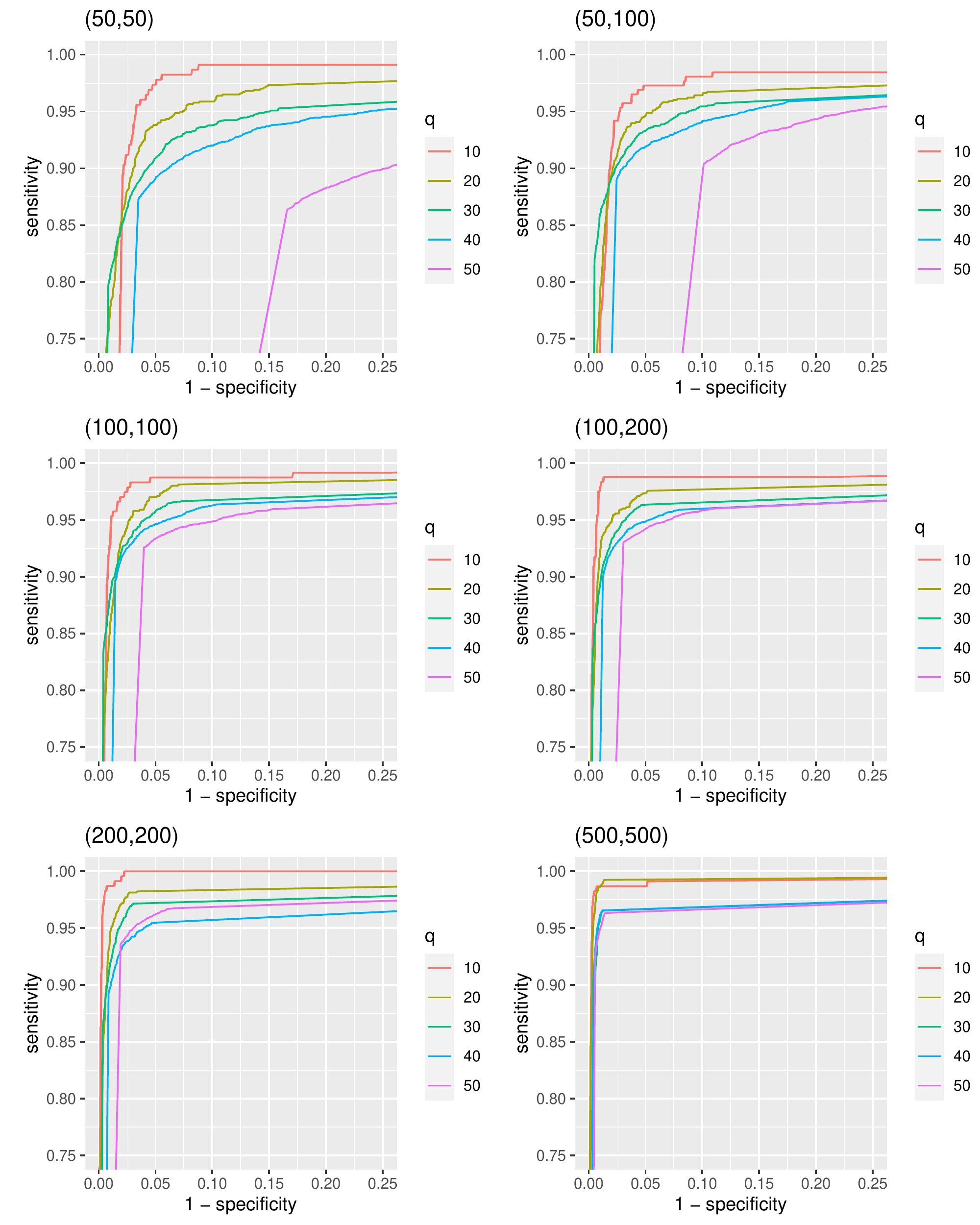}
\caption{ROC for $\xi=0.1$. Each plot is for different sample sizes $(n_1,n_2)$. Different colors represent different DAG sizes, $q$.}
\label{fig:sim_roc1}
\end{figure}

\begin{figure}
\centering
\includegraphics[width=\textwidth]{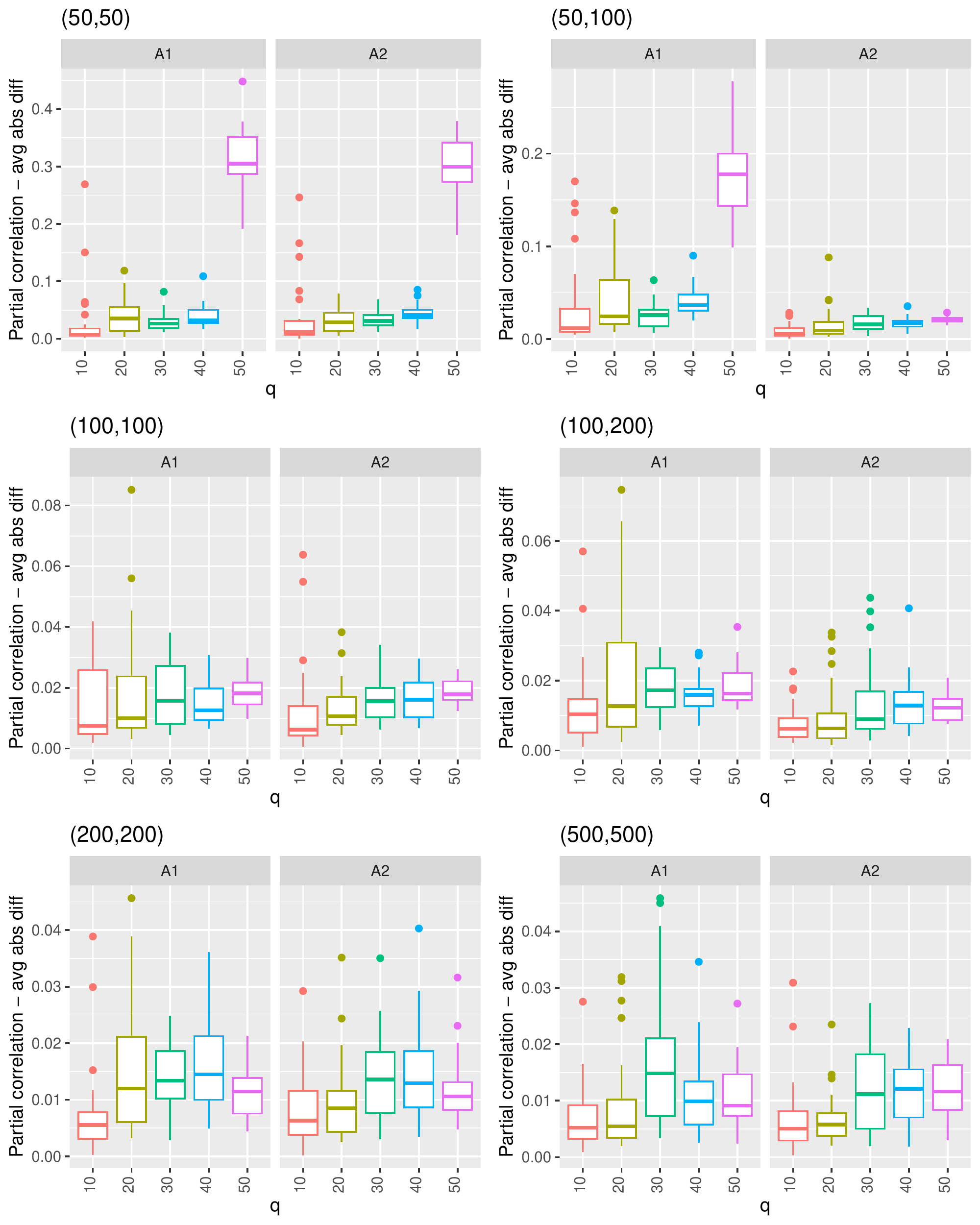}
\caption{Boxplot for the absolute difference between the estimated and true partial correlations defined in (\ref{eq:mcmc_end_partial_abs}) for $T=5000$ iterations and $\xi=0.1$. As the sample size increases, the absolute difference decreases.}
\label{fig:sim_Partial_meanabs_edge0.1}
\end{figure}

\section{Real Data}
\label{sec:real_data}
In this section we apply our method on two well-known real datasets. The first dataset, introduced by \cite{desmedt2007strong}, is the breast cancer gene data, studied in many causality papers and clinical cancer researches such as \cite{rueda2019dynamics}, \cite{momenzadeh2020using}, \cite{bertucci2020therapeutic}, \cite{poirion2021deepprog} and \cite{miao2022elevated}. The network of gene expression data of a set of genes is
expected to be sparse (\cite{cai2016joint}), which makes it sutable for the proposed method. We also validate our results with several clinical studies.

For the second dataset, we study the effect of airborne particles on the cardiovascular mortality rate (CRM) avalible in \cite{Annual_PM25}. This data is also studied by \cite{bahadori2022end}. Similar data wasalso used to check the air quality of Los Angeles County using counterfactual evaluation \cite{chen2021abatement}.

\begin{figure}
\centering
\includegraphics[width=\textwidth]{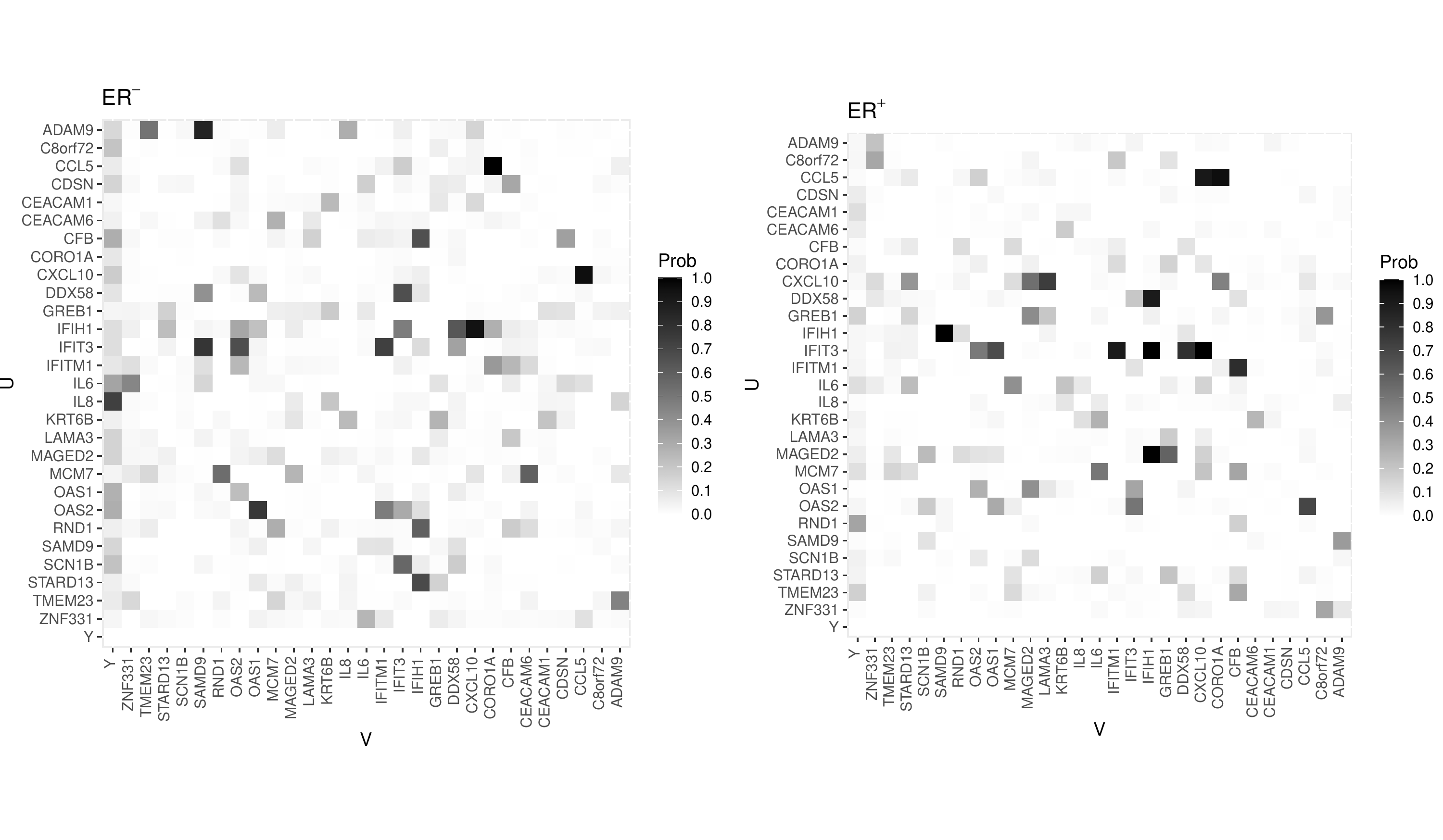}
\caption{Heat map with estimated marginal posterior probabilities of edge inclusion for each edge $u \to v$ for $A^{(1)}$ (left) and $A^{(2)}$ (right). The left plot is for ER$^-$ breast cancers and the right plot is for ER$^+$. IL-8 is highly expressed in ER$^-$ breast cancers \cite{todorovic2013interleukin}, a clear difference between the left and right plots with a direct effect on $Y$.}
\label{fig:real_data1}
\end{figure}

\subsection{Breast Cancer Data}
\label{sec:breast_cancer}
Breast cancer is the second leading cause of cancer-related mortality among women worldwide \cite{mansoori2019mir}. Finding the underlying gene's causal graph is one of the important topics in recent cancer research. For example, \cite{si2021identifying} conducted a research on identifying causality and genetic correlation on breast and ovarian cancers. The autors expressed that identifying genetic correlations can provide useful etiological insights and help prioritise likely causal relationships and can be used to identify direct causal relations and shared genetic risks for an exposure–outcome pair. Furethemore, evaluating the gene's causal effect on metastasis due to a hypothetical intervention on a specific gene may help understand which genes are more relevant.

Recently, a 76-gene prognostic signature able to predict distant metastases in patients with breast cancer was reported and the outcomes were independently validated with clinical risk assessment \cite{desmedt2007strong}. Gene expression profiling of frozen samples from 198 systemically untreated patients was performed at the Bordet Institute, blinded to genomic risk. Survival analyses, done by an independent statistician, were performed with the genomic risk and adjusted for the clinical risk. The data can be downloaded from the National Center for Biotechnology Information website \cite{Desmedt2007Data}. We denote by the binary response variable $Y$ as the occurrence of distant metastasis, i.e., $Y=1$ if the cancer is metastatic and $Y=0$ otherwise. The data were also studied by \cite{castelletti2021bayesiana} with one DAG on $q = 28$ genes. We also use the same set of genes. This datset is suitable for this study because the precision matrix for gene expression data is expected to be sparse (\cite{cai2016joint}).

If breast cancer cells have estrogen receptors, it is called ER positive breast cancer (ER$^+$), otherwise it is called ER negative (ER$^-$). When the estrogen and progesterone hormones attach to these receptors, they induce tumor-cell growth. \cite{saha2012role} showed that ER signaling contributes to metastasis, and explored possible therapeutic targets to block ER-driven metastasis. They expressed that deregulation of ER coregulators or ER extranuclear signaling has potential to promote metastasis in ER$^+$ breast cancer cells. Recentely, \cite{bertucci2020therapeutic} analyzed gene expression data from 5,342 clinically-proven breast cancer data and concluded that 
the expression profiles were very different in ER$^+$/HER2$^-$ and ER$^-$ Basal subtypes. So, we divide the data into two groups, ER$^-$ and ER$^+$. Table \ref{tbl:breast_sample_size} summerized the sample size based on the response variable $Y$ and ER status.

\begin{table}
\caption{The number of the observations per group for the breast cancer data.}
\centering
\begin{tabular}{|l | c c|} 
 \hline
  & ER negative (ER$^-$) & ER positive (ER$^+$) \\
 \hline
  cancer cell is not metastatic, $Y=0$ & 41 & 106 \\ 
 cancer cell is metastatic, $Y=1$ & 23 & 28 \\
 \hline
\end{tabular}
\label{tbl:breast_sample_size}
\end{table}

\begin{figure}
\centering
\includegraphics[width=\textwidth]{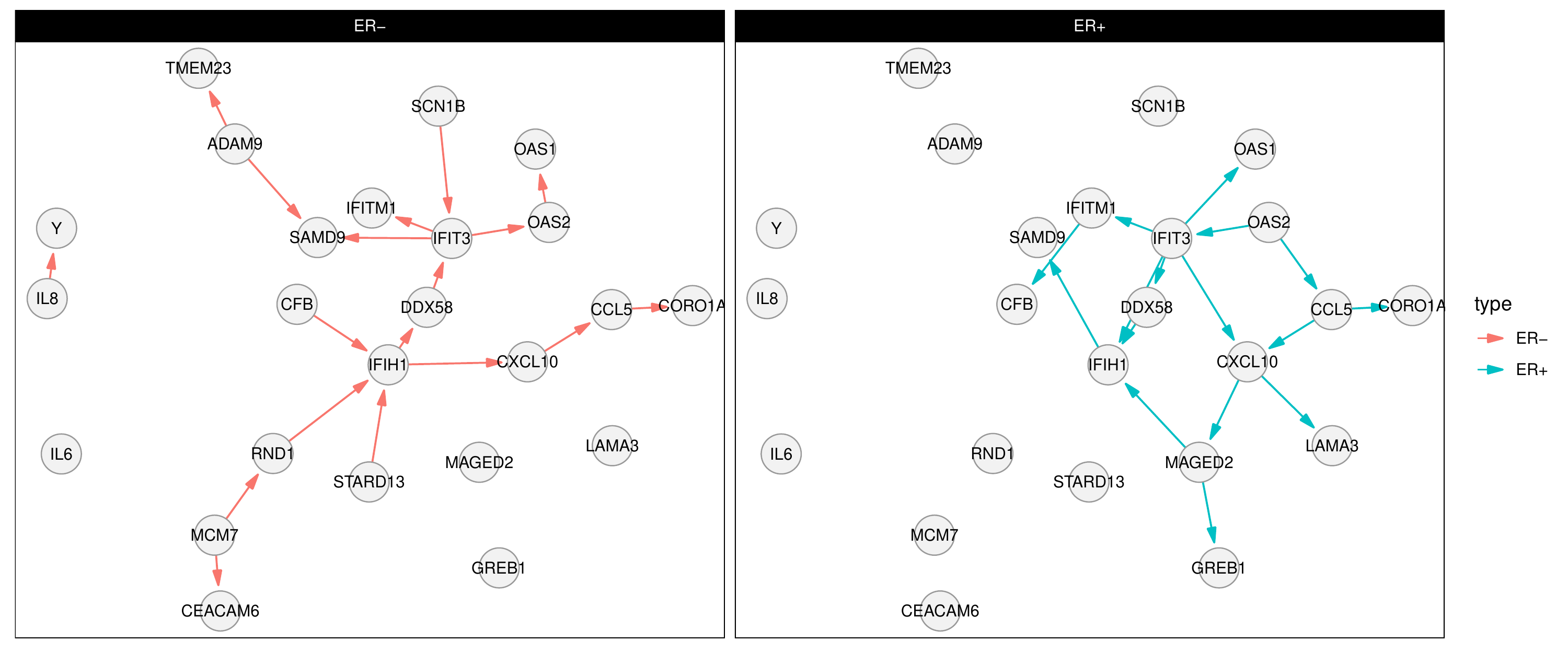}
\caption{Graph of the gene network for breast cancer. Left plot is for the ER$^-$ breast cancers and the right plot is for ER$^+$, which shows obvious differences in the gene network connections. Genes IFIH1, SAMD9, IFITM1, IFIT3, OAS1, OAS2, CXCL10, CCL5 and CORO1A are common connecting nodes between the two groups.}
\label{fig:real_data1_N}
\end{figure}

\begin{figure}
\centering
\includegraphics[width=\textwidth]{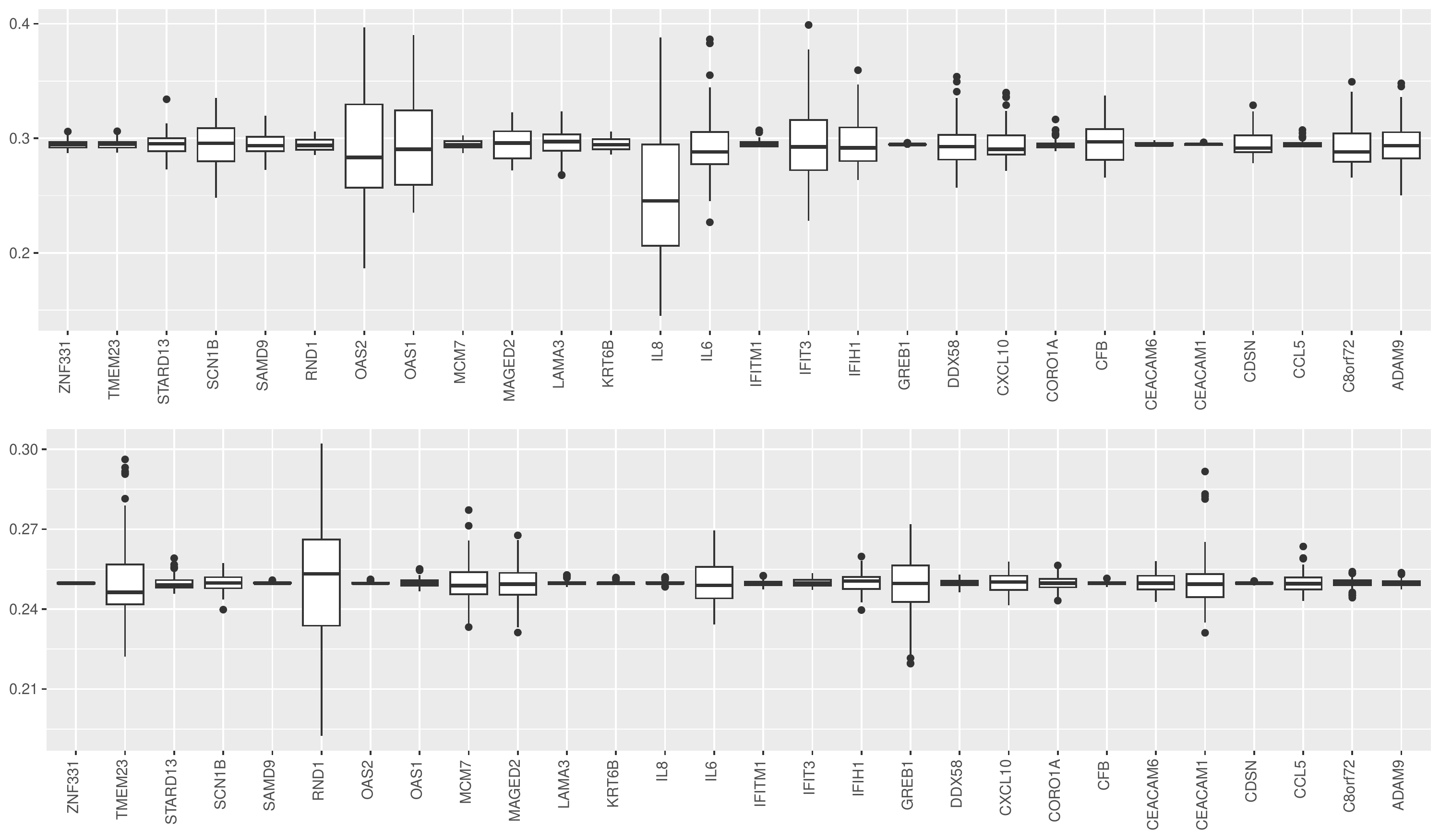}
\caption{Box-plots of BMA estimate of causal effect. The top plot is for ER$^-$ breast cancers and the bottom plot is for ER$^+$. IL-8 is highly expressed in ER$^-$ breast cancers \cite{todorovic2013interleukin}, a clear diffenrce between left and right plots.}
\label{fig:real_data1_causal_effect}
\end{figure}

\begin{figure}
\centering
\includegraphics[width=\textwidth]{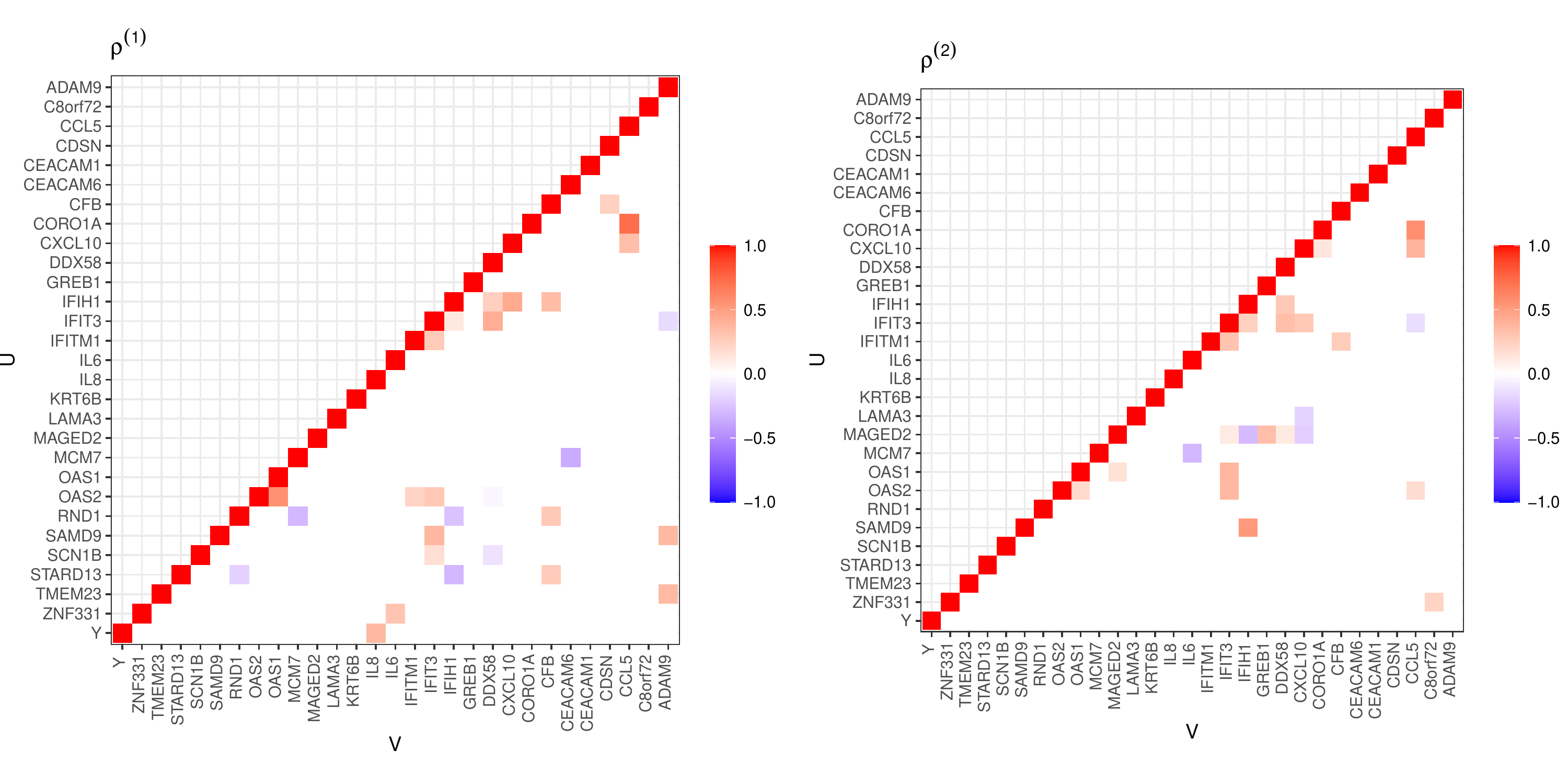}
\caption{Heat map for estimated partial correlations  $\rho^{(1)}$ and $\rho^{(2)}$. Left plot is for the  ER$^-$ breast cancers and the right plot is for ER$^+$.}
\label{fig:real_data1_Partial}
\end{figure}

We ran the proposed algorithm \ref{alg:mcmc} with $TT=200,000$ iterations and 50,000 burn-in. The heat map with estimated marginal posterior probabilities of edge inclusion for each edge $u \to v$ is plotted in figure (\ref{fig:real_data1}). The left plot is for ER$^-$ breast cancer cells and the right plot is for ER$^+$. We converted the heatmap into a DAG using threshold $0.5$ shown in figure \ref{fig:real_data1_N}. To make the DAG bigger, we removed a few isolated nodes. The causal effect of each gene on $Y$ is shown in figure \ref{fig:real_data1_causal_effect} using equation (\ref{eq:post_intervention_Y}). Genes can have positive or negative effects on other genes or on the responce variable, too. For example, gene $A$ can influence the expression of gene $B$ but not otherwise and we showed this relationship by adding the edge $A\to B$ in the corresponding DAG. But this influence can be in the positive or negative direction, meaning that the increase in expression of $A$ can increase the expression of $B$ (positive effect), or increase in the expression of $A$ can lead to decrease in the expression of $B$ (negative effect). Since the direction of the gene's effect on the responce variable $Y$ can not be inferred from the causal effect plot, we also computed the partial correlations (\ref{eq:partial}) and plotted them in figure \ref{fig:real_data1_Partial}.

In order to validate our causal gene graph, we summerize our results and compared them with the top tier clinical and research breast cancer studies in the following paragraphs.

As can be seen in figure \ref{fig:real_data1_causal_effect}, IL-8 is highly expressed in ER$^-$ breast cancer cells but not in ER$^+$. This gene has also no direct edge to the node $Y$ for ER$^+$ group, a clear difference between the left and right plots in figure \ref{fig:real_data1_N}. \cite{todorovic2013interleukin} showed that the IL-8 gene is highly expressed in ER$^-$ breast cancer cells. Such gene experesions in different groups such as different ethnicities or different cells, are important in the Simpson's paradox context. The proposed method can provide a powerful tool to check these changes and can provide more data driven guidance for biologists and other practical sicences.

IFIH1, SAMD9, IFITM1, IFIT3, OAS1, OAS2, CXCL10, CCL5 and CORO1A genes are common between the two groups. This observation is consistent with the results of research done by \cite{magbanua2015serial}. They expressed that the interferon signaling genes, such as IFIT2, IFIT1, IFITM1, IFIH1 and EML2 are associated with the recurrence-free survival (RFS) in breast tumors which was also confirmed by path analysis.

As another example, there ia a strong probability of edge inclusion from gene IL-8 to IL-6 in the research by \cite{castelletti2021bayesiana}, but we could not see this edge when we grouped the cells based on the ER$^-$ and ER$^+$ in figure \ref{fig:real_data1}. \cite{weitkamp2002interleukin} also  mentioned that the overall correlation between IL-6 and IL-8 is $\rho = 0.50$ (p--value $<0.001$), using a sample size of 182. $\rho = 0.50$ is not a strong correlation, and their study was not on breast cancer cells. The aim of their study was to evaluate the correlation of IL-6 and IL-8 values during neonatal clinical infection and to assess whether IL-8 would be a more beneficial infection marker. Besides the data quality and sample size, One possible explanation can be related to Simpson's paradox. Since the goal of this study is not to check the Simpson's paradox, we refer the readers to \cite{wagner1982simpson} for more information on Simpson's paradox.


\cite{mouly2019rnd1} showed the RND1 gene is involved in oncogenesis and response to cancer therapeutics. \cite{okada2015rho} identified the RND1 gene as a candidate metastasis suppressor in basal-like and triple-negative breast cancer through bioinformatics analysis. Triple negative breast cancer is ER$^-$, progesterone receptor--negative and HER2--negative. Inactivation of RND1 in mammary epithelial cells induced highly undifferentiated and invasive tumors in mice. Although there is no direct edge connecting RND1 to $Y$ in figure \ref{fig:real_data1_N}, this gene is connected to other genes in ER$^-$ DAG while it is isolated in ER$^+$. The partial correlation between this gene and the connected genes (MCM7 and IFIH1) for ER$^-$ in figure \ref{fig:real_data1_Partial} is nagative, which can possibly explain why removing RND1 from mice results in more tumors, although we do not know the effectdirection of MCM7 and IFIH1 on $Y$ and this hypothesis needs more research, we thought it is good to illustrate the concept of nagative effects of one gene on other genes or on the response variable.

In a recent paper published in 2022, the authors showed that the Melanoma-associated antigen D2 (MAGED2) gene positively regulated breast cancer cell metastasis \cite{thakur2022deletion}. This gene also affects both ER$^+$ and ER$^-$ breast cancer groups (Figure \ref{fig:real_data1_causal_effect}). 
\cite{jia2019prognostic} also showed that the MAGE gene family, such as MAGED2, is influential on breast cancer.


\subsection{Cardiovascular Mortality Rate}
\label{sec:cmr}
For the second real data analysis, we study the impact of PM2.5 particle level on the cardiovascular mortality rate (CMR). The PM2.5 particle level and the mortality rate are measured by $\mu g /m^3$ and the number of annual deaths due to cardiovascular conditions per 100,000 people, respectively. The data comprises 2,132 counties and is provided by the US National Studies on Air Pollution and Health and is publicly available under U.S. Public Domain license \cite{Annual_PM25}. To simplify the experiment setup, we use only the data for 2010. A similar approach was taken by \cite{bahadori2022end}. The data includes 10 variables such as poverty rate, population, educational attainment, vacant housing units and household income, which we use as confounders. For the response variable $Y$, we transformed the CMR data into a high and low binary variable by settig the thereshold $250$. We divided the data into two groups based on the population size. We labeled a county as small if the population was less than $40,000$ people and big otherwise. We also removed \texttt{SES\_index} (Socioeconomic status index) because it has a high correlation with \texttt{pctfam\_pover} (percentage of families below poverty level) and \texttt{femaleHH} (percentage of female household, no husband present). We also included the number of establishments in each county (\texttt{establishments}) in our model. The establishments data is included in the \cite{Annual_PM25} database too. For more information on the county business pattern establishments, see \cite{uscb2020}.

\begin{table}
\caption{The number of the observations per group for the CMR data.}
\centering
\begin{tabular}{|c | c c|} 
 \hline
  & Small city $A^{(1)}$ & Big city $A^{(2)}$ \\
 \hline
  Low CRM ($Y=0$) & 432 & 577 \\ 
 High CRM ($Y=1$) & 520 & 397 \\
 \hline
\end{tabular}
\end{table}

\begin{figure}
\centering
\includegraphics[width=\textwidth]{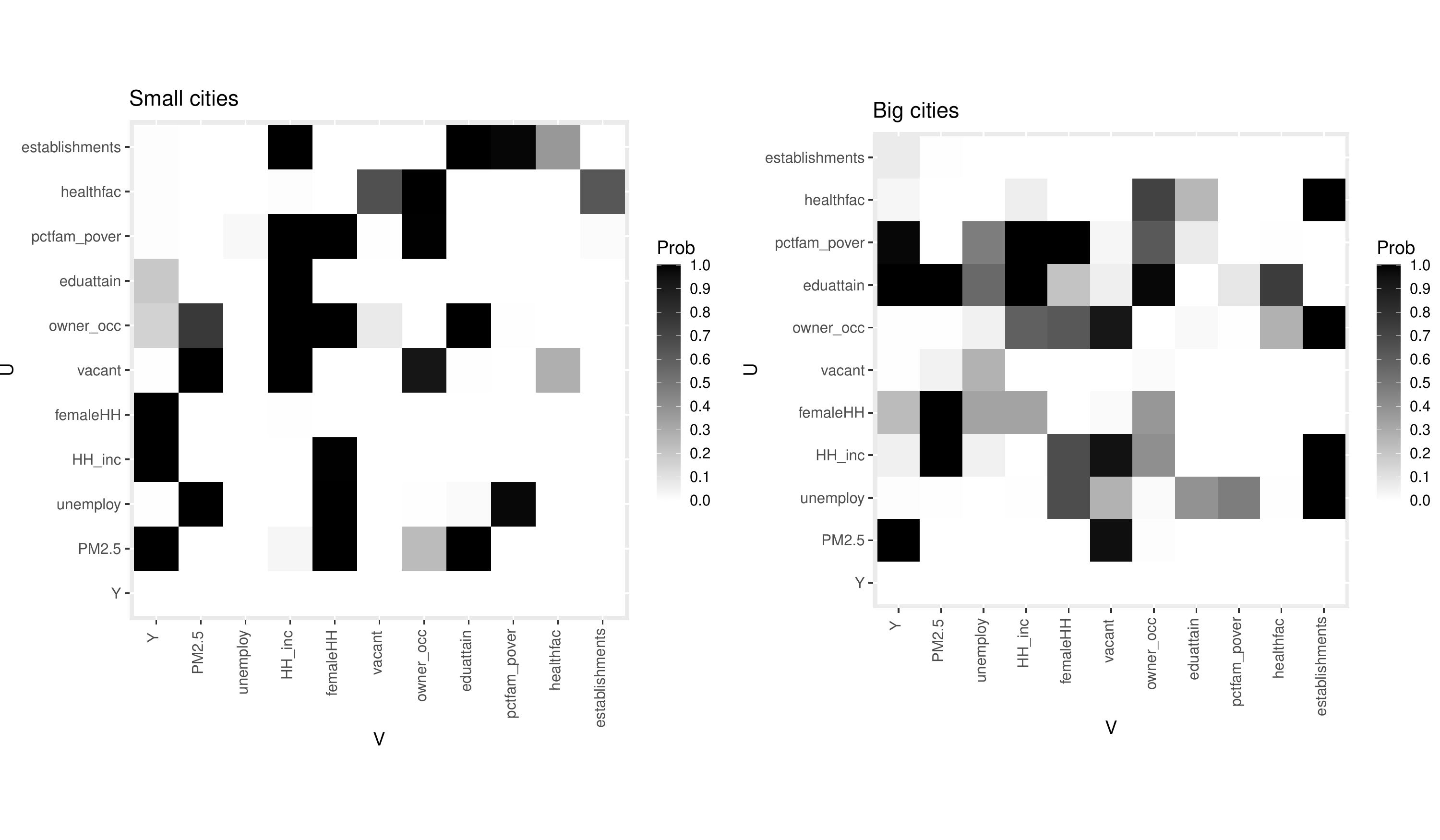}
\caption{Heat map with estimated marginal posterior probabilities
of edge inclusion for each edge $u \to v$ for $A^{(1)}$ (left) and $A^{(2)}$ (right).}
\label{fig:real_data2}
\end{figure}

\begin{figure}
\centering
\includegraphics[width=\textwidth]{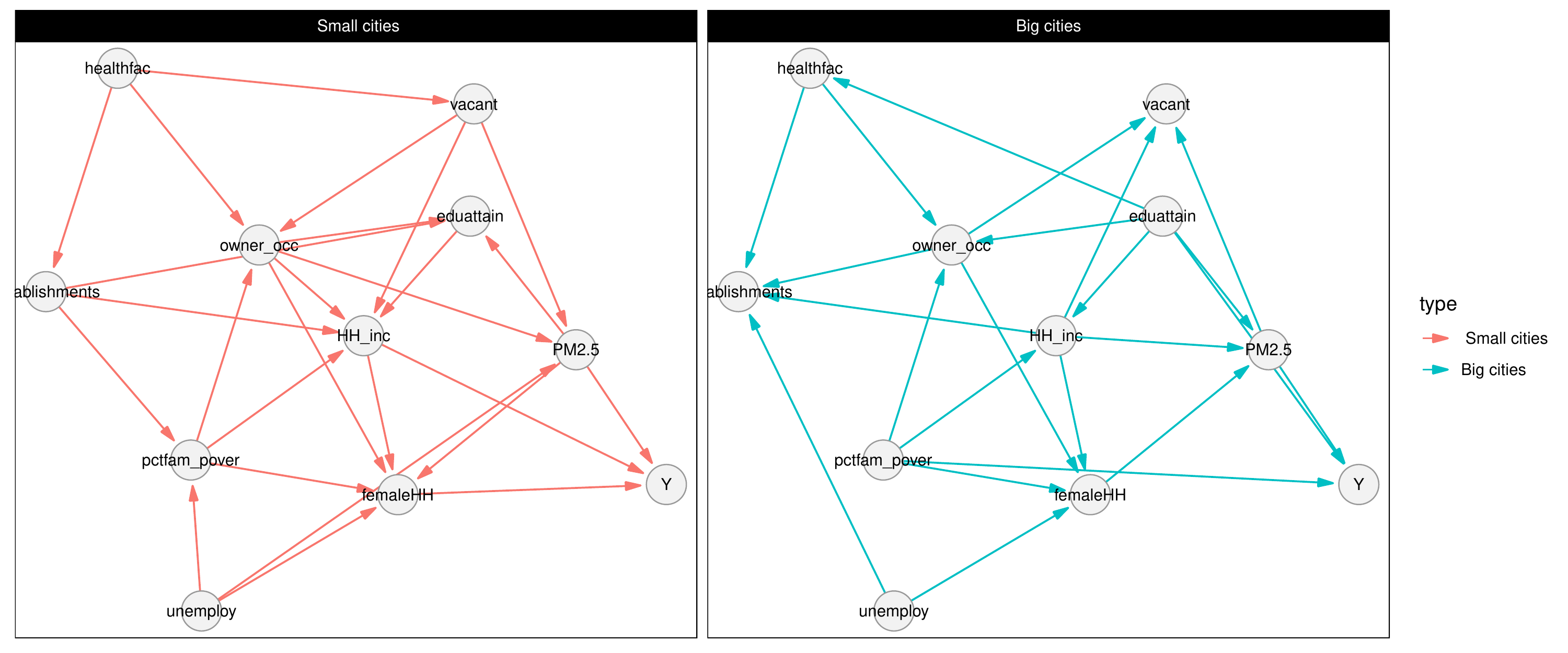}
\caption{Graph of the nodes. Left plot is for the small cities and the right plot is for the big cities.}
\label{fig:real_data2_N}
\end{figure}

\begin{figure}
\centering
\includegraphics[width=\textwidth]{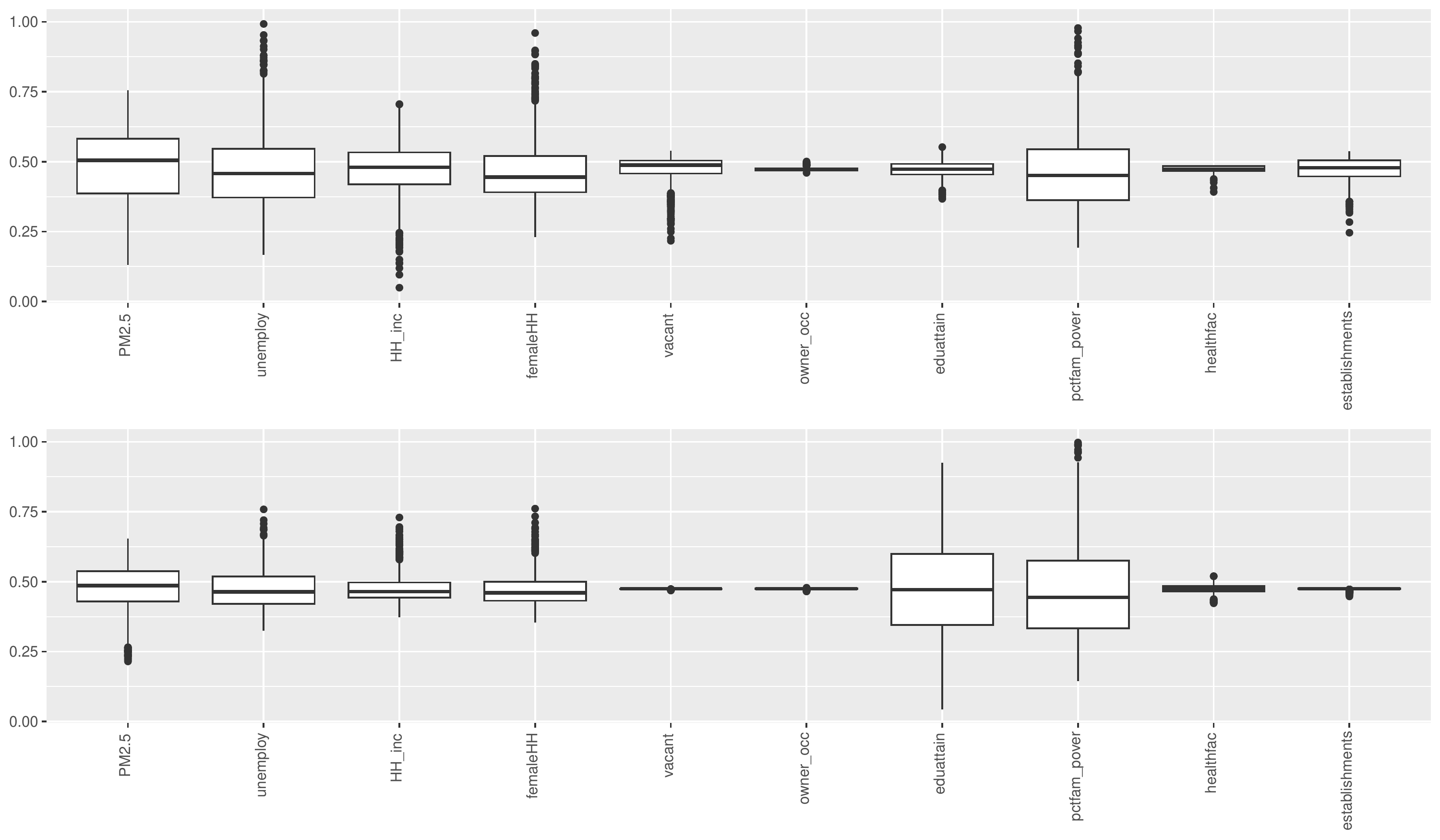}
\caption{Boxplots of BMA estimate of causal effect.}
\label{fig:real_data2_causal_effect}
\end{figure}

PM2.5 is the main cause of CMR in both small and big population counties, a clear difference in the heat map of the edge prediction in figures \ref{fig:real_data2} and \ref{fig:real_data2_N} with big effect size (figure \ref{fig:real_data2_causal_effect}). 
Household income (\texttt{HH\_inc}) and \texttt{femaleHH} are directely connected to CMR for small cities with moderate effect size, while those are indirectley connected to CMR through PM2.5.

Healthcare facility density per 1,000 individuals (\texttt{healthfac}) has no direct effect on CMR. \texttt{pctfam\_pover} has a strong relation with \texttt{HH\_inc} in both groups, meaning poverty leads to less household income. Educational attainment (\texttt{eduattain}) has a direct effect on the \texttt{HH\_inc} in both groups. Unemployment rate (\texttt{unemploy}) is negatively correlated with \texttt{HH\_inc} in big cities.

\begin{figure}
\centering
\includegraphics[width=\textwidth]{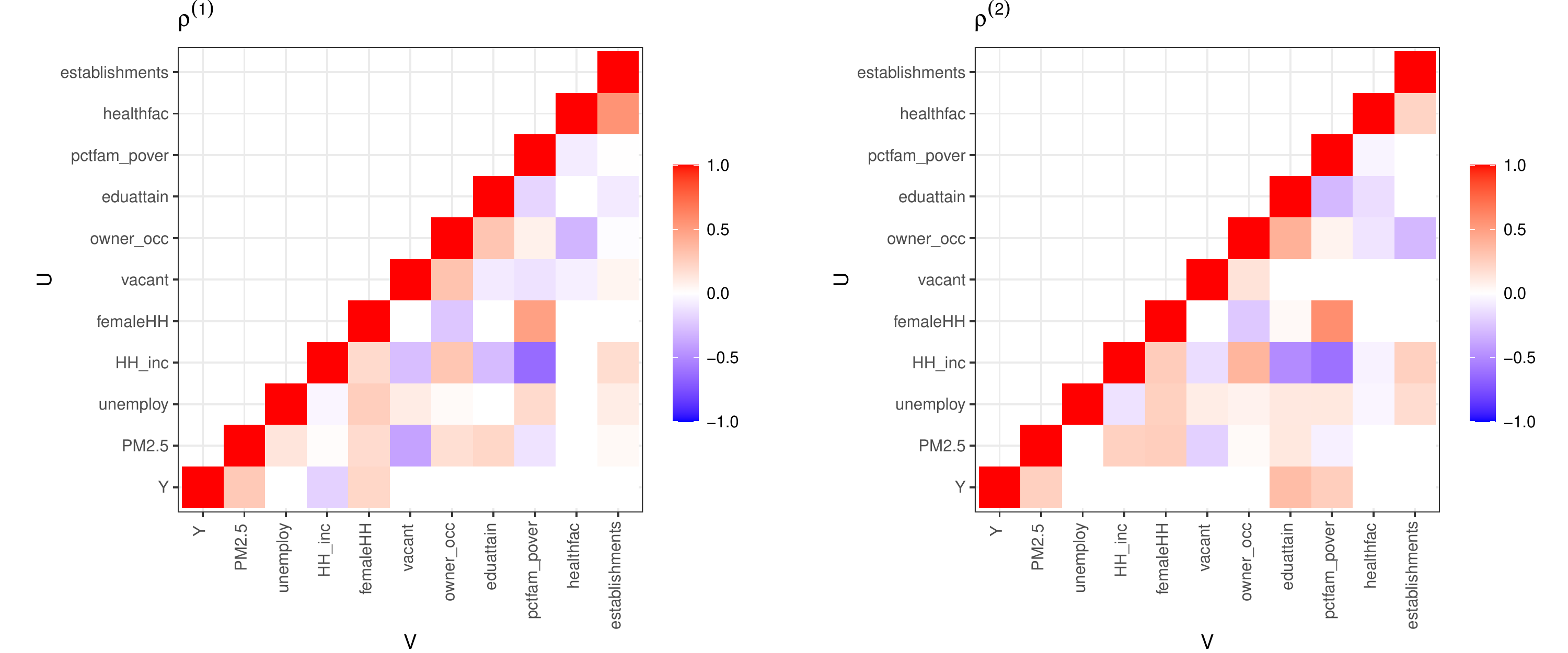}
\caption{Heat map for estimated partial correlations  $\rho^{(1)}$ and $\rho^{(2)}$. Left plot is for the small cities and the right plot is for the big cities.}
\label{fig:real_data2_Partial}
\end{figure}

\section{Conclusion}
\label{sec:conclusion}

We considered modeling a binary response variable together with a set of independent variables for two groups  under observational data. This work can  be extended for groups with more than two elements simply be using $k>2$ in all equations. We could speed up the MCMC convergence by picking better initial graphs for the DAGs using modified Cholesky decomposition.

In this work, we assumed equal edge inclusion of the edges while we can improve it by adding some weights to the edge inclusion. For example, we can add more weights to the edges coming into node 1, or we can decrease the probability of node inclusion inversely proportionate to the degree of its parent node. This method is useful if someone wants to keep the proposed graph as a sparse DAG at each MCMC iteration.

There are many real examples where there are more than one binary variables in data. This method can model just one binary response variable. Categorical or nominal response variables also have many applications which are not under the scope of this paper and need more research. Normality assumption of the independent variables is also a limitation of this method which may produce biased outputs. 

Regardless of those limitations which are not under the proposed model assumptions, we were able to show that this method could estimate the DAG structure and the model parameters accurately. We also demonstrated its value in real well-known datasets, especially using two DAGs.


\bibliographystyle{johd}
\bibliography{bib}

\appendix
\section*{Appendix}
\renewcommand{\thesubsection}{\Alph{subsection}}

\subsection{Full posterior distribution}
\label{sec:appendix_full_posterior}
The full posterior distribution is
\begin{align}
    & f\left( \calD^{(1)}, \calD^{(2)}, \boldsymbol{L}^{(1)}, \boldsymbol{L}^{(2)}, \boldsymbol{D}, \boldsymbol{X}_1^{(1)}, \boldsymbol{X}_1^{(2)}, \theta \big| \boldsymbol{y}^{(1)}, \boldsymbol{y}^{(2)}, \boldsymbol{X}_{-1}^{(1)}, \boldsymbol{X}_{-1}^{(2)} \right) \nonumber \\
    & \propto f\left( \boldsymbol{y}^{(1)}, \boldsymbol{y}^{(2)}, \boldsymbol{X}_{-1}^{(1)}, \boldsymbol{X}_{-1}^{(2)}, \boldsymbol{X}_1^{(1)}, \boldsymbol{X}_1^{(2)} \Big| \calD^{(1)}, \calD^{(2)}, \boldsymbol{L}^{(1)}, \boldsymbol{L}^{(2)}, \boldsymbol{D}, \theta \right)
    f\left( \calD^{(1)}, \calD^{(2)}, \boldsymbol{L}^{(1)}, \boldsymbol{L}^{(2)}, \boldsymbol{D}, \theta \right) \nonumber \\
    & = f\left( \boldsymbol{y}^{(1)}, \boldsymbol{y}^{(2)}, \boldsymbol{X}^{(1)}, \boldsymbol{X}^{(2)} \Big| \calD^{(1)}, \calD^{(2)}, \boldsymbol{L}^{(1)}, \boldsymbol{L}^{(2)}, \boldsymbol{D}, \theta \right)
    f\left(\boldsymbol{L}^{(1)}, \boldsymbol{L}^{(2)}, \boldsymbol{D} \big|  \calD^{(1)}, \calD^{(2)}, \theta \right)
    f\left( \calD^{(1)} \right) f\left( \calD^{(2)} \right) f\left( \theta \right) \nonumber \\
    & = \prod_{k=1}^2 f\left( \boldsymbol{y}^{(k)},  \boldsymbol{X}^{(k)} \Big| \calD^{(k)}, \boldsymbol{L}^{(k)}, \boldsymbol{D}, \theta \right)
    f\left(\boldsymbol{L}^{(k)} \big| \boldsymbol{D},  \calD^{(k)} \right) f\left( \calD^{(k)} \right) \times f\left(\boldsymbol{D} \big|  \calD^{(1)}, \calD^{(2)} \right) f\left( \theta \right),
\label{eq:full_post_proof_calD}
\end{align}
where $f\left( \theta \right)\propto 1$, too.

\subsection{Acceptance Probability for  $\calD^{'(k)}$}
\label{sec:proof_acceptance_probability_calD}

We summerize the PAS algorithm here at first. Consider $N$ distinct models, $\mathcal{M}_1, \dots, \mathcal{M}_N$, each one with set of parameters $\boldsymbol{\theta}_{\nu}$ and assume the true data generating model is one of the $N$ models. The associated likelihood function for each model $\mathcal{M}_\nu$ is $f(\boldsymbol{X} | \boldsymbol{\theta}_\nu, \mathcal{M}_\nu)$, where $\boldsymbol{X}=(X_1\dots, X_q)$. Consider a move from the current model $\mathcal{M}_\nu$ to a new model $\mathcal{M}_{\nu'}$. Suppose 
\begin{enumerate}[label=\alph*)]
    \item there exists a subvector $(\boldsymbol{\theta}_{\nu'})_\mathcal{U}$ of the parameter $\boldsymbol{\theta}_{\nu'}$ for a new model $\mathcal{M}_{\nu'}$ such that $f\left((\boldsymbol{\theta}_{\nu'})_\mathcal{U} \,\big| \, (\boldsymbol{\theta}_{\nu'})_{-\mathcal{U}}, \mathcal{M}_{\nu'}, \boldsymbol{X} \right)$ is available in closed form,
    \item in the current model $\mathcal{M}_{\nu}$, there exists an equivalent subset of parameters $(\boldsymbol{\theta}_{\nu})_{-\mathcal{U}}$ with the same dimension as $(\boldsymbol{\theta}_{\nu'})_{-\mathcal{U}}$.
\end{enumerate}
The PAS algorithm is
\begin{enumerate}
    \item Propose $\mathcal{M}_{\nu'} \sim q(\mathcal{M}_{\nu'} | \mathcal{M}_\nu)$ and set $(\boldsymbol{\theta}_{\nu'})_{-\mathcal{U}} = (\boldsymbol{\theta}_{\nu})_{-\mathcal{U}}$.

    \item Accept $\mathcal{M}_{\nu'}$ with probability $\alpha = \min\{1, r_\nu\}$, where
    \begin{equation}
        r_\nu = \frac{f\left(\mathcal{M}_{\nu'} \,\big| \, (\boldsymbol{\theta}_{\nu'})_{-\mathcal{U}}, \boldsymbol{X} \right) q(\mathcal{M}_{\nu} |\mathcal{M}_{\nu'})}
        {f\left(\mathcal{M}_{\nu} \,\big| \, (\boldsymbol{\theta}_{\nu})_{-\mathcal{U}}, \boldsymbol{X} \right) q(\mathcal{M}_{\nu'} | \mathcal{M}_{\nu})}
    \label{eq:PAS_alg_rk}
    \end{equation}
    and
    \begin{equation}
    f\left(\mathcal{M}_{\nu'} \,\big| \, (\boldsymbol{\theta}_{\nu'})_{-\mathcal{U}}, \boldsymbol{X} \right) = \int f\left(\mathcal{M}_{\nu'}, (\boldsymbol{\theta}_{\nu'})_{\mathcal{U}} \,\big| \, (\boldsymbol{\theta}_{\nu'})_{-\mathcal{U}}, \boldsymbol{X} \right)  d(\boldsymbol{\theta}_{\nu'})_{\mathcal{U}}.
    \label{eq:proof_PAS_int}
    \end{equation}
    
    \item If $\mathcal{M}_{\nu'}$ is accepted, generate $(\boldsymbol{\theta}_{\nu'})_\mathcal{U} \sim f\left((\boldsymbol{\theta}_{\nu'})_\mathcal{U} \,\big| \, (\boldsymbol{\theta}_{\nu'})_{-\mathcal{U}}, \mathcal{M}_{\nu'}, \boldsymbol{X} \right)$, Otherwise, generate \\ $(\boldsymbol{\theta}_{\nu})_\mathcal{U} \sim f\left((\boldsymbol{\theta}_{\nu})_\mathcal{U} \,\big| \, (\boldsymbol{\theta}_{\nu})_{-\mathcal{U}}, \mathcal{M}_{\nu}, \boldsymbol{X} \right)$.

    \item Update the parameters $\boldsymbol\theta_{\nu'}$ if $\mathcal{M}_{\nu'}$ is accepted using standard MCMC steps. Otherwise, update the parameters $\boldsymbol{\theta}_{\nu}$ using standard MCMC steps.
\end{enumerate}

To adapt this algorithm to the proposed method, we need to find which edges been have changed in the new DAGs $\mathcal{D'}^{(1)}$ and $\mathcal{D'}^{(2)}$ to define the set $\boldsymbol{\theta}_{-\mathcal{U}}$ and them to compute equation (\ref{eq:proof_PAS_int}) for them, accordingly. So, we provide the proofs for the $\texttt{Insert}(i\to j)$, $\texttt{Delete}(i\to j)$ and $\texttt{Reverse}(i\to j)$ operators, separetely in the following subsections.

\subsubsection{One Parent Node is Changed}
\label{sec:proof_acceptance_probability_calD_1parent}


In algorithm \ref{alg:mcmc}, new $\calD^{'(1)}$ and $\calD^{'(2)}$ are  proposed seperately, so we need to compute the integral in equation (\ref{eq:proof_PAS_int}) for $k\in \{1,2\}$ independentey. For the operastors $\texttt{Insert}(i\to j)$ and $\texttt{Delete}(i\to j)$, the difference between the proposed DAG $\calD^{'(k)}$ and $\calD^{(k)}$ is just one parent node, denoted by node $j$, so $\mathcal{U} = (\sigma^2_j$, $\boldsymbol{L}_{\preceq j \succ})$ and the set of parameters in PAS algorithm is $\boldsymbol{\theta}=(\boldsymbol{D}^{(k)}, \boldsymbol{L}^{(k)})$. For $j>1$ and  using Bayes theorem, equation (\ref{eq:proof_PAS_int}) becomes
\begin{align}
    & f\left(\calD^{(k)} \,|\, \boldsymbol{X}^{(k)}, \boldsymbol{\theta}_{-\mathcal{U}} \right) 
    \propto f\left(\boldsymbol{X}^{(k)}, \boldsymbol{\theta}_{-\mathcal{U}} \,|\, \calD^{(k)} \right) f\left( \calD^{(k)} \right) \nonumber \\
    & = f\left( \calD^{(k)} \right) \int_{\mathbb{R}^{|\Pa(j)|}} \int_0^\infty   f\left(\boldsymbol{X}^{(k)}, \boldsymbol{D}^{(k)}, \boldsymbol{L}^{(k)} | \calD^{(k)} \right) \, d\boldsymbol{L}^{(k)}_{\preceq j \succ} \, d\sigma^2_j \nonumber \\
    & = f\left( \calD^{(k)} \right) \prod_{r\neq j} f\left(\boldsymbol{X}^{(k)}_r, \sigma^2_r, \boldsymbol{L}^{(k)}_{\preceq r \succ} \Big| \calD^{(k)} \right)
    \times \int \int f\left(\boldsymbol{X}^{(k)}_j  \Big| \sigma^2_j, \boldsymbol{L}^{(k)}_{\preceq j \succ} \right)
    f\left(\boldsymbol{L}^{(k)}_{\preceq j \succ} | \sigma^2_j \right) 
    f\left(\sigma^2_j | \calD^{(k)}\right) \, d\boldsymbol{L}^{(k)}_{\preceq j \succ} \, d\sigma^2_j,
\label{eq:proof_calD_one_parent_PAS}
\end{align}
where $f\left(\sigma^2_j | \calD^{(k)}\right)$ is the prior for $\sigma^2_j$ assuming there is just one DAG $\calD^{(k)}$, i.e., $\sigma^2_j | \calD^{(k)}\sim \mathrm{I\text{-}Ga}\left(\frac{a_j^{(k)}}{2},\frac{g_k}{2}\right)$.

To simplify the equations, we define
\begin{align}
    m(\boldsymbol{X}^{(k)}_j | \boldsymbol{X}^{(k)}_{\Pa(j)}, \calD^{(k)}) := \int \int f\left(\boldsymbol{X}^{(k)}_j  \Big| \sigma^2_j, \boldsymbol{L}^{(k)}_{\preceq j \succ} \right)
    f\left(\boldsymbol{L}^{(k)}_{\preceq j \succ} | \sigma^2_j \right) 
    f\left(\sigma^2_j | \calD^{(k)}\right) \, d\boldsymbol{L}^{(k)}_{\preceq j \succ} \, d\sigma^2_j.
\label{eq:proof_calD_one_parent_PAS_mj}
\end{align}
The integrand in (\ref{eq:proof_calD_one_parent_PAS_mj}) can be written as
\begin{align}
    &f\left( \boldsymbol{X}^{(k)}_j \Big| {\sigma^{2}_j}, \boldsymbol{L}^{(k)}_{\preceq j \succ} \right)
    f\left( \boldsymbol{L}^{(k)}_{\preceq j \succ} \Big| {\sigma^{2}_j} \right) f\left({\sigma^{2}_j} | \calD^{(k)} \right) \nonumber \\
    & =\frac{1}{(2\pi)^\frac{n_k}{2}}{|\sigma^2_j\boldsymbol{I}_{n_k}|}^{-\frac{1}{2}} \exp\left\{-\frac{1}{2\sigma^2_j} \left(\boldsymbol{X}^{(k)}_j + \boldsymbol{X}^{(k)}_{\Pa(j)}\boldsymbol{L}^{(k)}_{\preceq j \succ}\right)' \left(\boldsymbol{X}^{(k)}_j + \boldsymbol{X}^{(k)}_{\Pa(j)}\boldsymbol{L}^{(k)}_{\preceq j \succ}\right)\right\} \nonumber \\
    & \times \frac{1}{{(2\pi)^\frac{{|\Pa^{(k)}(j)|}}{2}}} \Big|\frac{1}{g_k}\sigma^2_j\boldsymbol{I}_{|\Pa^{(k)}(j)|}\Big|^{-\frac{1}{2}} \exp\left\{-\frac{g_k}{2\sigma^2_j}(\boldsymbol{L}^{(k)}_{\preceq j \succ})' \boldsymbol{L}^{(k)}_{\preceq j \succ}\right\} \nonumber \\
    & \times C \,.\, (\sigma^2_j)^{-\frac{a^{(k)}_j}{2}-1} \exp\left\{-\frac{\frac{g_k}{2}}{\sigma^2_j}\right\},
\label{eq:proof_calD_posterior_Xk_likelihood}
\end{align}
where $C = {(\frac{g_k}{2}) }^{\frac{a^{(k)}_j}{2}}/ \Gamma(\frac{a^{(k)}_j}{2})$.

By rearranging the terms containing $\boldsymbol{L}_{\preceq j \succ}^{(k)}$ in (\ref{eq:proof_calD_posterior_Xk_likelihood}), it is easy to show that
\begin{align}
    \boldsymbol{L}_{\preceq j \succ}^{(k)} | \sigma^2_j, \boldsymbol{X}^{(k)} 
    \sim \calN_{|\Pa^{(k)}(j)|} \left(-\hat{\boldsymbol{L}}'^{(k)}_j, \sigma^2_j (\bar{\boldsymbol{T}}^{(k)}_j)^{-1} \right),
\end{align}
so, the inner integral becomes
\begin{align}
\int f\left(\boldsymbol{X}^{(k)}_j  \Big| \sigma^2_j, \boldsymbol{L}^{(k)}_{\preceq j \succ} \right)
    f\left(\boldsymbol{L}^{(k)}_{\preceq j \succ} | \sigma^2_j \right) \, d\boldsymbol{L}^{(k)}_{\preceq j \succ} = (2\pi)^{\frac{|\Pa^{(k)}(j)|}{2}} \,\,
    {\left| \frac{\bar{ \boldsymbol{T}}_j^{(k)}}{\sigma^2_j}\right|}^{-\frac{1}{2}}
    \exp\left\{\frac{1}{2\sigma^2_j} \hat{\boldsymbol{L}}^{'(k)}_j \bar{\boldsymbol{T}}^{(k)}_j \hat{\boldsymbol{L}}^{(k)}_j \right\},
\label{eq:proof_one_parent_inner_int}
\end{align}
which is the constant terms in a random variable that follows $\calN_{|\Pa^{(k)}(j)|} \left(-\hat{\boldsymbol{L}}'^{(k)}_j, \sigma^2_j (\bar{\boldsymbol{T}}^{(k)}_j)^{-1} \right)$. Furthermore, using (\ref{eq:proof_one_parent_inner_int}) and other terms containg $\sigma^2_j$ in (\ref{eq:proof_calD_posterior_Xk_likelihood}) together with the prior distribition for $\sigma^2_j | \calD^{(k)}$, we can show that
\begin{align}
    f\left(\sigma^2_j | \boldsymbol{X}^{(k)}_j, \calD^{(k)} \right)
    & \propto ({\sigma^2_j})^{-\frac{n_k}{2}-\frac{a^{(k)}_j}{2} - 1 -\frac{|\Pa^{(k)}(j)|}{2} +\frac{|\Pa^{(k)}(j)|}{2} } \nonumber\\
    & \times \exp \left\{\frac{-\frac{1}{2}(g_k + {\boldsymbol{X}^{'(k)}_j} \boldsymbol{X}^{(k)}_j -\hat{\boldsymbol{L}}^{'(k)}_j \bar{\boldsymbol{T}}^{(k)}_j \hat{\boldsymbol{L}}^{(k)}_j)}{\sigma^2_j}\right\},
\end{align}
where we substitute ${\left| \frac{ \bar{ \boldsymbol{T}}_j^{(k)}}{\sigma^2_j} \right|}^{-\frac{1}{2}} =  {\sigma^2_j}^{\frac{|\Pa^{(k)}(j)|}{2}}  {\left| \bar{\boldsymbol{T}}_j^{(k)} \right|}^{-\frac{1}{2}}$. It is easy to see that the posterior is
\begin{align}
    \sigma^2_j | \boldsymbol{X}^{(k)}_j,  \calD^{(k)} \sim \mathrm{I-Ga}\left(\frac{a^{(k)}_j}{2} +\frac{n_k}{2} , \frac{1}{2} K^{(k)} \right),
\end{align}
where $K^{(k)} = g_k + {\boldsymbol{X}^{'(k)}_j} \boldsymbol{X}^{(k)}_j -\hat{\boldsymbol{L}}^{'(k)}_j \bar{\boldsymbol{T}}^{(k)}_j \hat{\boldsymbol{L}}^{(k)}_j$.

Finally by bringing back the constant terms from (\ref{eq:proof_calD_posterior_Xk_likelihood}), the outer integral in (\ref{eq:proof_calD_one_parent_PAS_mj}) becomes
\begin{align}
    m(\boldsymbol{X}^{(k)}_j | \boldsymbol{X}^{(k)}_{\Pa(j)}, \calD^{(k)})
    & = \int_0^\infty f\left(\sigma^2_j | \boldsymbol{X}^{(k)}_j, \calD^{(k)} \right) \, d\sigma^2_j \nonumber \\
    & = (2\pi)^{-\frac{n_k}{2}} \,
    \frac{{|\boldsymbol{T}}_j^{(k)}|^{1/2}}{|\bar{ \boldsymbol{T}}_j^{(k)}|^{1/2}} \,
    \frac{\Gamma(\frac{a^{(k)}_j}{2} +\frac{n_k}{2})}{\Gamma(\frac{a^{(k)}_j}{2})}\,
    {(\frac{g_k}{2}) }^{a^{(k)}_j} \,
    \left[ \frac{1}{2} (g_k + {\boldsymbol{X}^{'(k)}_j} \boldsymbol{X}^{(k)}_j -\hat{\boldsymbol{L}}^{'(k)}_j \bar{\boldsymbol{T}}^{(k)}_j \hat{\boldsymbol{L}}^{(k)}_j) \right]^{-(a^{(k)}_j+n_k)/2 }.
\end{align}


For $j=1$, since $\sigma^2_j=1$ then ther would be just one integral, so
\begin{align}
    m(\boldsymbol{X}^{(k)}_1 | \boldsymbol{X}^{(k)}_{\Pa(1)}, \calD^{(k)}) := \int f\left(\boldsymbol{X}^{(k)}_1  \Big|  \boldsymbol{L}^{(k)}_{\preceq 1 \succ} \right)
    f\left(\boldsymbol{L}^{(k)}_{\preceq 1 \succ} \right) 
     \, d\boldsymbol{L}^{(k)}_{\preceq 1 \succ},
\label{eq:proof_calD_one_parent_PAS_m1}
\end{align}
which we computed in (\ref{eq:proof_one_parent_inner_int}). Bringing back the constant terms from (\ref{eq:proof_calD_posterior_Xk_likelihood}) we have
\begin{align}
    m(\boldsymbol{X}^{(k)}_1 \,|\, \boldsymbol{X}^{(k)}_{\Pa(1)}, \calD^{(k)}) = (2\pi)^{-\frac{n_k}{2}}
    \frac{|\boldsymbol{T}_1^{(k)}|^{1/2}}{|\bar{\boldsymbol{T}}_1^{(k)}|^{1/2}} \,.\, 
    \exp\left\{ -\frac{1}{2} \left( \boldsymbol{X}'^{(k)}_1 \boldsymbol{X}^{(k)}_1 - \hat{\boldsymbol{L}}'^{(k)}_1 \bar{\boldsymbol{T}}^{(k)}_1 \hat{\boldsymbol{L}}^{(k)}_1 \right) \right\}.
\end{align}
Substituting $f\left(\calD^{(k)} \,|\, \boldsymbol{X}^{(k)}, \boldsymbol{\theta}_{-\mathcal{U}} \right)$ into equation (\ref{eq:PAS_alg_rk}) completes the proof.

\subsubsection{Two Parent Nodes are Changed}
\label{sec:proof_acceptance_probability_calD_2parent}

As we explaned beforre, for the $\texttt{Reverse}(i\to j)$ operator, both $\Pa(i)$ and $\Pa(j)$ will change. So, we need to account for both nodes $i$ and $j$. In this case, $\mathcal{U} = (\sigma^2_i, \sigma^2_j, \boldsymbol{L}_{\preceq i \succ}, \boldsymbol{L}_{\preceq j \succ})$ and the set of parameters in PAS algorithm is $\boldsymbol{\theta}=(\boldsymbol{D}^{(k)}, \boldsymbol{L}^{(k)})$. For $j>1$, equation (\ref{eq:proof_PAS_int}) becomes
\begin{align}
    & f\left(\calD^{(k)} \,|\, \boldsymbol{X}^{(k)}, \boldsymbol{\theta}_{-\mathcal{U}} \right) 
    \propto f\left(\boldsymbol{X}^{(k)}, \boldsymbol{\theta}_{-\mathcal{U}} \,|\, \calD^{(k)} \right) f\left( \calD^{(k)} \right) \nonumber \\
    & = f\left( \calD^{(k)} \right) \int \int \int \int   f\left(\boldsymbol{X}^{(k)}, \boldsymbol{D}^{(k)}, \boldsymbol{L}^{(k)} | \calD^{(k)} \right) \, d\boldsymbol{L}^{(k)}_{\preceq i \succ} \,d\boldsymbol{L}^{(k)}_{\preceq j \succ} \, d\sigma^2_i  d\sigma^2_j \nonumber \\
    & = f\left( \calD^{(k)} \right) \prod_{r\neq i,j} f\left(\boldsymbol{X}^{(k)}_r, \sigma^2_r, \boldsymbol{L}^{(k)}_{\preceq r \succ} \Big| \calD^{(k)} \right)
    \times \int \int f\left(\boldsymbol{X}^{(k)}_i  \Big| \sigma^2_i, \boldsymbol{L}^{(k)}_{\preceq i \succ} \right)
    f\left(\boldsymbol{L}^{(k)}_{\preceq i \succ} | \sigma^2_i \right) 
    f\left(\sigma^2_i | \calD^{(k)}\right) \, d\boldsymbol{L}^{(k)}_{\preceq i \succ} \, d\sigma^2_i \nonumber \\
    & \times \int \int f\left(\boldsymbol{X}^{(k)}_j  \Big| \sigma^2_j, \boldsymbol{L}^{(k)}_{\preceq j \succ} \right)
    f\left(\boldsymbol{L}^{(k)}_{\preceq j \succ} | \sigma^2_j \right) 
    f\left(\sigma^2_j | \calD^{(k)}\right) \, d\boldsymbol{L}^{(k)}_{\preceq j \succ} \, d\sigma^2_j \nonumber \\
    & = f\left( \calD^{(k)} \right) \prod_{r\neq i,j} f\left(\boldsymbol{X}^{(k)}_r, \sigma^2_r, \boldsymbol{L}^{(k)}_{\preceq r \succ} \Big| \calD^{(k)} \right) \times 
    m(\boldsymbol{X}^{(k)}_i | \boldsymbol{X}^{(k)}_{\Pa(i)}, \calD^{(k)}) \,
    m(\boldsymbol{X}^{(k)}_j | \boldsymbol{X}^{(k)}_{\Pa(j)}, \calD^{(k)}),
\label{eq:proof_calD_two_parent_PAS}
\end{align}
where $m(.)$ is defined in (\ref{eq:proof_calD_one_parent_PAS_mj}). 

For $j=1$, we apply the same process that we used before to compute $m(\boldsymbol{X}^{(k)}_1 \,|\, \boldsymbol{X}^{(k)}_{\Pa(1)}, \calD^{(k)})$, which is defined in (\ref{eq:proof_calD_one_parent_PAS_m1}). Substituting $f\left(\calD^{(k)} \,|\, \boldsymbol{X}^{(k)}, \boldsymbol{\theta}_{-\mathcal{U}} \right)$ in equation (\ref{eq:PAS_alg_rk}) completes the proof for the $\texttt{Reverse}(i\to j)$ operator.

\subsection{Posteriors for $\boldsymbol{D}$ and $\boldsymbol{L}^{(k)}$ }
\label{sec:proof_posterior_D_L}

Assuming $\calD^{(1)}$ and $\calD^{(2)}$ are given, to compute the posteriors for $\boldsymbol{D}$ and $\boldsymbol{L}^{(k)}$, we need to start with the joint probability distribution
\begin{align}
    f\left(\boldsymbol{X}^{(1)}, \boldsymbol{X}^{(2)}, \boldsymbol{D}, \boldsymbol{L}^{(1)},\boldsymbol{L}^{(2)}\right)
    & = f\left(\boldsymbol{X}^{(1)}, \boldsymbol{X}^{(2)} \Big| \boldsymbol{D}, \boldsymbol{L}^{(1)},\boldsymbol{L}^{(2)}\right) f\left(\boldsymbol{L}^{(1)},  \boldsymbol{L}^{(2)} | \boldsymbol{D}\right) f\left(\boldsymbol{D}\right) \nonumber \\ 
    & = \prod_{j=1}^q f\left( \boldsymbol{X}^{(1)}_j, \boldsymbol{X}^{(2)}_j \Big| {\sigma^{2}_j}, \boldsymbol{L}^{(1)}_{\preceq j \succ},\boldsymbol{L}^{(2)}_{\preceq j \succ} \right) f\left( \boldsymbol{L}^{(1)}_{\preceq j \succ},\boldsymbol{L}^{(2)}_{\preceq j \succ} \Big| {\sigma^{2}_j} \right) f\left({\sigma^{2}_j} \right),
\end{align}
and because of the Markov property of Gaussian DAG's, we just need to find the posterior for node $j$. So
\begin{align}
    &f\left( \boldsymbol{X}^{(1)}_j, \boldsymbol{X}^{(2)}_j \Big| {\sigma^{2}_j}, \boldsymbol{L}^{(1)}_{\preceq j \succ},\boldsymbol{L}^{(2)}_{\preceq j \succ} \right) f\left( \boldsymbol{L}^{(1)}_{\preceq j \succ},\boldsymbol{L}^{(2)}_{\preceq j \succ} \Big| {\sigma^{2}_j} \right) f\left({\sigma^{2}_j} \right) \nonumber \\
    & =\frac{1}{(2\pi)^\frac{n_1}{2}}{|\sigma^2_j\boldsymbol{I}_{n_1}|}^{-\frac{1}{2}} \exp\left\{-\frac{1}{2\sigma^2_j} \left(\boldsymbol{X}^{(1)}_j + \boldsymbol{X}^{(1)}_{\Pa(j)}\boldsymbol{L}^{(1)}_{\preceq j \succ}\right)' \left(\boldsymbol{X}^{(1)}_j + \boldsymbol{X}^{(1)}_{\Pa(j)}\boldsymbol{L}^{(1)}_{\preceq j \succ}\right)\right\} \nonumber \\
    & \times \frac{1}{(2\pi)^\frac{n_2}{2}}{|\sigma^2_j\boldsymbol{I}_{n_2}|}^{-\frac{1}{2}} \exp\left\{-\frac{1}{2\sigma^2_j}\left(\boldsymbol{X}^{(2)}_j + \boldsymbol{X}^{(2)}_{\Pa(j)}\boldsymbol{L}^{(2)}_{\preceq j \succ}\right)' \left(\boldsymbol{X}^{(2)}_j + \boldsymbol{X}^{(2)}_{\Pa(j)}\boldsymbol{L}^{(2)}_{\preceq j \succ}\right)\right\} \nonumber \\
    & \times \frac{1}{{(2\pi)^\frac{{|\Pa^{(1)}(j)|}}{2}}} \Big|\frac{1}{g_1}\sigma^2_j\boldsymbol{I}_{|\Pa^{(1)}(j)|}\Big|^{-\frac{1}{2}} \exp\left\{-\frac{g_1}{2\sigma^2_j}(\boldsymbol{L}^{(1)}_{\preceq j \succ})' \boldsymbol{L}^{(1)}_{\preceq j \succ}\right\} \nonumber \\
    & \times \frac{1}{{(2\pi)^\frac{{|\Pa^{(2)}(j)|}}{2}}} \Big|\frac{1}{g_2}\sigma^2_j\boldsymbol{I}_{|\Pa^{(2)}(j)|}\Big|^{-\frac{1}{2}} \exp\left\{-\frac{g_2}{2\sigma^2_j}(\boldsymbol{L}^{(2)}_{\preceq j \succ})' \boldsymbol{L}^{(2)}_{\preceq j \succ}\right\} \nonumber \\
    & \times C \,.\, (\sigma^2_j)^{-\frac{a^{(1)}_j+a^{(2)}_j}{2}-1} \exp\left\{-\frac{\frac{g_1+g_2}{2}}{\sigma^2_j}\right\},
\label{eq:proof_posterior_X_D_L}
\end{align}
where $C = {(\frac{g_1+g_2}{2}) }^{\frac{a^{(1)}_j+a^{(2)}_j}{2}}/ \Gamma(\frac{a^{(1)}_j+a^{(2)}_j}{2})$, which is not a function of $\boldsymbol{D}^{(k)}$ and $\boldsymbol{L}^{(k)}$.

\subsubsection{Posterior for $ \boldsymbol{L}^{(k)}_{\preceq j \succ} | \sigma^2_j, \boldsymbol{X}^{(k)}$}
\label{sec:proof_posterior_L}

For each $k \in \{1,2\}$, the terms with $\boldsymbol{L}^{(k)}_{\preceq j \succ}$ in equation (\ref{eq:proof_posterior_X_D_L}) are
\begin{align}
    A^{(k)} := \exp\left\{-\frac{1}{2\sigma^2_j} \left(\boldsymbol{X}^{'(k)}_j \boldsymbol{X}^{(k)}_{\Pa(j)} \boldsymbol{L}^{(k)}_{\preceq j \succ} +{\boldsymbol{L}^{'(k)}_{\preceq j \succ}} \boldsymbol{X}^{'(k)}_{\Pa(j)} \boldsymbol{X}^{(k)}_j
    + \boldsymbol{L}^{'(k)}_{\preceq j \succ} \boldsymbol{X}^{'(k)}_{\Pa(j)} \boldsymbol{X}^{(k)}_{\Pa(j)} \boldsymbol{L}^{(k)}_{\preceq j \succ}
    + g_k \boldsymbol{L}^{'(k)}_{\preceq j \succ} \boldsymbol{L}^{(k)}_{\preceq j \succ} \right) \right\}.
\label{eq:proof_L_exp}
\end{align}
Using notations defined in (\ref{eq:Tj}), the exponent term in $A^{(k)}$ can be rewritten as 
\begin{align}
    & -\frac{1}{2\sigma^2_j} \left[ \boldsymbol{L}^{'(k)}_{\preceq j \succ} \left( \boldsymbol{X}^{'(k)}_{\Pa(j)} \boldsymbol{X}^{(k)}_{\Pa(j)} + g_k \boldsymbol{I}_{|\Pa^{(k)}(j)|} \right) \boldsymbol{L}^{(k)}_{\preceq j \succ}
    + \boldsymbol{L}^{'(k)}_{\preceq j \succ} \bar{\boldsymbol{T}}^{(k)}_j \left({\bar{\boldsymbol{T}}^{(k)}_j}\right)^{-1} \boldsymbol{X}^{'(k)}_{\Pa(j)} \boldsymbol{X}^{(k)}_j
    + \boldsymbol{X}^{'(k)}_j \boldsymbol{X}^{(k)}_{\Pa(j)} \left({\bar{\boldsymbol{T}}^{'(k)}_j}\right)^{-1} \bar{\boldsymbol{T}}^{(k)}_j  \boldsymbol{L}^{(k)}_{\preceq j \succ}
    \right] \nonumber \\
    & = -\frac{1}{2\sigma^2_j} \left[ \boldsymbol{L}^{'(k)}_{\preceq j \succ} \bar{\boldsymbol{T}}^{(k)}_j \boldsymbol{L}^{(k)}_{\preceq j \succ}
    + \boldsymbol{L}^{'(k)}_{\preceq j \succ} \bar{\boldsymbol{T}}^{(k)}_j \hat{\boldsymbol{L}}^{(k)}_j
    + \hat{\boldsymbol{L}}^{'(k)}_j \bar{\boldsymbol{T}}^{'(k)}_j  \boldsymbol{L}^{(k)}_{\preceq j \succ}
    \right],
\end{align}
which is proportional to the exponent of a normal distribution, i.e.,
\begin{align}
    A^{(k)} \propto \exp\left\{ -\frac{1}{2\sigma^2_j}
    \left( \boldsymbol{L}^{(k)}_{\preceq j \succ} + \hat{\boldsymbol{L}}^{(k)}_j \right)'
    \bar{\boldsymbol{T}}^{(k)}_j
    \left( \boldsymbol{L}^{(k)}_{\preceq j \succ} + \hat{\boldsymbol{L}}^{(k)}_j \right)
    \right\}.
\label{eq:Ak_normal}
\end{align}
In other words, equation (\ref{eq:proof_L_exp}) can be written as
\begin{align}
    \boldsymbol{L}_{\preceq j \succ}^{(k)} | \sigma^2_j, \boldsymbol{X}^{(k)} \sim \calN_{|\Pa^{(k)}(j)|} \left(-\hat{\boldsymbol{L}}'^{(k)}_j, \sigma^2_j (\bar{\boldsymbol{T}}^{(k)}_j)^{-1} \right),
\end{align}
which completes the proof.

For case when $j=1$, because we set $\sigma^2_1=1$, we have
\begin{align}
    \boldsymbol{L}_{\preceq 1 \succ}^{(k)} |  \boldsymbol{X}^{(k)} \sim \calN_{|\Pa^{(k)}(1)|} \left(-\hat{\boldsymbol{L}}'^{(k)}_1,  (\bar{\boldsymbol{T}}^{(k)}_1)^{-1} \right).
\end{align}

\subsubsection{Posterior for $\sigma^2_j| \boldsymbol{X}^{(1)}_j, \boldsymbol{X}^{(2)}_j$}
\label{sec:proof_posterior_D}

For $j=1$, we set $\sigma^2_1=1$, so we assume $j>1$ in this section. The terms with $\sigma^2_j$ in equation (\ref{eq:proof_posterior_X_D_L}) are
\begin{align}
    & (\sigma^2_j)^{-\frac{n_1}{2}-\frac{n_2}{2}-\frac{a^{(1)}_j}{2}-\frac{a^{(2)}_j}{2}-1-\frac{|\Pa^{(1)}(j)|}{2}-\frac{|\Pa^{(2)}(j)|}{2}} \nonumber \\
    & \times \exp\left\{-\frac{1}{2{\sigma^2_j}} \left({\boldsymbol{X}^{'(1)}_j} \boldsymbol{X}^{(1)}_j+{\boldsymbol{X}^{'(2)}_j} \boldsymbol{X}^{(2)}_j+g_1+g_2\right) \right\} \nonumber \\
    & \times \exp\left\{-\frac{1}{2\sigma^2_j} \left(\boldsymbol{X}^{'(1)}_j \boldsymbol{X}^{(1)}_{\Pa(j)} \boldsymbol{L}^{(1)}_{\preceq j \succ} +{\boldsymbol{L}^{'(1)}_{\preceq j \succ}} \boldsymbol{X}^{'(1)}_{\Pa(j)} \boldsymbol{X}^{(1)}_j + \boldsymbol{L}^{'(1)}_{\preceq j \succ} \boldsymbol{X}^{'(1)}_{\Pa(j)} \boldsymbol{X}^{(1)}_{\Pa(j)} \boldsymbol{L}^{(1)}_{\preceq j \succ} + g_1\boldsymbol{L}^{'(1)} \boldsymbol{L}^{(1)}_{\preceq j \succ} \right) \right\}  \nonumber \\
    & \times \exp\left\{-\frac{1}{2\sigma^2_j} \left(\boldsymbol{X}^{'(2)}_j \boldsymbol{X}^{(2)}_{\Pa(j)} \boldsymbol{L}^{(2)}_{\preceq j \succ} +{\boldsymbol{L}^{'(2)}_{\preceq j \succ}} \boldsymbol{X}^{'(2)}_{\Pa(j)} \boldsymbol{X}^{(2)}_j + \boldsymbol{L}^{'(2)}_{\preceq j \succ} \boldsymbol{X}^{'(2)}_{\Pa(j)} \boldsymbol{X}^{(2)}_{\Pa(j)} \boldsymbol{L}^{(2)}_{\preceq j \succ} + g_2\boldsymbol{L}^{'(2)} \boldsymbol{L}^{(2)}_{\preceq j \succ} \right) \right\}
\label{eq:exp_sigma}
\end{align}

Lets denote by $A^{(k)}$ the last two terms in the above equations for $k=1,2$, which are a function of $\boldsymbol{L}^{(k)}_{\preceq j \succ}$. To obtain the marginal posterior for $\sigma^2_j$, we need to integrate $A^{(k)}$ with respect to $\boldsymbol{L}^{(k)}_{\preceq j \succ}$. We showed that $A^{(k)}$ is proportional to the p.d.f of a normal distribution in equation (\ref{eq:Ak_normal}), so
\begin{equation}
    \int_{\mathbb{R}^{|\Pa^{(k)}(j)|}} {A^{(k)}} \,\, d\boldsymbol{L}^{(k)}_{\preceq j \succ} 
    = (2\pi)^{\frac{|\Pa^{(k)}(j)|}{2}} \,\,
    {\left| \frac{\bar{ \boldsymbol{T}}_j^{(k)}}{\sigma^2_j}\right|}^{-\frac{1}{2}}
    \exp\left\{\frac{1}{2\sigma^2_j} \hat{\boldsymbol{L}}^{'(k)}_j \bar{\boldsymbol{T}}^{(k)}_j \hat{\boldsymbol{L}}^{(k)}_j \right\}.
\label{eq:int_normal}
\end{equation}
Substituting (\ref{eq:int_normal}) in  (\ref{eq:exp_sigma}) and using the fact ${\left| \frac{ \bar{ \boldsymbol{T}}_j^{(k)}}{\sigma^2_j} \right|}^{-\frac{1}{2}} =  {\sigma^2_j}^{\frac{|\Pa^{(k)}(j)|}{2}}  {\left| \bar{\boldsymbol{T}}_j^{(k)} \right|}^{-\frac{1}{2}}$, we have
\begin{align}
    f(\sigma^2_j| \boldsymbol{X}^{(1)}_j, \boldsymbol{X}^{(2)}_j) & \propto ({\sigma^2_j})^{-\frac{n_1}{2}-\frac{n_2}{2}-\frac{a^{(1)}_j}{2}-\frac{a^{(2)}_j}{2}-1-\frac{|\Pa^{(1)}(j)|}{2}-\frac{|\Pa^{(2)}(j)|}{2}+\frac{|\Pa^{(1)}(j)|}{2}+\frac{|\Pa^{(2)}(j)|}{2}} \nonumber\\&
    \times \exp\left\{\frac{-\frac{1}{2}(g_1+g_2+ {\boldsymbol{X}^{'(1)}_j} \boldsymbol{X}^{(1)}_j+ {\boldsymbol{X}^{'(2)}_j} \boldsymbol{X}^{(2)}_j -\hat{\boldsymbol{L}}^{'(1)}_j \bar{\boldsymbol{T}}^{(1)}_j \hat{\boldsymbol{L}}^{(1)}_j- \hat{\boldsymbol{L}}^{'(2)}_j \bar{\boldsymbol{T}}^{(2)}_j \hat{\boldsymbol{L}}^{(2)}_j)}{\sigma^2_j}\right\}.
\end{align}
Therefore the posterior for $\sigma^2_j| \boldsymbol{X}^{(1)}_j, \boldsymbol{X}^{(2)}_j$ becomes
\begin{equation}
    \sigma^2_j| \boldsymbol{X}^{(1)}_j, \boldsymbol{X}^{(2)}_j \sim \mathrm{I-Ga}\left(\frac{a^{(1)}_j}{2} +\frac{a^{(2)}_j}{2} +\frac{n_1}{2} +\frac{n_2}{2}, \frac{1}{2}K\right),
\end{equation}
where $K=\frac{1}{2}(g_1+g_2+ {\boldsymbol{X}^{'(1)}_j} \boldsymbol{X}^{(1)}_j+ {\boldsymbol{X}^{'(2)}_j} \boldsymbol{X}^{(2)}_j -\hat{\boldsymbol{L}}^{'(1)}_j \bar{\boldsymbol{T}}^{(1)}_j \hat{\boldsymbol{L}}^{(1)}_j- \hat{\boldsymbol{L}}^{'(2)}_j \bar{\boldsymbol{T}}^{(2)}_j \hat{\boldsymbol{L}}^{(2)}_j)$.

\subsection{Posterior for $\theta \big| \boldsymbol{y}^{(1)}, \boldsymbol{y}^{(2)}, \boldsymbol{X}^{(1)}_{-1}, \boldsymbol{X}^{(2)}_{-1}, \boldsymbol{D}, \boldsymbol{L}^{(1)}, \boldsymbol{L}^{(2)}, \calD^{(1)}, \calD^{(2)}$ }
\label{sec:proof_posterior_theta}

The posterior for $\theta$ can be written as
\begin{equation}
    f\left(\theta \big| \boldsymbol{y}^{(1)}, \boldsymbol{y}^{(2)}, \boldsymbol{X}^{(1)}_{-1}, \boldsymbol{X}^{(2)}_{-1}, \boldsymbol{D}, \boldsymbol{L}^{(1)}, \boldsymbol{L}^{(2)}, \calD^{(1)}, \calD^{(2)} \right)
    \propto  f\left(\boldsymbol{y}^{(1)}, \boldsymbol{y}^{(2)}, \boldsymbol{X}^{(1)}_{-1}, \boldsymbol{X}^{(2)}_{-1} \big| \boldsymbol{D}, \boldsymbol{L}^{(1)}, \boldsymbol{L}^{(2)}, \calD^{(1)}, \calD^{(2)}, \theta\right) f(\theta),
\label{eq:proof_theta_full_posterior}
\end{equation}
where $f(\theta)\propto 1$.
To simplify the equations, we omit $\calD^{(1)}$ and $\calD^{(2)}$. By seperating $\boldsymbol{X}^{(k)}_1$, the full likelihood function can be written as
\begin{align}
    & f\left(\boldsymbol{y}^{(1)}, \boldsymbol{y}^{(2)}, \boldsymbol{X}^{(1)}, \boldsymbol{X}^{(2)} \big| \boldsymbol{D}, \boldsymbol{L}^{(1)}, \boldsymbol{L}^{(2)}, \theta \right) \nonumber \\
    & = \prod_{j=2}^q f_{\calN_{n_1}} \left(\boldsymbol{X}^{(1)}_j| - \boldsymbol{X}^{(1)}_{\Pa(j)} \boldsymbol{L}^{(1)}_{\preceq j \succ}, \sigma^2_{j} \boldsymbol{I}_{n_1} \right)
    . f_{\calN_{n_1}} \left(\boldsymbol{X}^{(1)}_1| - \boldsymbol{X}^{(1)}_{\Pa(1)} \boldsymbol{L}^{(1)}_{\preceq 1 \succ},  \boldsymbol{I}_{n_1} \right)
    . \prod_{i=1}^{n_1} \mathbbm{1}\left(\theta_{y^{(1)}_i-1} < x^{(1)}_{i,1} \leq \theta_{y^{(1)}_i}\right) \nonumber \\
    & \times \prod_{j=2}^q f_{\calN_{n_2}} \left(\boldsymbol{X}^{(2)}_j| - \boldsymbol{X}^{(2)}_{\Pa(j)} \boldsymbol{L}^{(2)}_{\preceq j \succ}, \sigma^2_{j} \boldsymbol{I}_{n_2} \right)
    . f_{\calN_{n_2}} \left(\boldsymbol{X}^{(2)}_1 | - \boldsymbol{X}^{(2)}_{\Pa(1)} \boldsymbol{L}^{(2)}_{\preceq 1 \succ},  \boldsymbol{I}_{n_2} \right)
    . \prod_{i=1}^{n_2} \mathbbm{1}\left(\theta_{y^{(2)}_i-1} < x^{(2)}_{i,1} \leq \theta_{y^{(2)}_i}\right) \nonumber \\
    & = \prod_{k=1}^2 f\left(\boldsymbol{X}^{(k)}_{-1} \big| \boldsymbol{D}, \boldsymbol{L}^{(k)}\right)
    \, f_{\calN_{n_k}} \left(\boldsymbol{X}^{(k)}_1| - \boldsymbol{X}^{(k)}_{\Pa(1)} \boldsymbol{L}^{(k)}_{\preceq 1 \succ},  \boldsymbol{I}_{n_k} \right)
    . \prod_{i=1}^{n_k} \mathbbm{1}\left(\theta_{y^{(k)}_i-1} < x^{(k)}_{i,1} \leq \theta_{y^{(k)}_i}\right),
\label{eq:proof_post_theta_eq2}
\end{align}
where $f\left(\boldsymbol{X}^{(k)}_{-1} \big| \boldsymbol{D}, \boldsymbol{L}^{(k)}\right)$ does not depend on $\theta$.

In order to get the marginal distribution, we need to take integral from both sides of (\ref{eq:proof_post_theta_eq2}) with respect to $\boldsymbol{X}^{(1)}_{1}$ and $\boldsymbol{X}^{(2)}_{1}$, i.e.,
\begin{align}
    & \int \int f\left(\boldsymbol{y}^{(1)}, \boldsymbol{y}^{(2)}, \boldsymbol{X}^{(1)}, \boldsymbol{X}^{(2)} \big| \boldsymbol{D}, \boldsymbol{L}^{(1)}, \boldsymbol{L}^{(2)}, \theta \right) \,\, d\boldsymbol{X}^{(1)}_{1} \,\,d\boldsymbol{X}^{(2)}_{1} \nonumber \\
    & \propto \prod_{k=1}^2 \int  f_{\calN_{n_k}} \left(\boldsymbol{X}^{(k)}_1| - \boldsymbol{X}^{(k)}_{\Pa(1)} \boldsymbol{L}^{(k)}_{\preceq 1 \succ},  \boldsymbol{I}_{n_k} \right)
    . \prod_{i=1}^{n_k} \mathbbm{1}\left(\theta_{y^{(k)}_i-1} < x^{(k)}_{i,1} \leq \theta_{y^{(k)}_i}\right) \,\, d\boldsymbol{X}^{(k)}_{1} \nonumber \\
    & = \prod_{k=1}^2 \prod_{i=1}^{n_k} \int  f_{\calN}\left({x}^{(k)}_{i,1}| - \boldsymbol{x}^{(k)}_{\Pa(1)} \boldsymbol{L}^{(k)}_{\preceq 1 \succ}, 1 \right)
    \mathbbm{1}\left(\theta_{y^{(k)}_i-1} < x^{(k)}_{i,1} \leq \theta_{y^{(k)}_i}\right) dx^{(k)}_{i,1},
\end{align}
where $\theta_{-1}:=-\infty, \theta_0:=\theta$, $\theta_1:=\infty$ and
\begin{align*}
    \int f_{\calN}\left({x}^{(k)}_{i,1}| - \boldsymbol{x}^{(k)}_{\Pa(1)} \boldsymbol{L}^{(k)}_{\preceq 1 \succ}, 1 \right)
    \mathbbm{1}\left(\theta_{y^{(k)}_i-1} < x^{(k)}_{i,1} \leq \theta_{y^{(k)}_i}\right) dx^{(k)}_{i,1} = \int_{\theta_{y^{(k)}_i-1}}^{\theta_{y^{(k)}_i}} f_{\calN}\left({x}^{(k)}_{i,1}| - \boldsymbol{x}^{(k)}_{\Pa(1)} \boldsymbol{L}^{(k)}_{\preceq 1 \succ}, 1 \right)\, dx^{(k)}_{i,1}
\end{align*}
is either the CDF or the survival function of a $\calN(- \boldsymbol{x}^{(k)}_{\Pa(1)} \boldsymbol{L}^{(k)}_{\preceq 1 \succ}, 1)$ ditsribution for $y_i^{(k)}=0$ and $y_i^{(k)}=1$, respectively.

Therefore,
\begin{align}
    f\left( \theta \big| \boldsymbol{y}^{(1)}, \boldsymbol{y}^{(2)}, \boldsymbol{X}^{(1)}_{-1}, \boldsymbol{X}^{(2)}_{-1}, \calD^{(1)}, \calD^{(2)}\right)
    \propto \prod_{k=1}^2 \prod_{i=1}^{n_k}\Psi\big(y^{(k)}_i,\theta | -\boldsymbol{x}^{(k)}_{\Pa(1)} \boldsymbol{L}^{(k)}_{\preceq 1 \succ}, 1 \big),
\end{align}
where $\Psi\big(y,\eta|\mu,\sigma^2\big):= |y - \Phi(\eta|\mu,\sigma^2\big)|$. In fact, $\Psi\big(y,\eta|\mu,\sigma^2\big)$ is a notation for the CDF or the survival function of a $\calN(\mu,\sigma^2)$ ditsribution for $y=0$ and $y=1$, respectively.

\subsection{Proof of proposition 1}
\label{sec:proof_proposition1}

We first need to reorder the matrix $\boldsymbol{X}$ as $\boldsymbol{X}^* =(\boldsymbol{X}_{-1}, \boldsymbol{X}_1)$ and let $\text{cov}(\boldsymbol{X}^*) = \boldsymbol{\Omega}^{-1}$.
The matrix $\boldsymbol{\Omega}$ can also be partitioned as
\begin{align}
    \boldsymbol{\Omega} & = 
    \left( \begin{array}{cc}
        \boldsymbol{\Omega}_{11} & \boldsymbol{\Omega}_{12} \\
        \boldsymbol{\Omega}_{21} & \boldsymbol{\Omega}_{22}
    \end{array} \right).
\label{eq:proof_om1}
\end{align}
Using Cholesky decomposition, we can decompose $\boldsymbol{\Omega}$ into
\begin{align}
    \boldsymbol{\Omega} & = \boldsymbol{L} \boldsymbol{L}' =
    \left( \begin{array}{cc}
        \boldsymbol{L}_{11} & \boldsymbol{0} \\
        \boldsymbol{L}_{21} & \boldsymbol{L}_{22}
    \end{array} \right)
    \left( \begin{array}{cc}
        \boldsymbol{L}'_{11} & \boldsymbol{L}'_{21} \\
        \boldsymbol{0} & \boldsymbol{L}'_{22}
    \end{array} \right).
\label{eq:proof_om2}
\end{align}
By equating the submatrices in (\ref{eq:proof_om1}) and (\ref{eq:proof_om2}), we have $\boldsymbol{\Omega}_{11} = \boldsymbol{L}_{11} \boldsymbol{L}'_{11}$, $\boldsymbol{\Omega}_{12} = \boldsymbol{L}_{11} \boldsymbol{L}'_{21}$ and $\boldsymbol{\Omega}_{22} = \boldsymbol{L}_{21} \boldsymbol{L}'_{21} + \boldsymbol{L}_{22}\boldsymbol{L}'_{22}$.These equations can be written as 
\begin{align}
\boldsymbol{L}_{11} &= \boldsymbol{\Omega}_{11}^{1/2} \nonumber \\ \boldsymbol{L}'_{21} &= \boldsymbol{\Omega}_{11}^{'-1/2} \boldsymbol{\Omega}_{12} \nonumber \\ \boldsymbol{L}_{22} &= \left( \boldsymbol{\Omega}_{22} - \boldsymbol{\Omega}_{11}^{-1/2} \boldsymbol{\Omega}_{12}' \boldsymbol{\Omega}_{12} \boldsymbol{\Omega}_{11}^{-1/2} \right)^{1/2},
\end{align}
where the square root for the matrix $A$ is defined by $A=A^{1/2}A^{'1/2}$.
On the other hand, $\boldsymbol{\Omega}_{-1}:=\boldsymbol{\Omega}_{11} = \text{cov}(\boldsymbol{X}_{-1})$ can be decomposed to $\boldsymbol{L}_{-1} \boldsymbol{L}'_{-1}$, so $\boldsymbol{\Omega}_{11}^{1/2}=\boldsymbol{L}_{-1}$, which completes the proof that $\boldsymbol{L}_{11}=\boldsymbol{L}_{-1}$.

\subsection{Simulation results}
\label{sec:Appendix_Simulation_results}

\begin{figure}
\centering
\includegraphics[width=\textwidth]{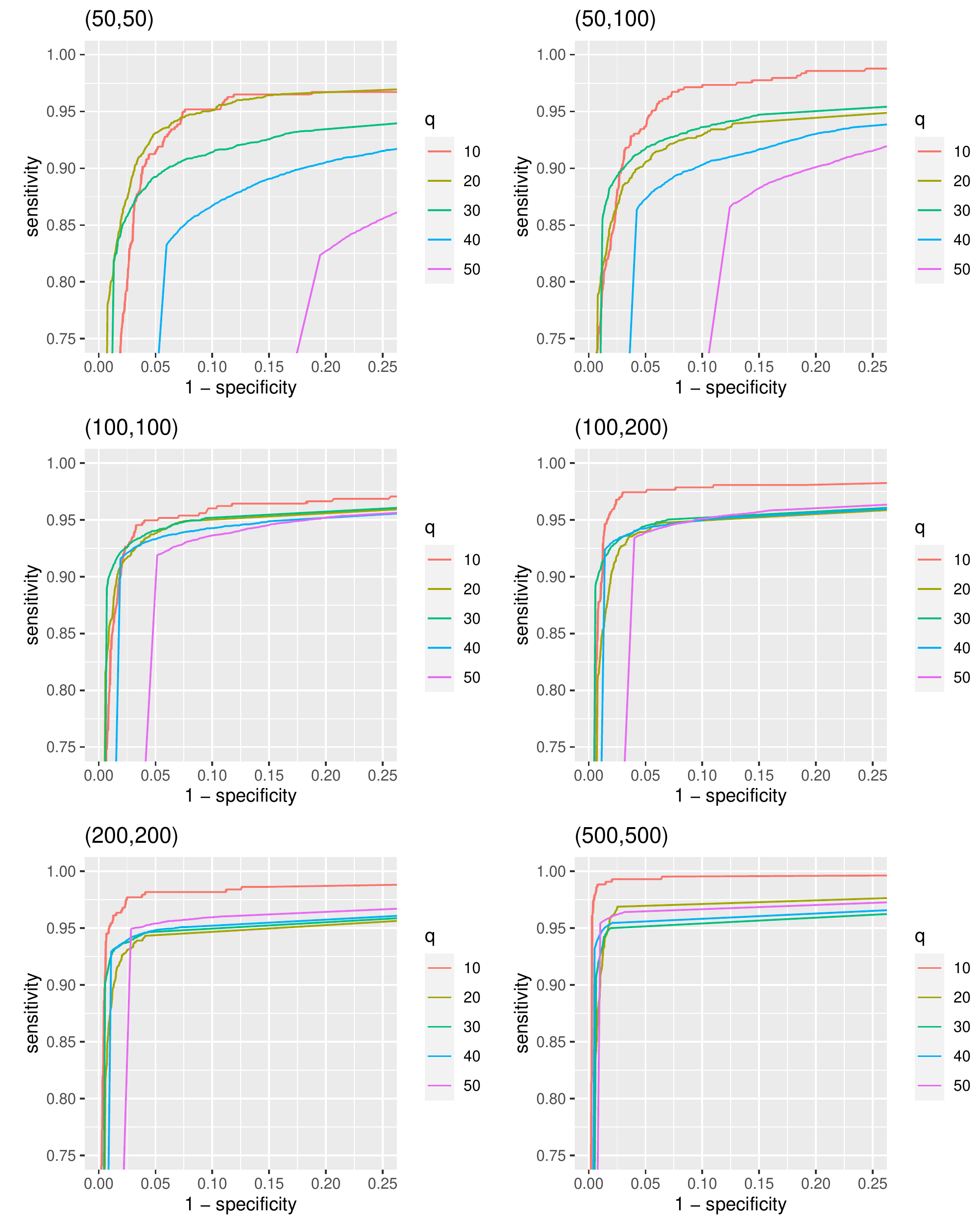}
\caption{ROC for $\xi=0.2$. Each plot is for different sample sizes $(n_1,n_2)$. Different colors represent different DAG sizes, $q$.}
\label{fig:sim_roc_2}
\end{figure}

\begin{table}
\centering
\caption{Area under the curve (AUC) for different  sample sizes and DAG sizes, $q$ and the probability of edge inclusion $\xi=0.2$.}
\begin{tabular}{|rr|ccccc|}
  \hline
  &  & \multicolumn{5} {c|} {$q$} \\ 
  $n_1$ & $n_2$ & 10 & 20 & 30 & 40 & 50 \\ 
  \hline
50 & 50 & 0.9679 & 0.9703 & 0.9488 & 0.9133 & 0.8307 \\ 
  50 & 100 & 0.9820 & 0.9574 & 0.9602 & 0.9345 & 0.8879 \\ 
  100 & 100 & 0.9740 & 0.9675 & 0.9686 & 0.9598 & 0.9442 \\ 
  100 & 200 & 0.9850 & 0.9661 & 0.9694 & 0.9651 & 0.9550 \\ 
  200 & 200 & 0.9895 & 0.9664 & 0.9689 & 0.9679 & 0.9638 \\ 
  500 & 500 & 0.9959 & 0.9802 & 0.9711 & 0.9744 & 0.9765 \\ 
  1000 & 500 & 0.9960 & 0.9795 & 0.9707 & 0.9746 & 0.9759 \\ 
  1000 & 1000 & 0.9971 & 0.9749 & 0.9695 & 0.9751 & 0.9789 \\ 
   \hline
\end{tabular}
\label{tbl:auc_xi2}
\end{table}

\begin{figure}
\centering
\includegraphics[width=\textwidth]{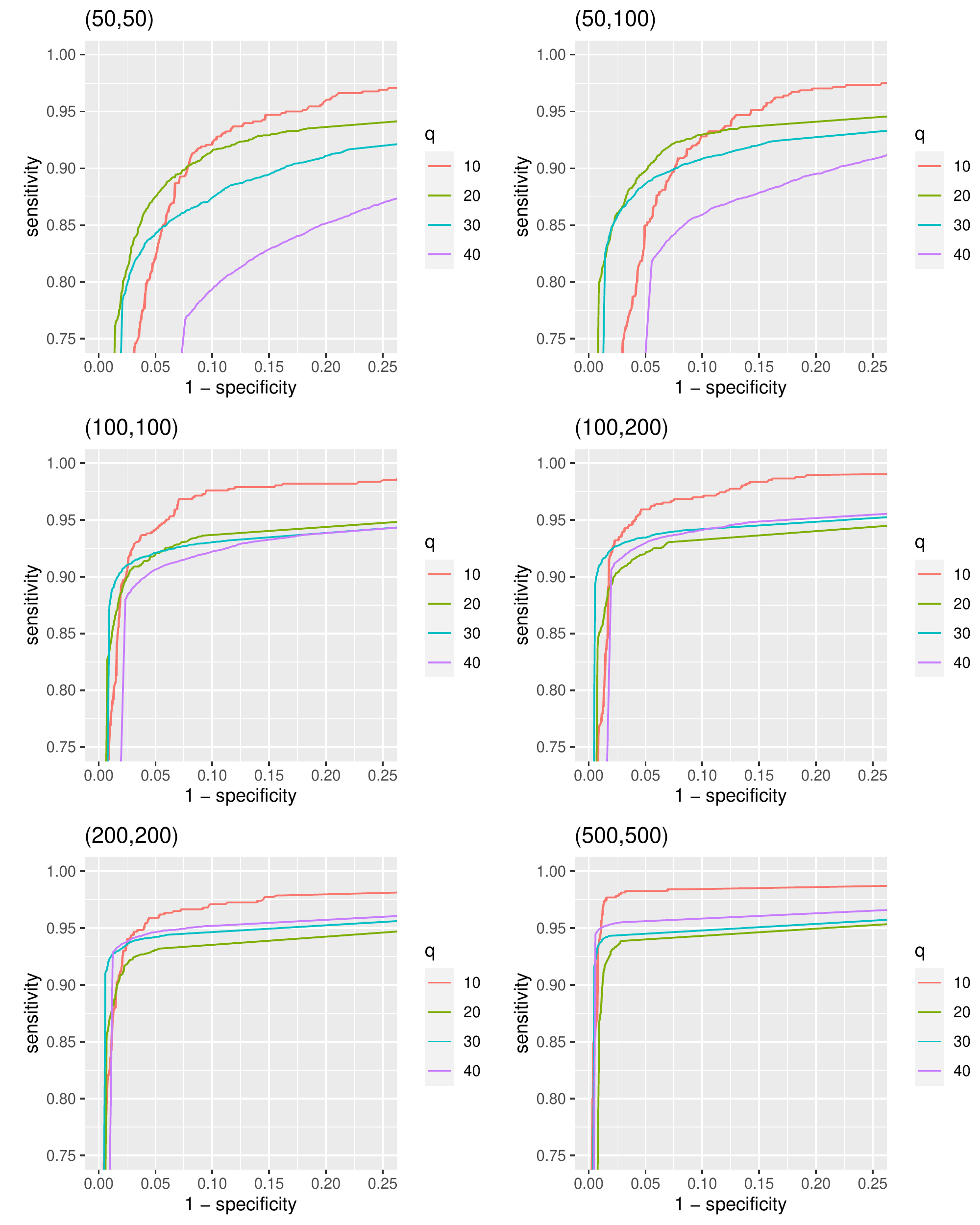}
\caption{ROC for $\xi=0.3$. Each plot is for different sample sizes $(n_1,n_2)$. Different colors represent different DAG sizes, $q$.}
\label{fig:sim_roc_3}
\end{figure}

\begin{table}
\centering
\caption{Area under the curve (AUC) for different  sample sizes and DAG sizes, $q$ and the probability of edge inclusion $\xi=0.3$.}
\begin{tabular}{|rr|cccc|}
  \hline
  &  & \multicolumn{4} {c|} {$q$} \\ 
  $n_1$ & $n_2$ & 10 & 20 & 30 & 40 \\ 
  \hline
50 & 50 & 0.9597 & 0.9473 & 0.9303 & 0.8784 \\ 
  50 & 100 & 0.9636 & 0.9550 & 0.9446 & 0.9097 \\ 
  100 & 100 & 0.9811 & 0.9591 & 0.9562 & 0.9484 \\ 
  100 & 200 & 0.9847 & 0.9573 & 0.9644 & 0.9589 \\ 
  200 & 200 & 0.9813 & 0.9598 & 0.9671 & 0.9670 \\ 
  500 & 500 & 0.9887 & 0.9635 & 0.9686 & 0.9740 \\ 
  1000 & 500 & 0.9857 & 0.9580 & 0.9669 & 0.9741 \\ 
  1000 & 1000 & 0.9847 & 0.9585 & 0.9680 & 0.9734 \\ 
   \hline
\end{tabular}
\label{tbl:auc_xi3}
\end{table}

\begin{figure}
\centering
\includegraphics[width=\textwidth]{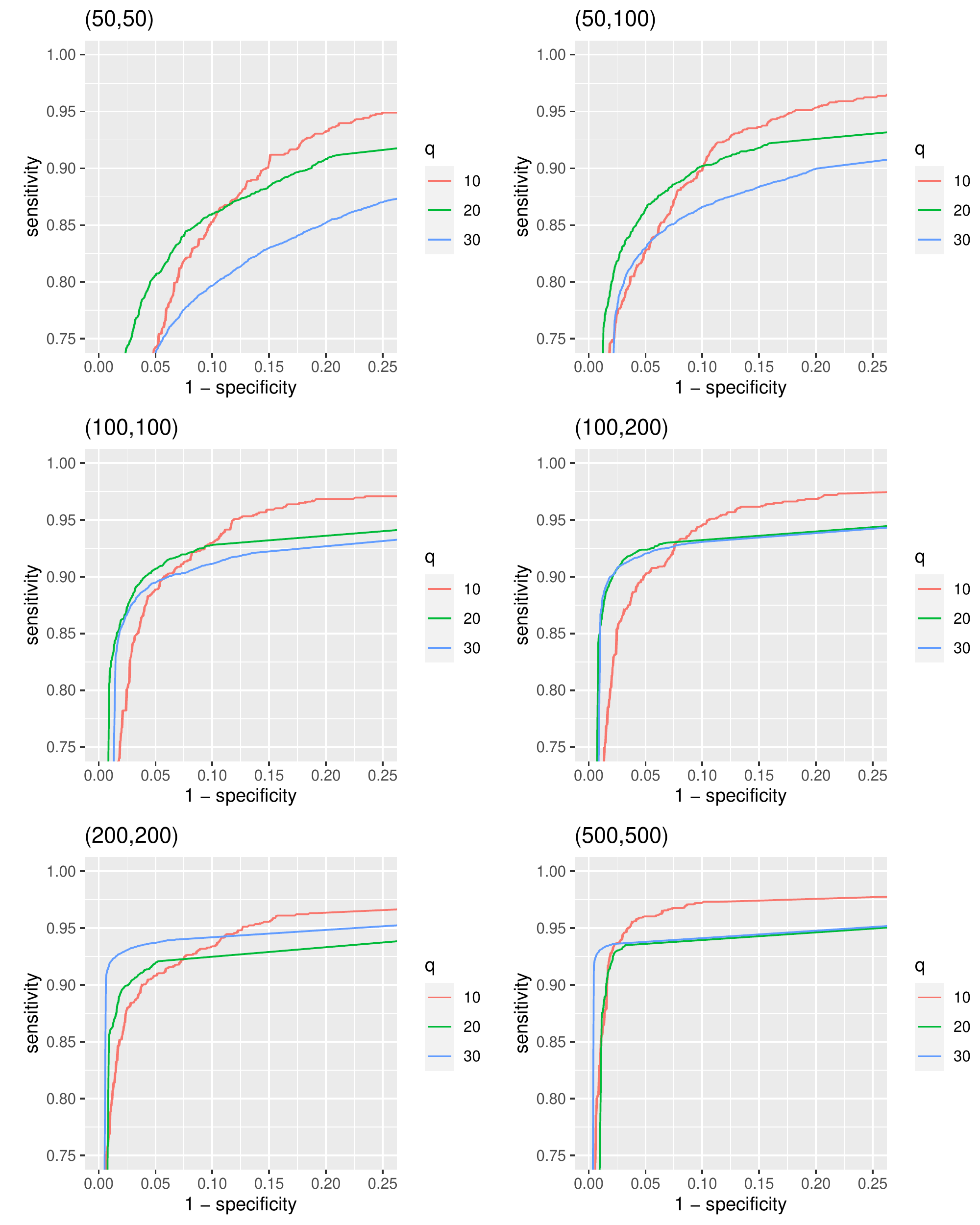}
\caption{ROC for $\xi=0.4$. Each plot is for different sample sizes $(n_1,n_2)$. Different colors represent different DAG sizes, $q$.}
\label{fig:sim_roc_4}
\end{figure}

\begin{table}
\centering
\caption{Area under the curve (AUC) for different  sample sizes and DAG sizes, $q$ and the probability of edge inclusion $\xi=0.4$.}
\begin{tabular}{|rr|ccc|}
  \hline
  &  & \multicolumn{3} {c|} {$q$} \\ 
  $n_1$ & $n_2$ & 10 & 20 & 30 \\ 
  \hline
50 & 50 & 0.9416 & 0.9259 & 0.8880 \\ 
  50 & 100 & 0.9606 & 0.9418 & 0.9215 \\ 
  100 & 100 & 0.9667 & 0.9530 & 0.9449 \\ 
  100 & 200 & 0.9692 & 0.9572 & 0.9556 \\ 
  200 & 200 & 0.9671 & 0.9530 & 0.9644 \\ 
  500 & 500 & 0.9788 & 0.9605 & 0.9651 \\ 
  1000 & 500 & 0.9765 & 0.9569 & 0.9661 \\ 
  1000 & 1000 & 0.9855 & 0.9600 & 0.9705 \\ 
   \hline
\end{tabular}
\label{tbl:auc_xi4}
\end{table}

\begin{figure}
\centering
\includegraphics[width=\textwidth]{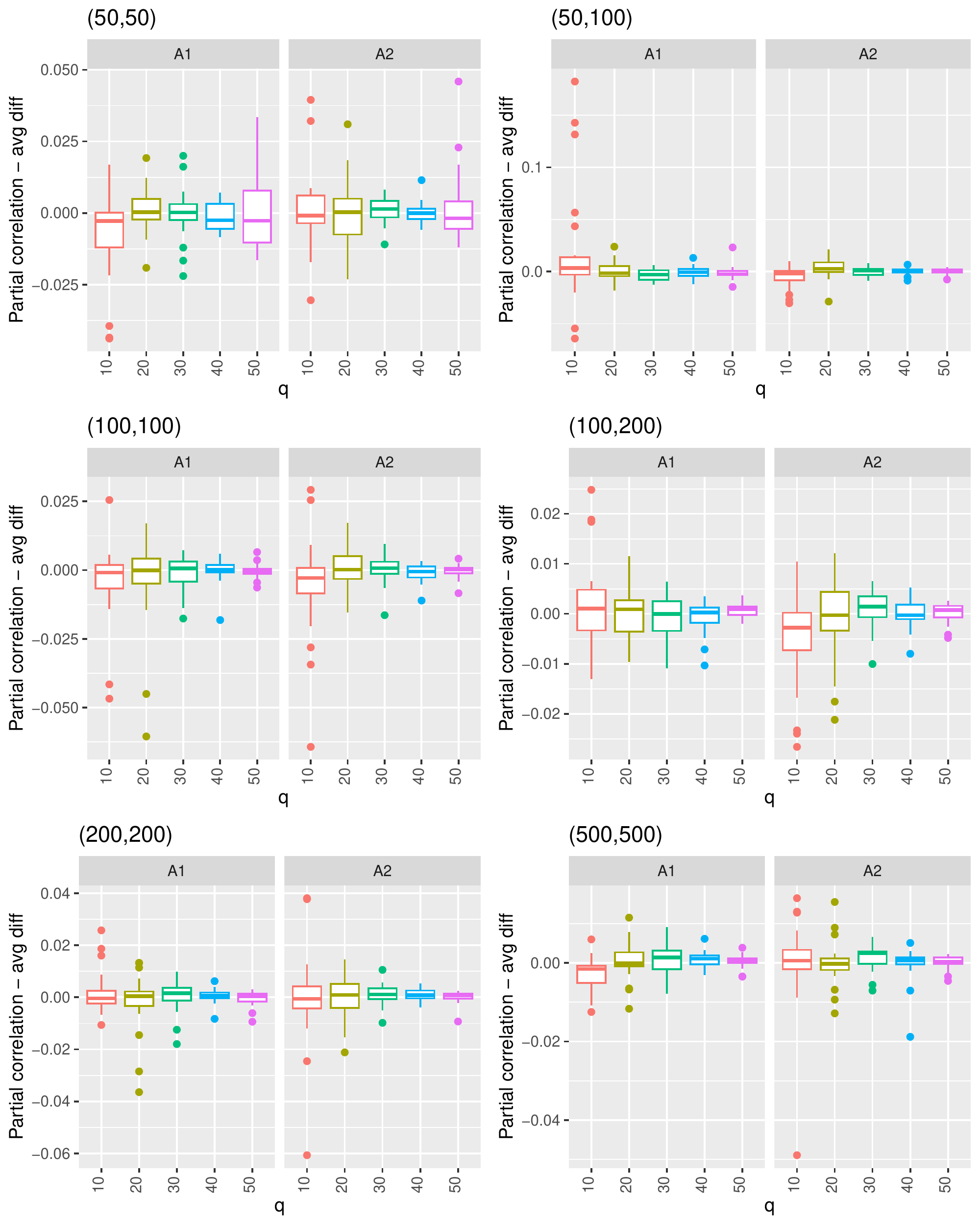}
\caption{Box plot for the difference between the estimated and true partial correlations defined in (\ref{eq:mcmc_end_partial}) for $T=5,000$ iterations and $\xi=0.2$.}
\label{fig:sim_Partial_mean_edge0.2}
\end{figure}

\begin{figure}
\centering
\includegraphics[width=\textwidth]{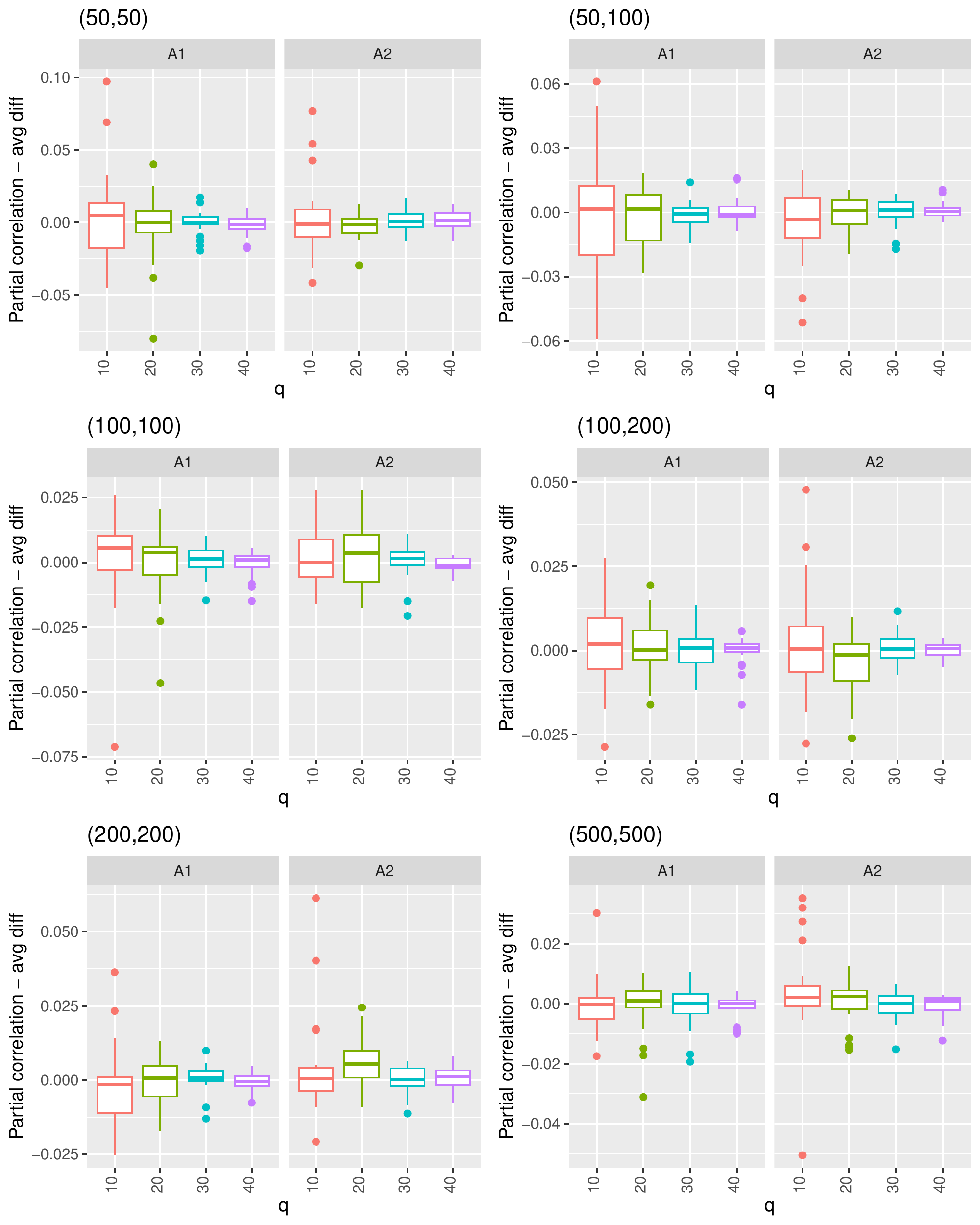}
\caption{Box plot for the difference between the estimated and true partial correlations defined in (\ref{eq:mcmc_end_partial}) for $T=5,000$ iterations and $\xi=0.3$.}
\label{fig:sim_Partial_mean_edge0.3}
\end{figure}

\begin{figure}
\centering
\includegraphics[width=\textwidth]{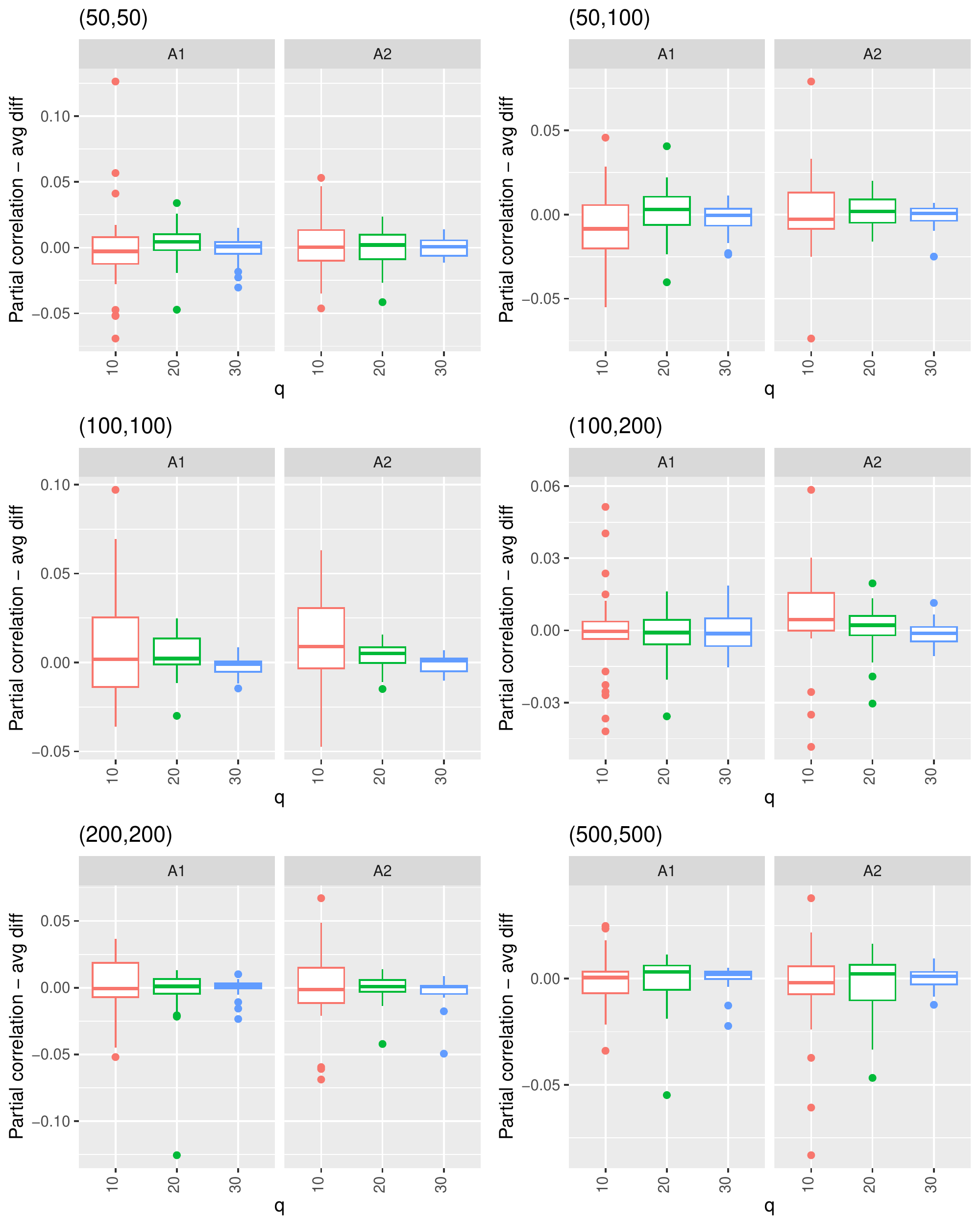}
\caption{Box plot for the difference between the estimated and true partial correlations defined in (\ref{eq:mcmc_end_partial}) for $T=5,000$ iterations and $\xi=0.4$.}
\label{fig:sim_Partial_mean_edge0.4}
\end{figure}


\begin{figure}
\centering
\includegraphics[width=\textwidth]{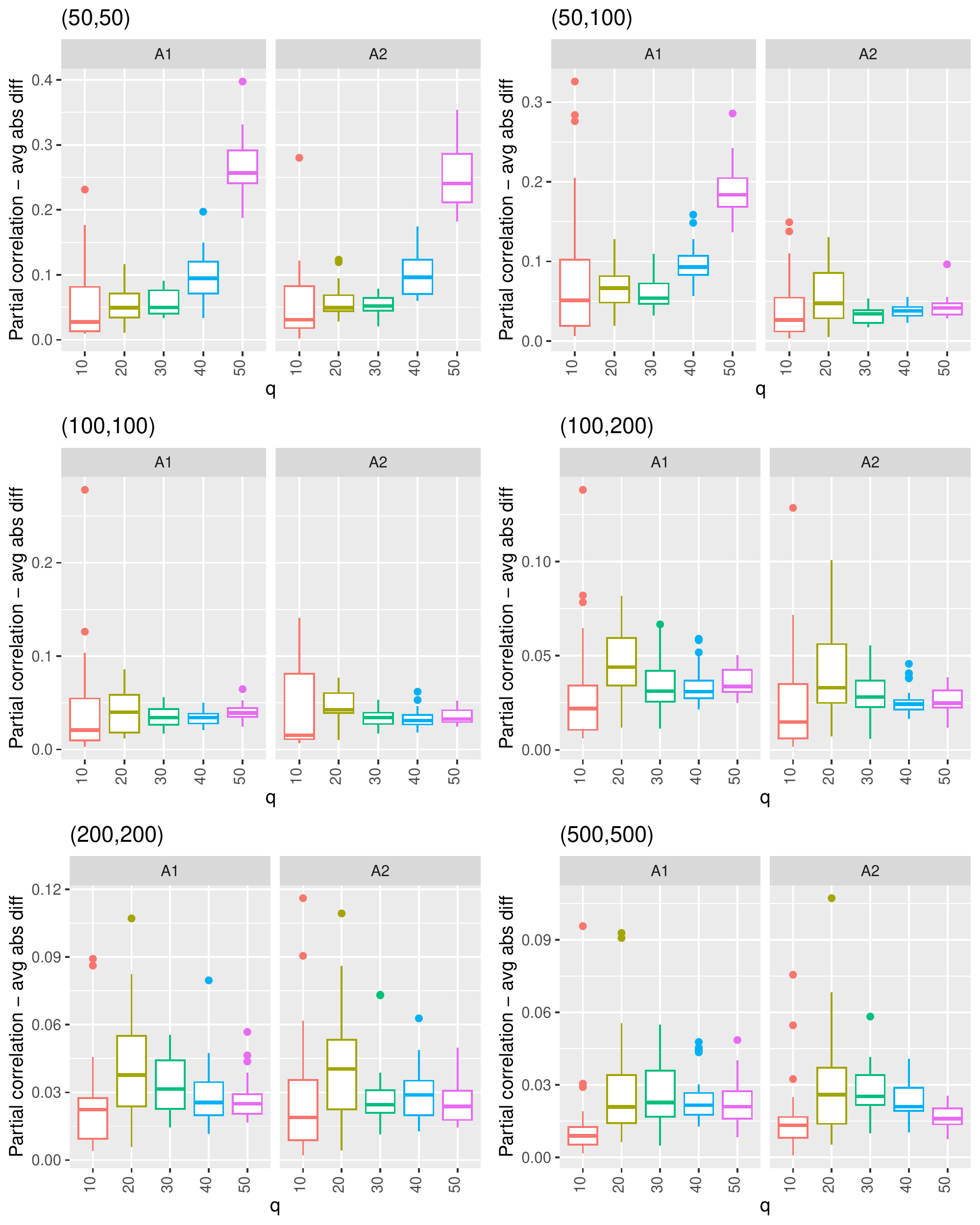}
\caption{Box plot for the absolute difference between the estimated and true partial correlations defined in (\ref{eq:mcmc_end_partial}) for $T=5,000$ iterations and $\xi=0.2$.}
\label{fig:sim_Partial_meanabs_edge0.2}
\end{figure}

\begin{figure}
\centering
\includegraphics[width=\textwidth]{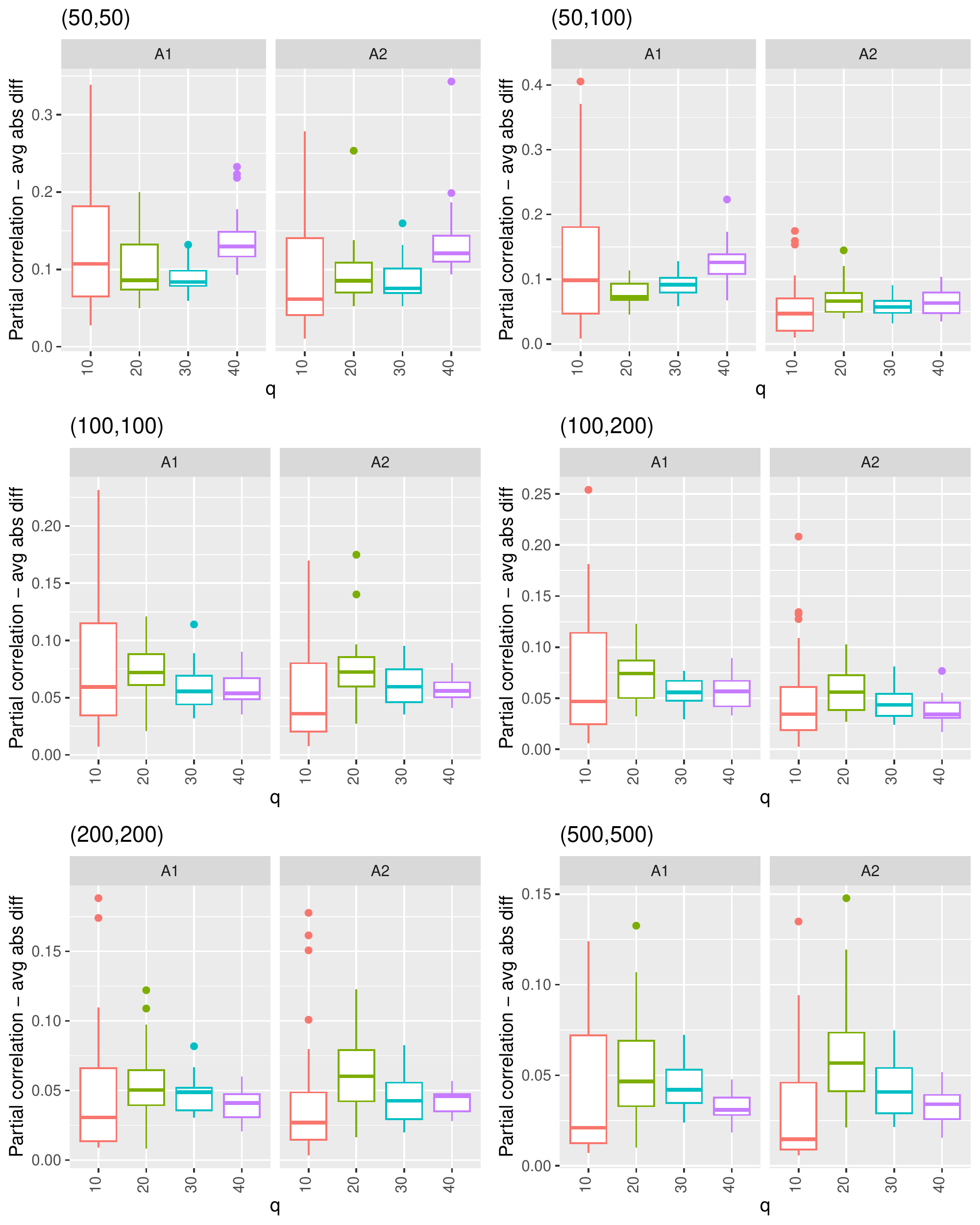}
\caption{Box plot for the absolute difference between the estimated and true partial correlations defined in (\ref{eq:mcmc_end_partial}) for $T=5,000$ iterations and $\xi=0.3$.}
\label{fig:sim_Partial_meanabs_edge0.3}
\end{figure}

\begin{figure}
\centering
\includegraphics[width=\textwidth]{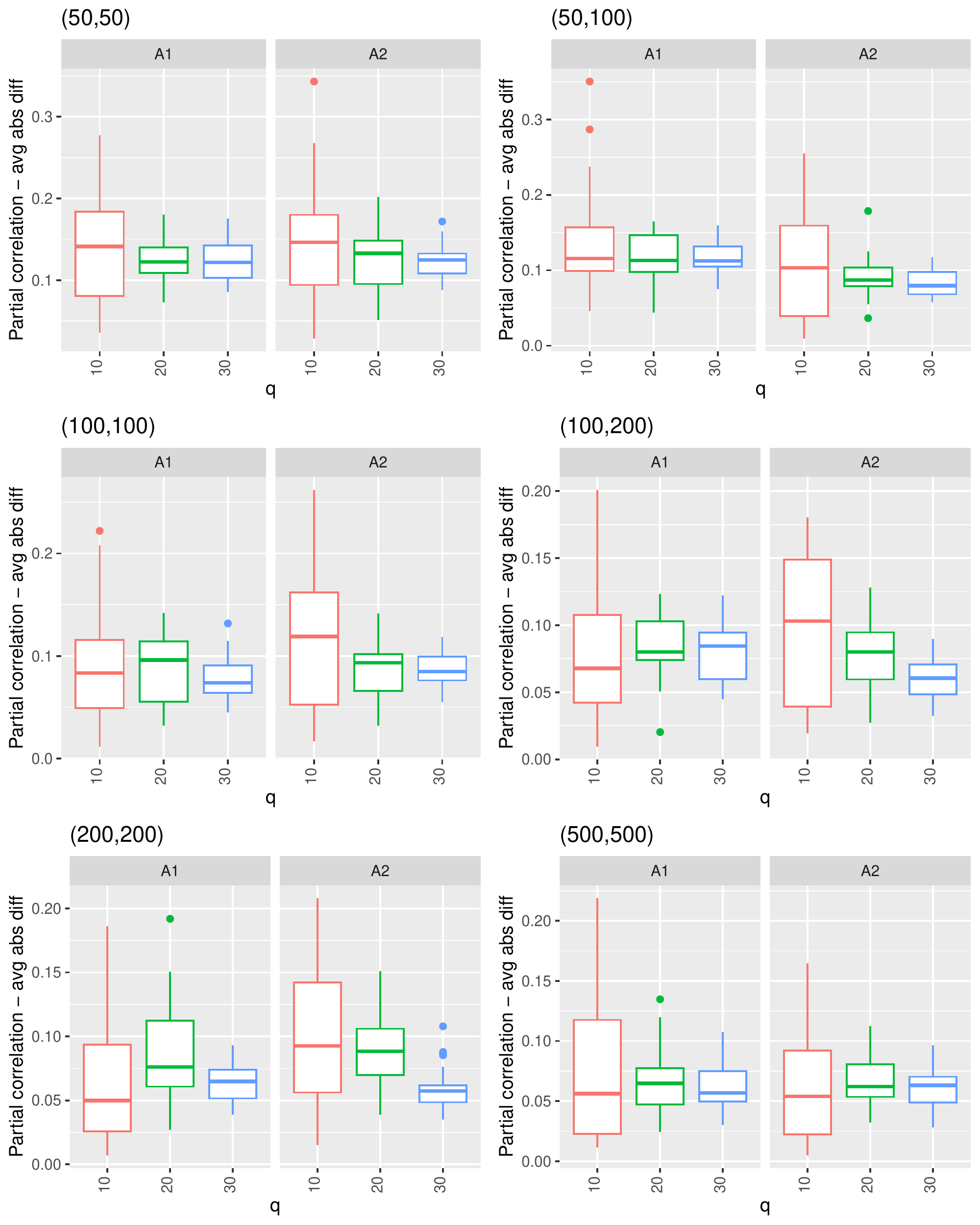}
\caption{Box plot for the absolute difference between the estimated and true partial correlations defined in (\ref{eq:mcmc_end_partial}) for $T=5,000$ iterations and $\xi=0.4$.}
\label{fig:sim_Partial_meanabs_edge0.4}
\end{figure}

\end{document}